\documentclass[twocolumn]{aastex62}

\usepackage{amsmath}
\graphicspath{{./}{figures/}}

\shorttitle{Systematics in atmospheric retrieval}
\shortauthors{Ih and Kempton}
\begin{document}

\title{Understanding the Effects of Systematics in Exoplanetary Atmospheric Retrievals}

\correspondingauthor{Jegug Ih}
\email{jegugih@astro.umd.edu}

\author[0000-0002-0786-7307]{Jegug Ih}
\affil{Department of Astronomy \\
University of Maryland \\
College Park, MD 20742, USA}

\author{Eliza M.-R. Kempton}
\affiliation{Department of Astronomy \\
University of Maryland \\
College Park, MD 20742, USA}



\begin{abstract}

Retrieval of exoplanetary atmospheric properties from their transmission spectra commonly assumes that the errors in the data are Gaussian and independent. However, non-Gaussian noise can occur due to instrumental or stellar systematics and merging discrete datasets. We investigate the effect of correlated noise and constrain the potential biases incurred in the retrieved posteriors. We simulate multiple noise instances of synthetic data and perform retrievals to obtain statistics of goodness-of-retrieval for varying noise models. We find that correlated noise allows for overfitting the spectrum, thereby yielding better goodness-of-fit on average but degrading the overall accuracy of retrievals. In particular, correlated noise can manifest as an apparent non-Rayleigh slope in the optical range, leading to an incorrect estimate of cloud/haze parameters. We also find that higher precision causes correlated results to be further off from the input values in terms of estimated errors.  As such, we emphasize that caution must be taken in analyzing retrieved posteriors and that estimated parameter uncertainties are best understood as lower limits.  Finally, we show that while correlated noise cannot be be reliably distinguished with \textit{HST} observations, inferring its presence and strength may be possible with \textit{JWST} observations.

\end{abstract}



\section{Introduction} \label{sec:intro}

Inverse modelling the transmission spectra of exoplanets allows for extracting information about various properties and processes in the atmosphere. This is commonly done by pairing a forward model, which generates a spectrum from atmospheric parameters, and a parameter estimation scheme, which samples the parameter space to compute the probability distribution of the set of parameters. This method of analyzing observed spectra, called atmospheric retrieval,  {originally developed for remote sensing in Solar System bodies,} \citep[e.g.][]{rod00} has become a standard tool in characterizing exoplanetary atmospheres and has allowed for measuring abundances of various species and identifying atmospheric phenomena such as the presence of clouds and thermal inversions \citep[e.g.][]{lin13, wal15, har16, mad18, zha19, ben19, bar20a}.

Recently, there has been a growing body of work that addresses the potential to be misled by incomplete physics or simplifying assumptions used in retrievals, often invoked to speed up the computation and make the retrieval feasible \citep[e.g.][]{pin18, cal19, cha19b, bar20b, lac20, tay20}. These studies are especially germane in preparation for interpreting \textit{James Webb Space Telescope} (\textit{JWST}) observations, the precision of which will now render the finer details of the model consequential. Such details include 3-dimensional atmospheric structure, host star effects, aerosol models, and disequilibrium chemistry. The general method used in the aforementioned works is to retrieve on a synthetic spectrum generated by a more complex and complete forward model and then inspect the retrieval result to quantify what bias may be incurred, with the aim of making judicious choices as to how and which model complexity and compromises should be introduced to the retrieval's forward model.

Adopting a similar approach, we focus on a separate but related issue in this work. A universal assumption made in atmospheric retrieval is that the reported errors in the observed spectrum are Gaussian and independent. This assumption is encoded into the cost function one tries to minimize during the parameter estimation, which is invariably a chi-squared statistic that does not take the covariance between the residuals into account \citep{and10, lin13,zha19}, While this assumption is a reasonable starting point for analyzing observations, as data quality reaches unprecedented precision and as retrievals incorporate increasingly sophisticated forward models and more rigorous statistical methods, it is necessary to understand the significance of the assumption of independent errors. 

\begin{figure}[h]
    \includegraphics[width=\linewidth]{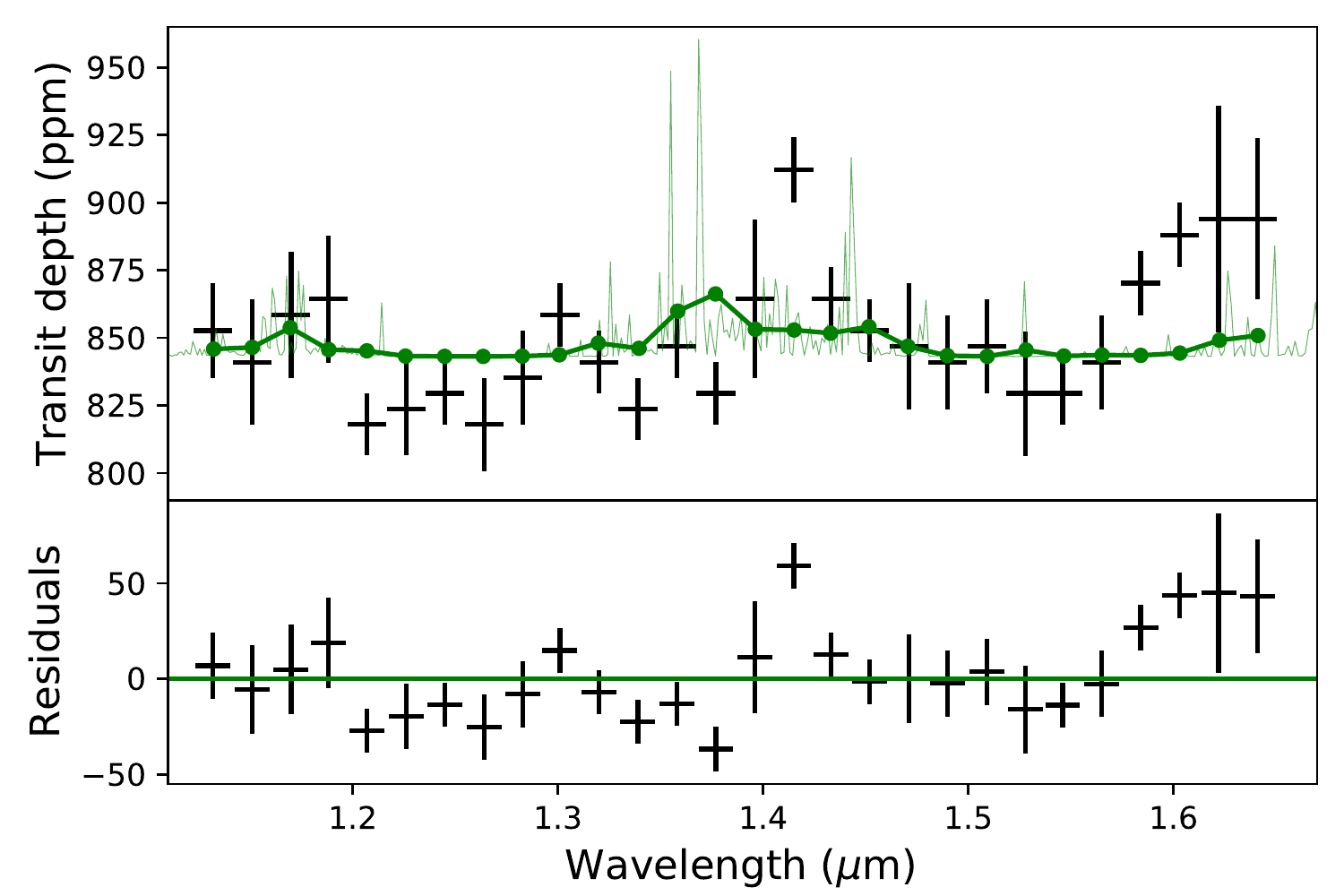}
    \caption{The observed WFC3 data of HD 97658b \citep{guo20} and the best-fit \textsc{PLATON} spectrum in full-resolution and binned (solid line). The residuals from the fit are also shown in the bottom panel. The best-fit $\chi^2$ is 56, and reduced $\chi^2_{\textrm{r}}=2.5$. An unusual upward trend in the residuals is present in the long wavelength limit, and the residuals appear to be correlated with wavelength.}
    \label{fig:HD97658b}
\end{figure}

As further motivation for this work, there have been observations that hint at the extent to which errors may be correlated with wavelength.  To pull one such example from the literature, we identify the observed spectrum of HD 97658b with the WFC3 instrument on \textit{Hubble} \citep{guo20}.  The retrieval on this dataset strongly favors either high-metallicity or cloudy atmospheres, corresponding with a nearly featureless transmission spectrum (a flat line).  However, none of the forward models considered by \citet{guo20} provide an excellent fit to the data.  For example, we show the best-fit \textsc{platon} \citep{zha19} spectrum in Figure~\ref{fig:HD97658b}, which yields a  reduced chi-squared of 2.5 and is ruled out by the data at 4.9-$\sigma$.  (We note that \citet{guo20} \textit{can} find models that provide a high-quality fit to their data by scaling their formal error bars by a factor $>$ 1 --- a practice that we weigh in on later in this paper and that we do not endorse.)  As can be seen in Figure~\ref{fig:HD97658b}, the best-fit spectrum produces residuals that are possibly correlated. One rudimentary method of quantifying correlation is to count the number of zero-crossings, which should follow a symmetric binomial distribution if the noise were independent. Additionally, an unusual upward slope is seen in the residual in the redmost edge. A similar behavior was seen in the observed spectrum of KELT-11b \citep[see][Figure 20]{col20}. This could be attributed to either complicated physics unaccounted for in the forward model or the presence of correlated noise in the data.  

%In addition, there have been observations that hint at the extent to which the errors may be correlated. One such example is the observed spectrum of HD 97658b \citep{guo20} in the WFC3 band. The retrieval strongly favors either high-metallicity or cloudy atmosphere, such that a nearly featureless spectrum (a flat line) is produced. However, with a best reduced chi-squared of 2.5, but yields correlated residuals (Figure \ref{fig:HD97658b}). This could be attributed to either complicated physics unaccounted for in the forward model or the presence of correlated noise in the data.

Potential sources of correlated noise are numerous and include faulty data reduction due to incorrect orbital parameters or incomplete subtraction of stellar contributions \citep{rac18, bru19}. Choices made during the data reduction, such as the choice of a model to remove visit-long trends, can potentially produce a wavelength-dependent correlated effect in the spectrum \citep[see][Figure 7]{guo20}. More fundamentally, the removal of instrumental systematics such as ramping or horizontal shifts in \textit{Hubble} data has intrinsic uncertainties and may potentially manifest themselves in wavelength-dependent manner \citep{tsi16}.

For space-based observation facilities, there are some reasons from observer experience to suspect that correlated noise is more likely in the case of a very bright host star, for which the instrumental systematics either behave differently or become more apparent due to lower photon noise. In \citet[][Figure 9]{ste19}, it can be seen that no observations with bright host stars of J-band magnitude $\leq 8.5$ meet the ideal precision per orbit, instead maxing out at $\sim$35 ppm. (Interestingly, our previously-discussed example case of HD 97658b fits into this category, with a J-magnitude of 6.2.) This effect has been attributed to unaccounted for wavelength-dependent systematics, that have  no guaranteed way of being completely modelled out. In the case of ground-based observations, the highly time-dependent telluric contamination may also be a potential source of correlated systematics.

%For space-based observation facilities, there is some reason from observer experience to suspect that this is more likely in case of a very bright host star, where the instrumental systematics behave differently. In \citet[Figure 9]{ste19}, it can be seen that no observations with bright host stars of J-band magnitude $\leq 8.5$ meet the ideal precision per orbit, instead maxing out $\sim 35$ppm. This has been attributed to unaccounted wavelength-dependent systematics, for which there is no guaranteed way of completely modelling out ( {maybe rewrite this more politely?}). In case of ground-based observations, the highly time-dependent telluric contamination may also be a potential source of correlated systematics.

A separate, but related cause of wavelength-dependent correlation in data errors arises when combining data from various instruments to gain a wider wavelength coverage. Each instrument has its own instrumental systematics and data reduction pipeline, leading to distinct noise statistics among datasets. More fundamentally, these observations are not simultaneous and are hence subject to differing conditions with respect to stellar and planet variability. An insufficient number of observed transits may admit such variability in the data, even if the desired signal-to-noise ratio is formally achieved. Some retrieval analyses have included \textit{ad hoc} ``offset'' parameters, which vertically shift all measured transit depths from one dataset by a variable amount, to fit for the discrepancies between datasets, but doing so can induce bias in the estimation of other parameters \citep{yip20}.

% Vestigial TODO: *maybe* add a paragraph or two on other types of biases: (1) previous work showing potential biases in retrievals due to missing physics (e.g. 3D forward model into 1D retrieval, see Caldas et al. 2019; or non-eqm mixing ratio into eqm retrieval) and (2) previous work showing that the combination of multiple noise-instance retrievals obviously converge to the noise-free retrieval posterior distribution, as expected from the central limit theorem, and that otherwise there are potential biases here.(copied from Heng) (Heng et al.  2018.: only had N=10 tho; and Lupu et al. 2016.: considered sensitivity and possible correlation; Henderson et al. 2018.: more sophisticated method for underestimated errors->doesn't solve correlation! ) eh

Another issue arises in how outliers in the observed data are interpreted in a retrieval. After fitting a spectrum to data, the presence of anomalous outliers in the residuals is certainly within expectation of what can happen, and statistical methods such as bootstrapping, though rarely used in retrieval studies, do offer objective criteria to exclude these outliers. However, the fact that we rely on the best-fit spectrum to determine outliers raises the question if a statistically equivalent datum could have been accommodated for as a detected feature if it happened to have occurred at where we expect one. This problem is especially pertinent in the context of retrievals with non-equilibrium models. It particularly affects resolution-limited observations and species with single, narrow features, e.g.\ atomic lines such as Na or K for which only one or two data points dictate the retrieved abundance.

Given the number of potential issues raised above, in the present work we aim to address the question: how reliable are our atmospheric retrievals and what are best practices in the face of these idiosyncratic data systematics? We perform atmospheric retrievals on simulated spectra, while varying the noise properties, and conduct a detailed analysis of the retrieved parameters. Such an analysis provides a statistical context in which one can assess the credibility of a retrieval beyond a raw retrieved posterior. In what follows, in \S \ref{sec:methods} we describe our planet parameters and noise models used; in \S \ref{sec:results}, we present our findings in how retrievals are affected;  in \S \ref{sec:if}, we test whether correlation can be retrieved during retrieval; in \S \ref{sec:discussion}, we discuss how we might be able to better understand the sources of these systematics and implications for future telescopes; in \S \ref{sec:summary} we summarize and conclude.

\section{Methods} \label{sec:methods}

\subsection{Framework for Statistics of Retrievals}
To study how non-Gaussian error can bias the retrieval results, we perform retrievals on simulated data generated with and without correlation in the noise. Here, by using the same forward model to create the synthetic spectrum and to retrieve on it, we remove model dependencies as much as possible and are able to examine the bias due to noise in isolation. To obtain statistics of retrievals, we use the following procedure (also shown schematically in Figure~\ref{fig:method_schema}):

\begin{figure*}[!ht]
    \centering
    \includegraphics[width=5in]{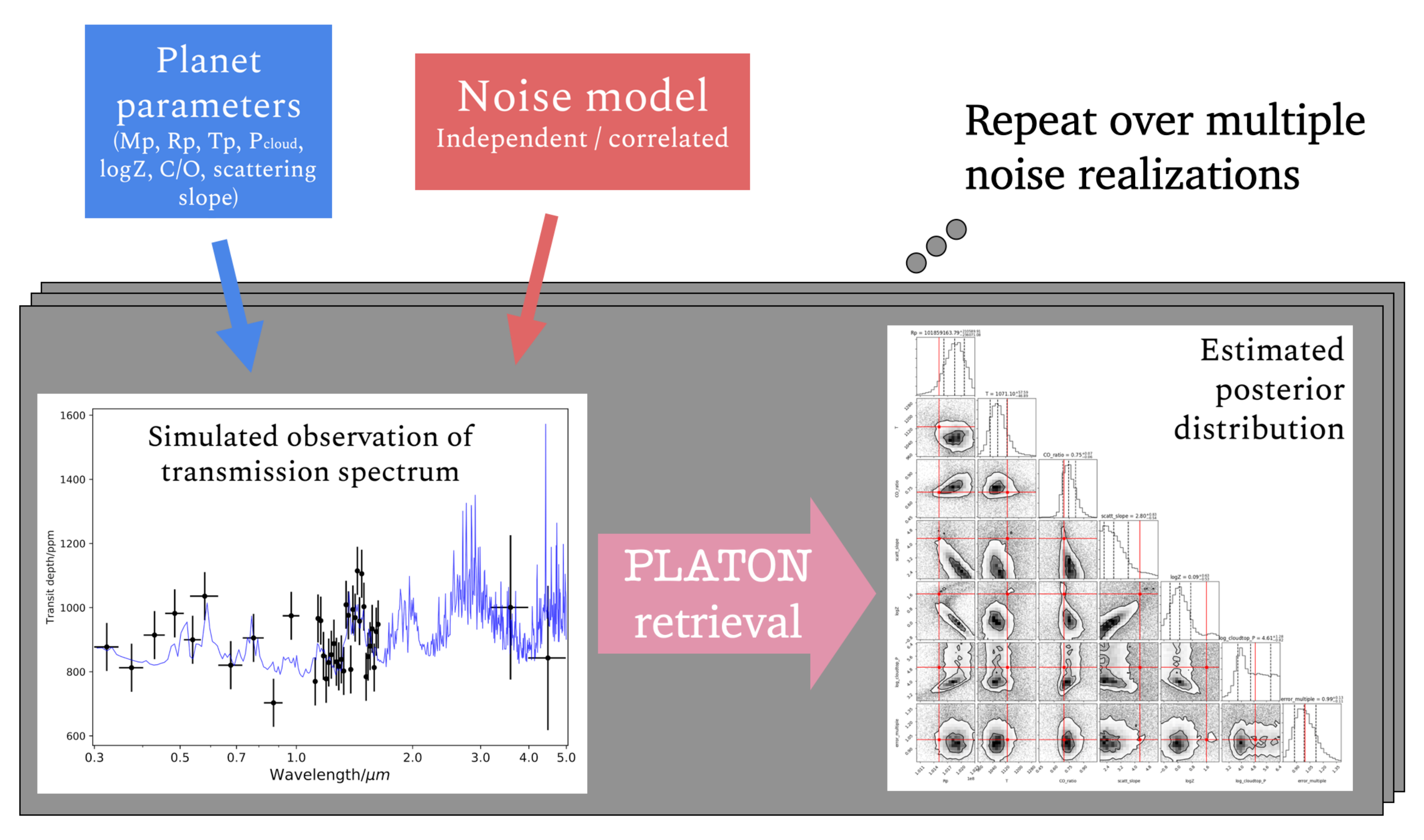}
    \caption{Schematic diagram of our method. We generate multiple ($\sim$500) observational instances of a given planetary scenario and noise model, and perform atmospheric retrieval on each spectrum using \textsc{platon}.}
    \label{fig:method_schema}
\end{figure*}

\begin{enumerate}

\item Choose the input parameters of a planet, such as radius, temperature, metallicity.

\item Run the forward \textsc{platon} \citep{zha19} model to produce the unpolluted spectrum of the atmosphere.

\item Bin the full-resolution unpolluted spectrum to the chosen wavelength bins.
 
\item Choose a noise model (independent or correlated) and noise parameters that simulate the noise of a real observation.
 
\item Sample  a  noise  realization  of  the  binned spectrum using the noise model.

\item Perform retrieval analysis on the simulated data.

\item Repeat steps 5 and 6 a sufficient number of times such that the retrieved parameters can be combined to generate reliable statistics. 

\end{enumerate}

%These steps are shown schematically in Figure \ref{fig:method_schema}. 

We use \textsc{platon} \citep[version 3.0;][]{zha19}, an open-source retrieval tool, as the forward and retrieval model for transmission spectra. \textsc{platon} has the advantage of being extremely fast for a retrieval code ($\lesssim 30$ minutes per run), which is suitable for our application as we perform hundreds of retrievals with randomly sampled noise realizations. We perform this process only on transmission spectroscopy, as the geometry allows for assuming an isothermal atmosphere, greatly reducing the number of free parameters in our model as well as the computation time per run.

We repeat the above procedure for five cases of observation: a clear hot Jupiter, a clear hot Jupiter with offsets between datasets, a cloudy hot Jupiter, a clear hot Jupiter at higher (\textit{JWST}-like) precision, and a warm Neptune. In what follows, we describe the forward model parameters, the noise model, and the retrieval setup. A summary of the input parameters and the assumed priors for each set of retrievals is provided in Table~\ref{tab:ground}. 

\subsection{Forward Model Parameters} \label{subsec:forward}
To best imitate retrievals on actual observations, we choose input parameters and spectrum bins similar to \textit{Hubble} and \textit{Spitzer} observations of the canonical hot Jupiter HD 209458b and exo-Neptune GJ436b which have reliable data and have been studied in the context of retrievals \citep{eva15, dem13, knu07, dem11, dit09, lot18}. We adopt the measured mass, radius, and temperature of each planet, and set the log-metallicity to 0.3 and carbon-to-oxygen ratio to the solar value of 0.53 \citep{asp09}.

We also choose to include cloud and hazes in our model. Such aerosols have been found to be ubiquitous in exoplanetary atmospheres \citep[e.g.][]{kre14, sin16} and produce a spectral signature that can be degenerate with other parameters \citep{dem17, mar13}. \textsc{platon} accounts for clouds and hazes using a parametric model. The user can specify a grey cloud-top pressure, the atmosphere absorption below which is truncated, and an amplitude and slope in the optical end of the spectrum to account for a non-Rayleigh slope caused by Mie scattering. In summary, the aerosol opacity $\kappa_{aer}$ is given as:
\begin{align*}
    P > P_{\textrm{cloud}} &: \kappa_{\textrm{aer}} = \infty \\
    P < P_{\textrm{cloud}} &: \kappa_{\textrm{aer}} \propto a \lambda^{-\gamma},
\end{align*}
where $a=1$ and $\gamma=4$ corresponds to Rayleigh scattering from the gaseous atmosphere. 

For our cloud-free simulations, we choose a low-altitude cloud-top pressure of 0.5 bar. For our cloudy simulations, we choose 0.1 mbar such that clouds obscure roughly half of the spectrum while preserving some molecular features. Similarly, we adopt a nearly-Rayleigh slope of 4.3 and amplitude of 1, indicating no excess Rayleigh scattering from aerosol particles. We stress that since we use the same forward model in the retrieval to isolate the effects of noise from model systematics, the specific choices in parameters are not of great importance as long as they can be correctly retrieved and as long as we select a set of forward models that span a representative set of exoplanet atmospheres.

\begin{figure}[h]
    \includegraphics[width=\columnwidth]{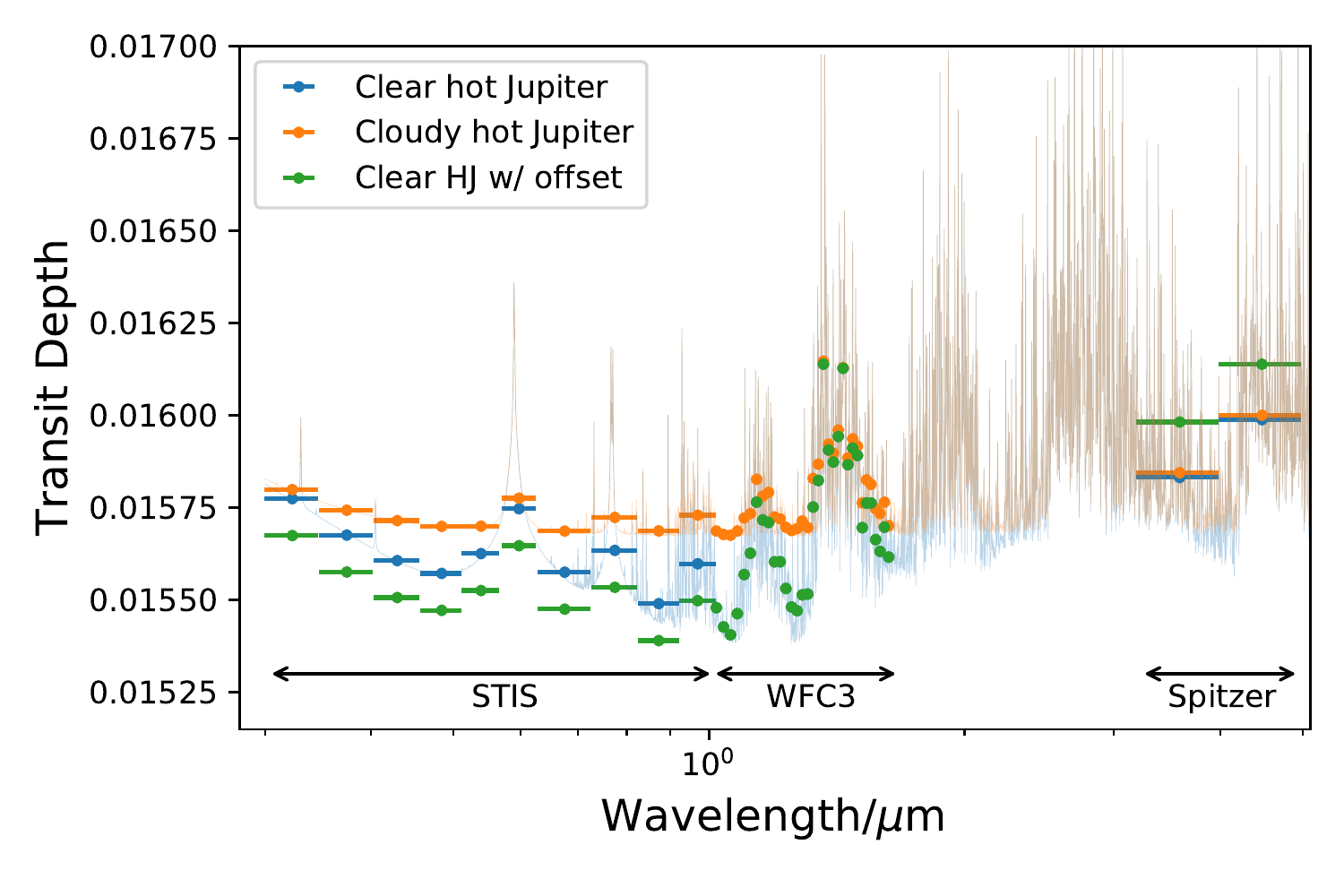}
    \caption{Simulated unpolluted spectra of the hot Jupiter cases. Both full-resolution and binned depths are  plotted. The offset case is identical to the clear case in the WFC3 band. The main effect of clouds is the truncation of the bottom-most depths compared to the clear case.}
    \label{fig:truth_HJ}
%\end{figure}

%\begin{figure}[h]
    \includegraphics[width=\columnwidth]{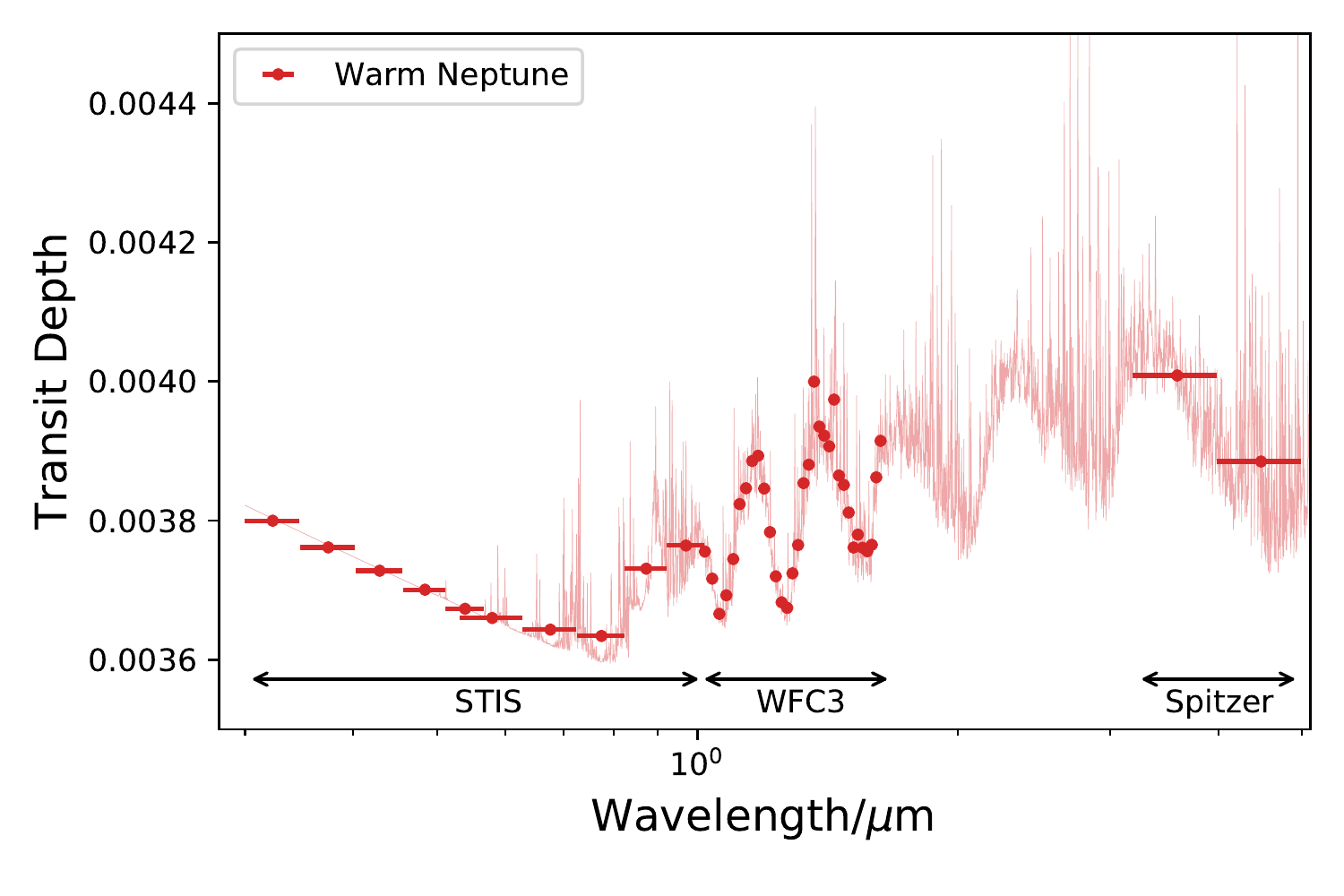}
    \caption{Same as Figure \ref{fig:truth_HJ}, but for the warm Neptune case.}
    \label{fig:truth_SN}
\end{figure}

We choose a wavelength range that spans observations from the \textit{Hubble Space Telescope} (\textit{HST}) spectrographs most commonly used for exoplanet atmospheric investigations --- specifically the Space Telescope Imaging Spectrograph (STIS) and Wide-Field Camera 3 (WFC3) --- as well as photometric observation from the \textit{Spitzer Space Telescope}. In the STIS wavelength range we follow the bins of \citet{knu07}; in the WFC3 wavelength we choose 33 equal sized bins between 1.01~$\mu$m and 1.64~$\mu$m \citep{gen18}; for Spitzer we include observations in the photometric bands of the IRAC instrument at 3.6~$\mu$m and 4.5~$\mu$m. The resulting wavelength bins are comparable to a complete panchromatic dataset from current space-based observations. The resulting simulated spectra are shown in Figure \ref{fig:truth_HJ} and Figure \ref{fig:truth_SN}.

% TODO: consider doing inset plots... also maybe put the error bar in the input spectrum

\begin{splitdeluxetable*}{lccccccBlcccccccc}
\tabletypesize{\scriptsize}
\tablewidth{0pt} 
\tablenum{1}
\tablecaption{Summary of equilibrium chemistry retrievals and the input parameters and priors used.}
\tablehead{
\colhead{} & \colhead{} & \multicolumn{2}{c}{Clear HJ} && \multicolumn{2}{c}{Clear HJ with offset} & \colhead{} & \multicolumn{2}{c}{Cloudy HJ} && \multicolumn{2}{c}{Clear HJ, high precision} && \multicolumn{2}{c}{Warm Neptune} \\
\cline{3-4} \cline{6-7} \cline{9-10} \cline{12-13} \cline{15-16}
\colhead{Parameter} & \colhead{Name} & \colhead{Truth value} & \colhead{Prior} && \colhead{Truth value} & \colhead{Prior} & \colhead{Parameter} & \colhead{Truth value} & \colhead{Prior} && \colhead{Truth value} & \colhead{Prior} && \colhead{Truth value} & \colhead{Prior}
}
% \colnumbers
\startdata \\
$M_\mathrm{p}$ & Planetary mass (M$_\mathrm{J}$) & 0.73 & 15\% && 0.73 & 15\% & $M_\mathrm{p}$ &  0.73 & 15\% &&  0.73 & 15\% && 0.0736 & 15\% \\
$R_\mathrm{p}$ & Planetary radius (R$_\mathrm{J}$) & 1.42 & 15\% &&  1.42 & 15\% & $R_\mathrm{p}$ & 1.42 & 15\% && 1.42 & 15\% && 0.395 & 15\% \\
$T$ & Temperature (K) & 1130  & [300, 1500] && 1130  & [300, 1500] & T & 1130  & [300, 1500] && 1130  & [300, 1500] && 700 & [300, 1500] \\
C/O & Carbon-to-Oxygen ratio & 0.53 & [0.2, 2.0] && 0.53 & [0.2, 2.0] & C/O & 0.53 & [0.2, 2.0] && 0.53 & [0.2, 2.0] && 0.53 & [0.2, 2.0] \\
$\log{a}$ & Log-Scattering factor & 0 & [-0.3, 0.3] && 0 & [-0.3, 0.3] & $\log{a}$ & 0 & [-3.0, 3.0] && 0 & [-0.3, 0.3] && 0 & [-2, 2]\\
$\gamma$ & Scattering slope & 4.3 & [2.0, 5.5] && 4.3 & [2.0, 5.5] & $\gamma$ & 4.3 & [2.0, 5.5] && 4.3 & [2.0, 5.5] && 4.3 & [, ] \\
$\log{Z}$ & Log-Metallicity & 0.3 & [-1, 3] && 0.3 & [-1, 3] & $\log{Z}$ & 0.3 & [-1, 3] && 0.3 & [-1, 3] && 0.3 & [-1, 3] \\
$\log{P_{\mathrm{cloud}}}$ & Cloudtop Pressure (log(Pa)) & 4.7 & [2.5, 6.5] && 4.7 & [2.5, 6.5] & $\log{P_{\mathrm{cloud}}}$ & 3.0 & [0.0, 6.5] && 4.7 & [2.5, 6.5] && 5.0 & [2.5, 6.5] \\
Offset 1 & STIS offset (ppm) & -- & -- && -100 & [-300, 300] & Offset1 & -- & -- && -- & -- && -- & --\\
Offset 2 & Spitzer offset (ppm) & -- & -- && 150 & [-300, 300] & Offset2 & -- & -- && -- & -- && -- & --\\
Error multiple & -- & Unity & [0.5, 2.0] && Unity & [0.5, 2.0] & Error multiple & Unity & [0.5, 2.0] && Unity & [0.5, 2.0] && Unity & [0.5, 2.0] \\
\hline
\textit{Other parameters} &&&&&& &\textit{Other parameters} &&& \\
$R_\mathrm{s}$ & Stellar radius (R$_\sun$) & \multicolumn{2}{c}{1.19} && \multicolumn{2}{c}{1.19} & $R_\mathrm{s}$ & \multicolumn{2}{c}{1.19} && \multicolumn{2}{c}{1.19} && \multicolumn{2}{c}{0.683} \\
$T_\mathrm{s}$ & Stellar temperature (K) & \multicolumn{2}{c}{6090} && \multicolumn{2}{c}{6090} & $T_\mathrm{s}$ & \multicolumn{2}{c}{6090} && \multicolumn{2}{c}{6090} && \multicolumn{2}{c}{4780}  \\
Data error & (ppm) & \multicolumn{2}{c}{75} && \multicolumn{2}{c}{75} & Data error & \multicolumn{2}{c}{75} && \multicolumn{2}{c}{10} && \multicolumn{2}{c}{75} \\
\# of runs & & \multicolumn{2}{c}{440} && \multicolumn{2}{c}{660} & \# of runs & \multicolumn{2}{c}{660} && \multicolumn{2}{c}{400} && \multicolumn{2}{c}{400} \\
\enddata
\tablecomments{The stellar parameters shown are not retrieved for. For mass and radius, the priors show the width of the Gaussian prior, relative to the input value. For other parameters, the priors are uniform in the interval. For log values, the priors are log-uniform in the interval. \label{tab:ground}}
\end{splitdeluxetable*}

\subsection{Noise Model} \label{subsec:noise}

To simulate observed data, we sample multiple noise instances centered around the unpolluted spectrum, treating the simulated unpolluted spectra as a multivariate Gaussian distribution with the unpolluted depths as the mean and the reported error at each bin as the width. In addition, we adopt a non-diagonal covariance matrix with an exponential kernel to simulate correlated noise, such that the matrix element that correlates wavelength bin at $\lambda_i$ and $\lambda_j$ is given by:

\begin{equation}
    K_{ij} = \epsilon_{ij} \sigma_{i} \sigma_{j} \exp{\left(-\frac{|\lambda_i - \lambda_j|}{l}\right)}, 
\end{equation}

\noindent where $\sigma_i$ is the reported error at wavelength $\lambda_i$, and $\epsilon_{ij}$ is 1 for pair of points from the same dataset and 0 otherwise. We choose the scaling factor $l$ to be the distance to the neighboring bin. We select this covariance matrix in particular because it allows for the best-fit spectrum to the WFC3 observations of HD 97658b in \citet{guo20} to yield a reduced chi-squared of $\sim 1$, as opposed to 2.5 when the errors are construed to be independent and Gaussian. We choose a random noise error of 75~ppm for all instruments, which represents a moderate quality data for STIS and WFC3 but is better than average for typical \textit{Spitzer} observation \citep{sin16,gar20}.  {We also assume a uniform transmission across the wavelength range of each Spitzer filter in both the forward and retrieval models.} In practice, we find that the two broadband \textit{Spitzer} points provide little constraint on the retrieved parameters, and using the same error for all instruments does not give undue importance to the \textit{Spitzer} points. For the high-precision hot Jupiter case, we use 10~ppm errors to match the best of current data quality \citep{col20}. 

\begin{figure*}[ht!]
    \centering
    \includegraphics[width=\textwidth]{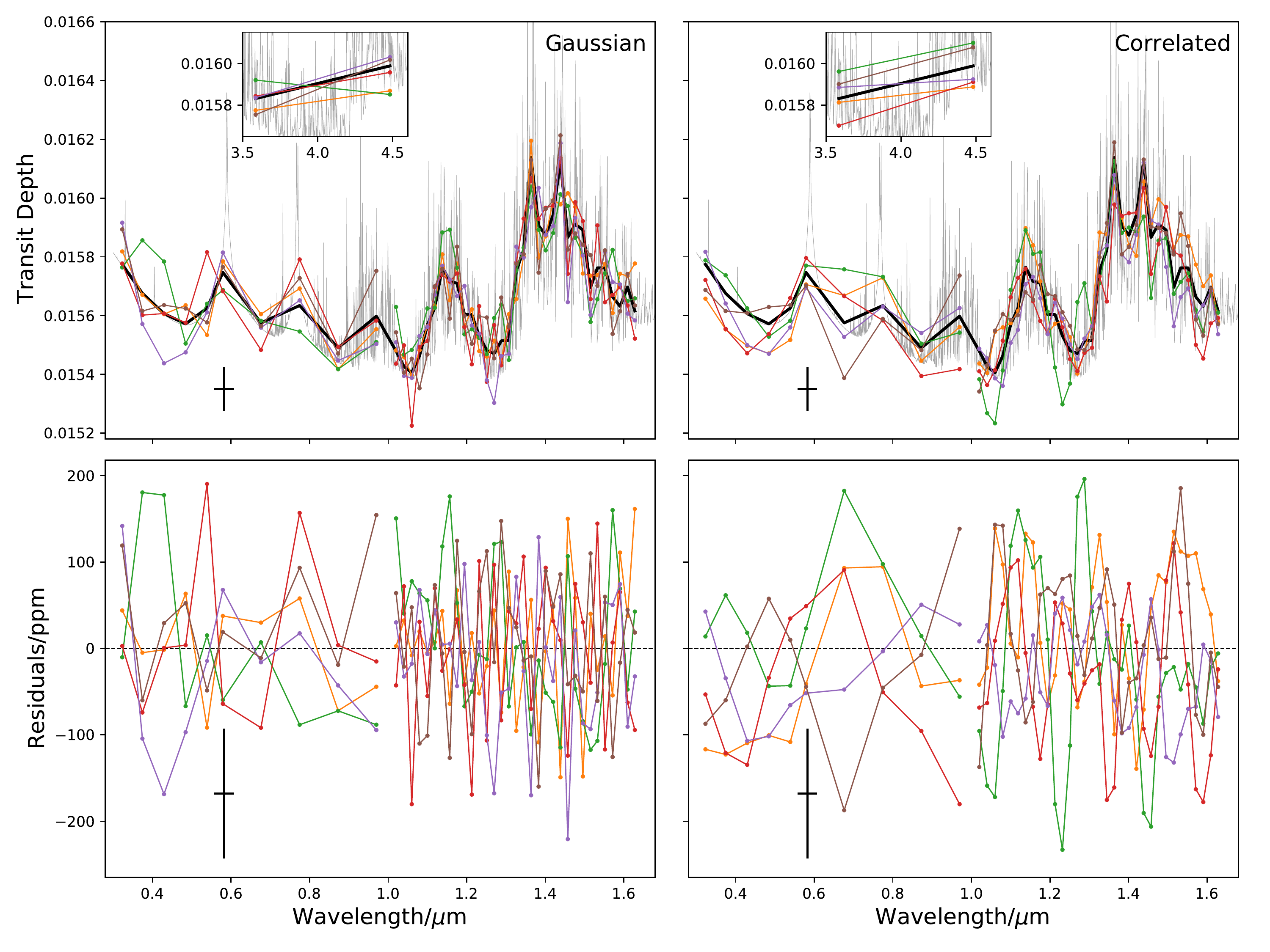}
    \caption{Comparison of Gaussian (left column) and correlated (right column) noise models for the clear hot Jupiter case. In the top row of panels, 5 randomly selected spectrum realizations are plotted in color and the unpolluted spectrum is plotted in black.  {The disconnection in lines indicate discrete instruments.} The \textit{Spitzer} data are shown separately in the inset plot. The assumed 75-ppm error bar is shown for scale. In the bottom row of panels, the residuals relative to the unpolluted spectrum are shown. The effect of wavelength-correlation is discernible in the slightly redder noise (i.e. the residuals appear subtly sparser  {due to less zero-crossings as neighboring residuals are more likely to have the same sign}) in the bottom right panel, but would not be discernible in the top right panel without prior knowledge of the ground-truth spectrum .}
    \label{fig:method_noise}
\end{figure*}

We show a few randomly selected noise realizations in Figure \ref{fig:method_noise} for the Gaussian and correlated noise models. It is discernible from the bottom row of panels that the correlated noise has slightly redder residuals compared to the Gaussian noise; textcolor{red}{that is, there are less zero crossings as neighboring points are correlated}. We also stress that overall this is a rather subtle effect; without knowing the unpolluted depths \textit{a priori} to produce the residuals, from the spectrum alone it is hardly obvious that there is a distinction between the two. 

Additionally, to examine the effects of including the offset between datasets as a retrieved parameter, we create spectra with and without a fixed offset between datasets. Namely, we add a vertical offset of $-100$ ppm to the STIS points and $+150$ ppm to Spitzer points, holding the WFC3 points constant. The specific amount of offset is a somewhat arbitrary choice. The relevant heuristic is that  the offset should be exactly retrieved in the absence of degeneracy.

\subsection{Retrieval Setup} \label{subsec:retrieval}
Using \textsc{platon}, setting up the retrieval involves choosing the priors and the statistical sampling method to be used. We choose the priors to be comparable to a real retrieval analysis. \textsc{platon} accepts either a Gaussian prior or a uniform prior in a user-specified interval. We set Gaussian priors for planet radius and mass, as these are often constrained via other methods such as radial velocity and transit measurements prior to observing the transmission spectrum. The Gaussian prior is centered around the input value and with a standard error of 15\%. This precision is comparable to or slightly overestimates the typical uncertainty in measurement of mass via the RV method and provides sufficient tolerance for the retrieved value to deviate from the input value, if necessary. For the instrumental offsets, to ensure that the prior is broad enough, we choose a uniform prior offset to be 2- and 3-folds of the offsets. For all other parameters we opt for uniform priors that are as wide and uninformative as possible and adopt the full parameter range supported.

 {For now, we only choose to do retrievals with equilibrium chemistry, where the composition of the atmosphere at a given temperature and pressure is dictated by the the global elemental abundances set by metallicity and carbon-to-oxygen ratio. While disequilibrium chemistry is indeed expected for planets below $T_{\mathrm{eq}} \leq 1200 \mathrm{K}$, most of its effects take place below the altitude that typically probed by transmission spectroscopy and have no easily discernible effect on the spectrum at the data precision of current instruments. The actual discrepancy in the relevant pressure range ($\sim 1$ mbar) is smaller than the uncertainties we can current obtain} \citep[e.g.][]{lin11, kem12, for20}.

\textsc{platon} supports either Markov chain Monte Carlo (MCMC) \citep{for13} or nested sampling methods for the posterior estimation \citep{fer08}. We note that, while both are statistically robust and widely used, we observe a minor discrepancy in the resulting posteriors between the two methods, in which the posteriors estimated by MCMC tend to be slightly broader than those by nested sampling. This does not pose a major issue for this work inasmuch as we are concerned with biases due to data idiosyncrasies, and we consistently use one algorithm across our analyses. Nevertheless, we draw attention to this point as it requires extra scrutiny when comparing retrieval results. We choose nested sampling as it is known to perform better in estimating multi-modal or oddly-shaped distributions.

\section{Results} \label{sec:results}

In this section we first present the overall effects of correlated systematics on our retrievals, using the clear hot Jupiter case as a baseline. We then examine which parameters in particular are affected. We finally show how the results for the baseline case also extend to the other retrieval cases, and point out additional effects that arise.

\subsection{Overfitting Due to Correlated Noise}

% \begin{figure*}[!t]
%     \centering
%     \includegraphics[width=5.0in]{figures/HJ_no_offset_compare_new_label.pdf}
%     \caption{Two-dimensional distribution comparing retrievals with Gaussian noise (black) and correlated noise (blue). The three parameters shown are: the chi-squared between the input value spectrum and the observation instance (True $\chi^2$); the reduced chi-squared between the best-fit spectrum and the observation instance (Fit $\chi_r^2$); and the number of standard errors the retrieved posterior rules out the input value parameters (Retrieval $\sigma_{acc}$), our ``accuracy-of-retrieval'' statistic. The expected $\chi_{9}$ distribution is also shown in red, with the 50-, 95-, and 99-percentiles demarked with dashed bars. The correlated noise allows for overfitting the spectrum, while simultaneously degrading the accuracy of the retrieval. We remind the reader that this corner plot is not showing a single retrieved posterior result, but a composite of multiple posteriors.}
%     \label{fig:compare_HJ}
% \end{figure*}

\begin{figure*}[!t]
    \centering
    \includegraphics[width=5.0in]{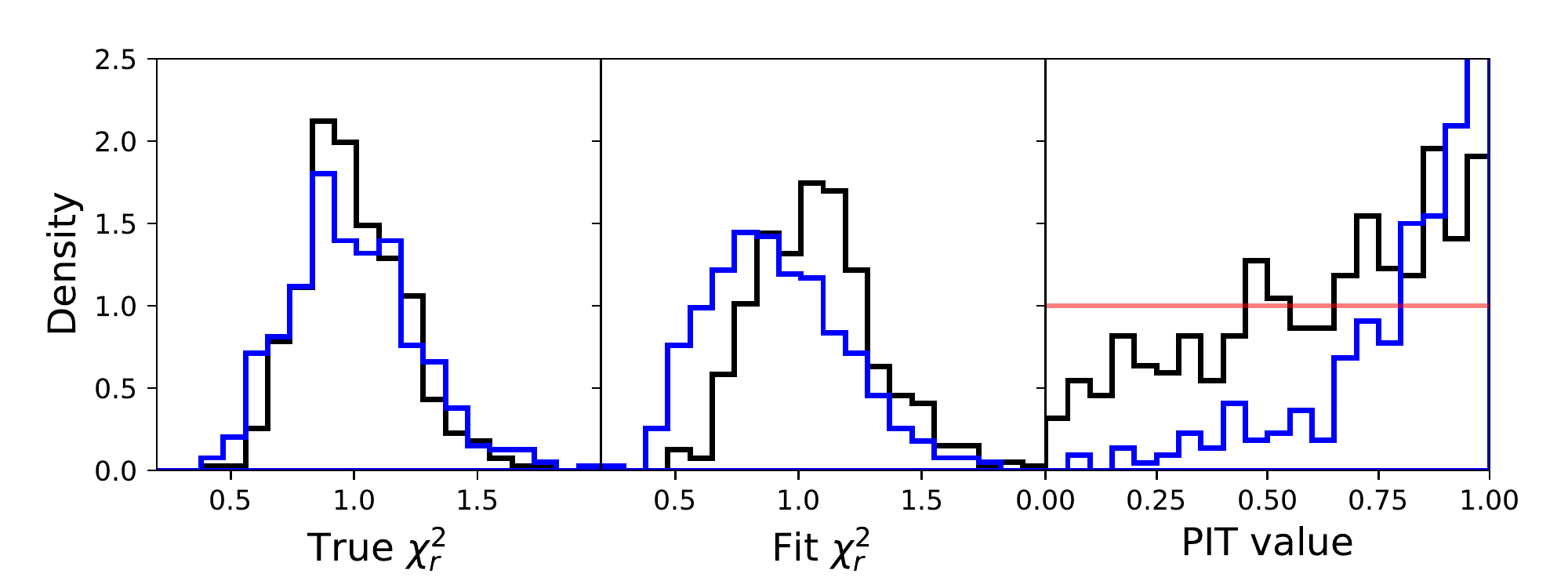}
    \caption{Two-dimensional distribution comparing retrievals with Gaussian noise (black) and correlated noise (blue). The three parameters shown are: the chi-squared between the unpolluted spectrum and the observation instance (True $\chi^2$); the reduced chi-squared between the best-fit spectrum and the observation instance (Fit $\chi_r^2$); and  {the PIT values for the retrieved posterior.} The correlated noise allows for overfitting the spectrum, while simultaneously degrading the accuracy of the retrieval. We remind the reader that this corner plot is not showing a single retrieved posterior result, but a composite of multiple posteriors.}
    \label{fig:compare_HJ}
\end{figure*}

Here we present the effects of correlation in data on the retrieval overall. To do this, we must reduce a retrieval result into a simpler metric. In retrievals on actual data this is commonly done by using the reduced chi-squared between the observed data and the best-fit or median spectrum. Here, as we begin from a known simulated ground truth, we can also compare the retrieved posterior directly to the input values to measure the accuracy of retrieval by using a  {probability integral transform (PIT), which is the cumulative distribution function evaluated at the input value.} To do this, in each retrieved posterior distribution, we draw an iso-likelihood contour of the input parameters and sum the  {relative} weights contained within, producing a confidence interval between 0 (the input was the most likely sample) and 1 (the least likely). { The distribution of PIT values should follow a uniform $\mathcal{U}(0,1)$ were the retrievals accurate. }

Our main finding is that, on average, correlation in the data allows for overfitting the spectrum, thereby weakening the overall accuracy of the retrieval. In other words, the best-fit spectrum is more likely to achieve a reduced chi-squared lower than unity in the presence of correlated noise, whilst simultaneously the retrieved posterior distribution rules out the input with higher significance.

\begin{figure}[h]
    \includegraphics[width=\linewidth]{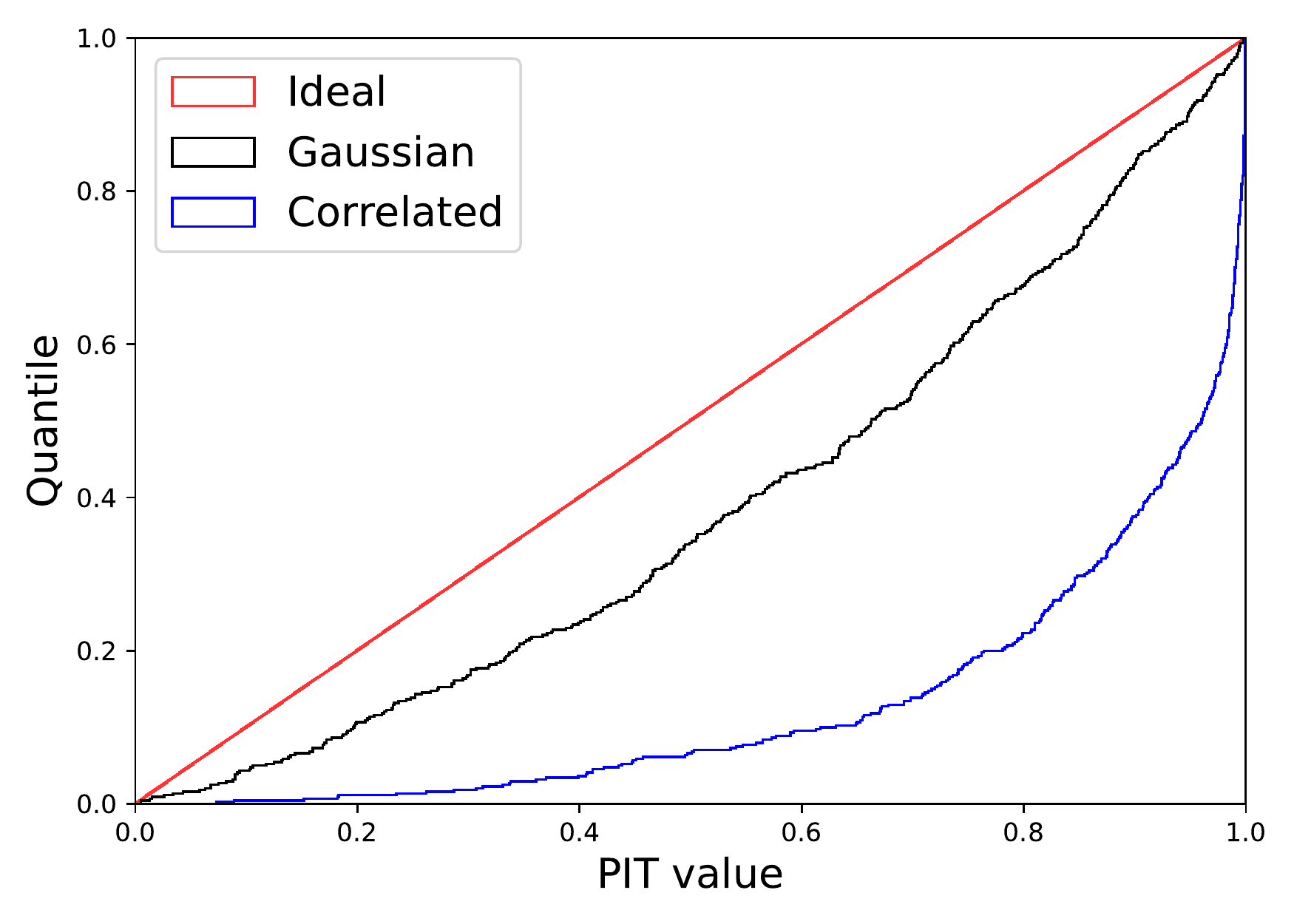}
    \caption{ {Same distribution as in the right panel of Figure \ref{fig:compare_HJ}, but shown as a cumulative distribution. The horizontal axis corresponds to the probability integral transform (PIT) values for each retrieval; the vertical axis corresponds to the cumulative fraction of all retrievals. } The K-S metric measures the maximum vertical discrepancy between two cumulative distributions, such as above, and values of this metric are reported in Table~\ref{tab:kstest}. The expected  {uniform}distribution (red), and our independent noise (black) and correlated noise (blue) simulation cases are shown. }
    \label{fig:compare_HJ_cumul}
\end{figure}

In Figure \ref{fig:compare_HJ}, we present the comparison between Gaussian and correlated noise for the clear hot Jupiter case. For both retrievals with Gaussian (black) and correlated noise (blue), we show the distributions of: $\chi^2$ between the unpolluted spectrum and the data instances; the reduced chi-squared between the best-fit spectrum and the data instances; and the PIT values showing the accuracy of retrieval, as described above. A few observations can be made:

\begin{itemize}
    \item The fit $\chi^2_{\textrm{r}}$, or the \textbf{goodness-of-fit} of the retrieval, on synthetic data is on average skewed to better than unity in the presence of correlated noise. In other words, it is more likely that the retrieval will overfit the data with forward models.
    \item  {The accuracy of the retrieval, shown as the PIT value, on the other hand, is \textit{worse} in the presence of correlated noise. We also show the cumulative distribution in Figure \ref{fig:compare_HJ_cumul} to demonstrate the worsening of the accuracy. The K-S statistic between the two cases is 0.19.}
    \item  {For the normal noise case, even in the absence of the correlated noise, the retrieval accuracy is close to but slightly worse than the expected uniform distribution. This minor discrepancy is likely due to the fact that the retrieved posterior distribution is already non-Gaussian for a few parameters (discussed in Subsection \ref{subsec:which}), as well as to degeneracy between certain parameters. }
    \item While the goodness-of-fit is on average better in the presence of correlated noise, it is not the case that the distributions of fit $\chi^2_{\textrm{r}}$ are so discrepant that one can deduce the presence of correlated noise from the goodness-of-fit alone. That is, a given overfit spectrum can plausibly be construed as either a consequence of correlated noise or as an unlucky instance of Gaussian noise that happens to lie at the tail of the $\chi^2_{\textrm{r}}$ distribution. As such, we stress that the effect of correlated noise is manifest statistically, and no individual value of $\chi^2_{\textrm{r}}$, good or bad, is uniquely diagnostic of correlated noise in a single retrieval instance.
\end{itemize}

\subsection{Which Parameters are Affected?} \label{subsec:which}

Given that the correlated noise degrades the overall accuracy of retrievals, it is necessary to then look at which parameters are affected. We do this by marginalizing the posterior distribution over each parameter, as is typically done in retrieval analyses. The effect on each marginalized distribution can be twofold --- the mean can be biased, the estimated error can be affected, or both. Either a shift in mean away from the input parameter or an underestimation of the error can worsen the accuracy of retrieval. As such, we examine the retrieved mean and the retrieved error separately for each parameter. 

\begin{figure*}[!h]
    \centering
    \includegraphics[width=0.8\textwidth]{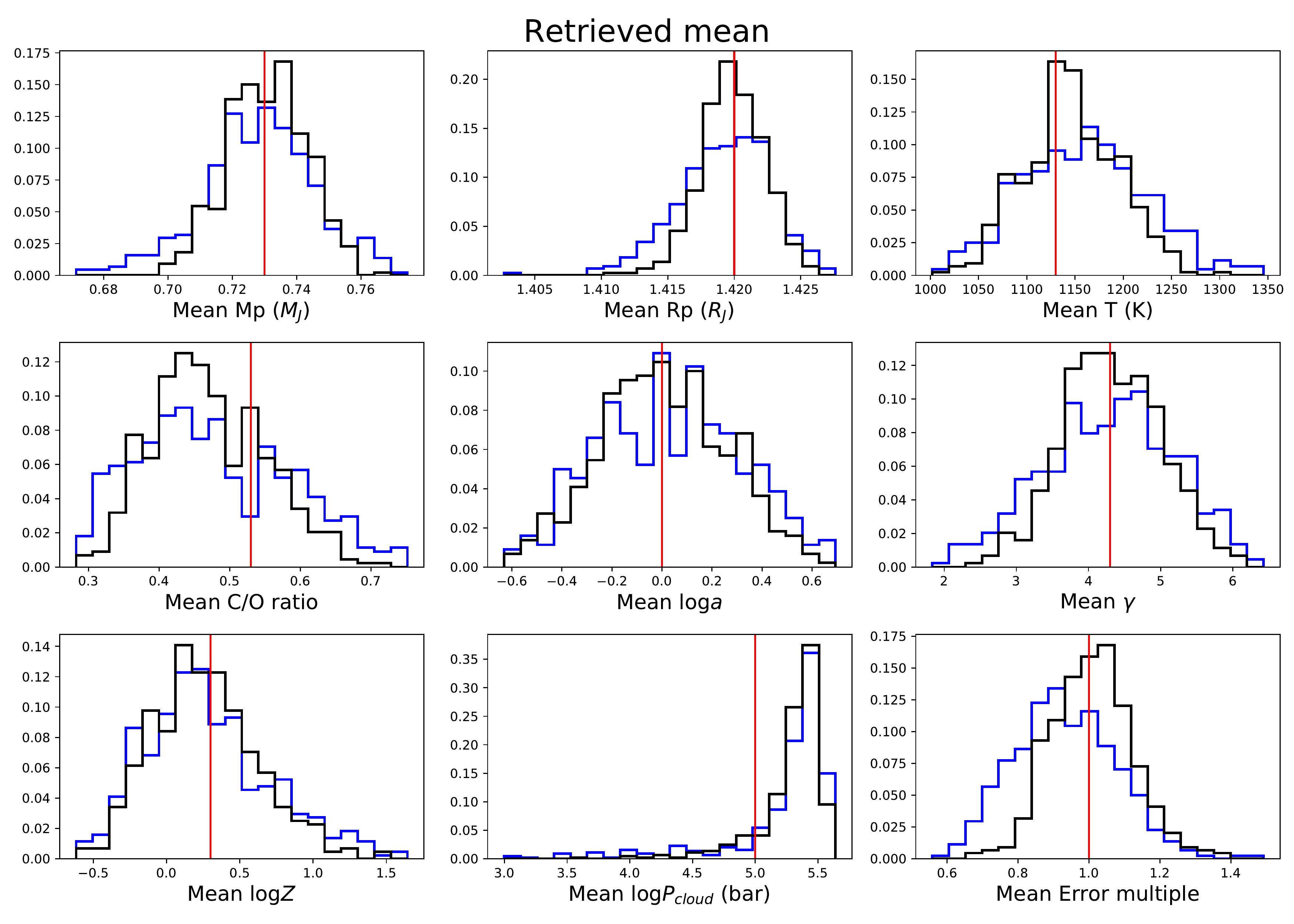}
    \caption{The distribution of retrieved means per each parameter, for independent noise (black) and for correlated noise (blue), for the baseline hot Jupiter case. The input values used to generate the spectrum are shown in red.}
    \label{fig:mean}
\end{figure*}

The distribution of retrieved means is shown in Figure~\ref{fig:mean}. For the case in which noise is independent (black), the retrieved means form clean normal distributions around the input values (red) for most parameters, as expected from the central limit theorem. The two exceptions are C/O ratio and the cloudtop pressure. This is most likely due to the fact that the retrieved distributions for these parameters are not Gaussian in the first place. For cloudtop pressure, the retrieved distribution is at best a flat distribution with a lower bound, ruling out a cloudy atmosphere as per the clear atmosphere in the input used. For C/O ratio, we suspect that the distribution is skewed due to the increasing influence the parameter has over its range. That is, as one sweeps through C/O ratio, the spectrum changes more rapidly over the range above the solar value of 0.53, and thus the means are naturally skewed to values lower than the solar value where there is a greater density of near-consistent solutions. 

The presence of correlated noise has a few interesting effects on the retrieved means. First off, the error multiple parameter is biased to less than unity. This intuitively follows from the global result that correlated noise allows for overfitting the spectrum, tricking the error multiple parameter to believe that the error bars are overestimated. This means that the error multiple parameter is more likely to behave pathologically in a situation where one may expect it to be useful, such as if the reported error bars truly were underestimated due to unknown and unaccounted systematics. The retrieval instead selects a less-than-unity value of the error multiple, incorrectly implying that the data precision is better than initially reported. This is possible if the domain of  {input parameters} and the forward model can still reproduce the spectrum polluted with systematics. 

% This means that correlated noise actually causes the error multiple parameter to behave pathologically, by taking on the opposite role of what it was meant for.  Instead of indicating that the errors bars were underestimated due to unknown and unaccounted for systematics, the retrieval selects a less-than-unity value of the error multiple, incorrectly implying that the data precision is better than what was initially reported. 
% %This means that in a situation where one may expect the error multiple parameter to be useful, such as if the reported error bars were truly underestimated due to unknown and unaccounted systematics, the error multiple parameter instead indicates the exact opposite, as the spectrum polluted with systematics still lie in the image of the forward model.  {hmm.. bad sentence}

In the correlated noise case, the retrieved means generally show a wider distribution to varying degrees for each parameter. Specifically, the radius, mass, and temperature are the most affected, while the effect is the least pronounced for metallicity and cloudtop pressure. This result may be explained by considering the wavelength-scale the former three parameters have on the spectrum. Mass and temperature affect the scale height of the atmosphere, which affects the overall vertical extent of the transmission spectrum. Radius affects the baseline transit depth as well as the scale height. These are ``global'' parameters in the sense the transit depths in all bins are affected together. As such, a wavelength-dependent correlation can bias these parameters. On the other hand, metallicity, while it also affects the scale height (via the mean molecular weight),  directly controls the individual transit depths. This has a more local effect in that it changes the actual shape of the spectrum.

\begin{figure*}[!h]
    \centering
    \includegraphics[width=0.8\textwidth]{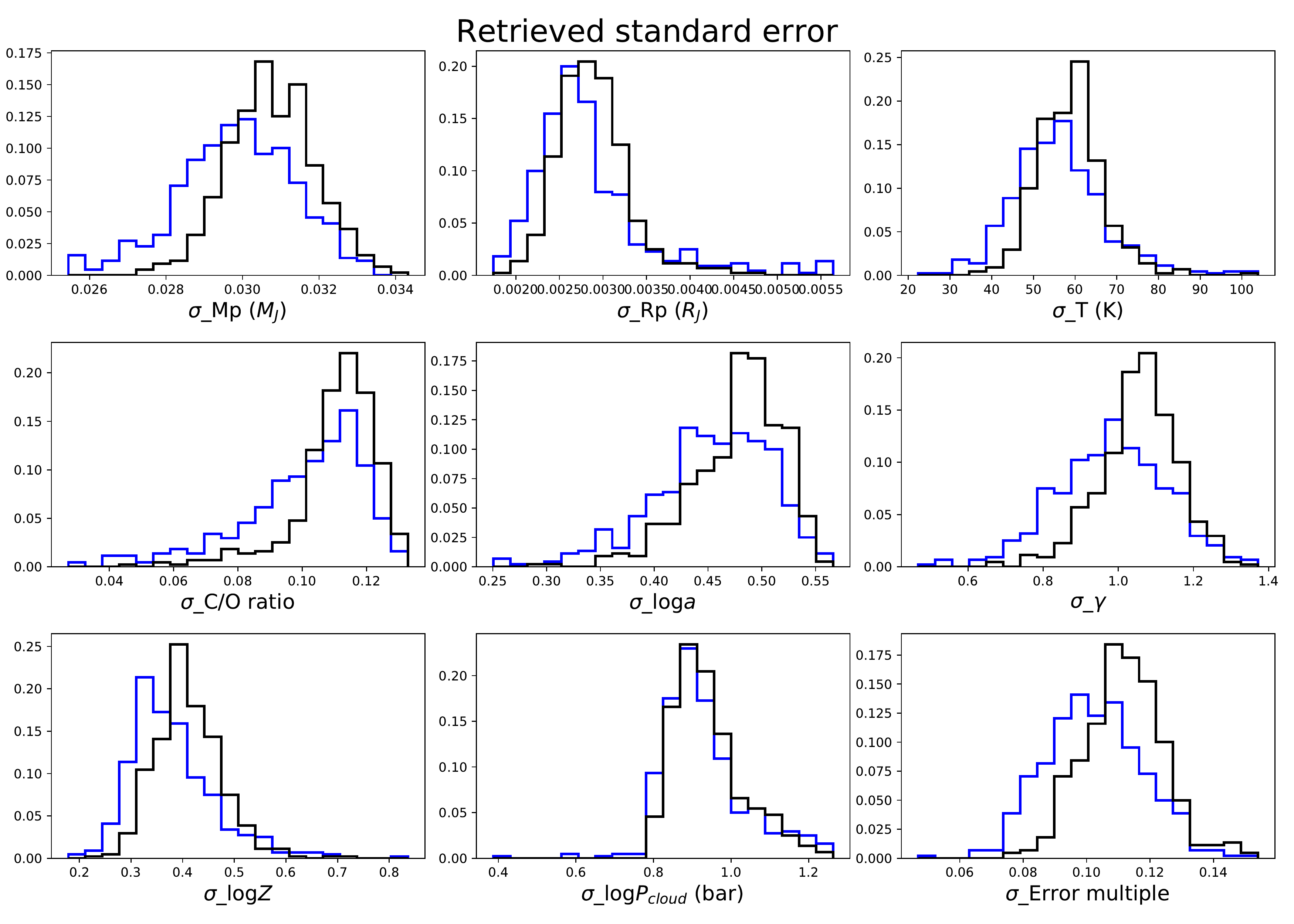}
    \caption{The distribution of retrieved standard errors per each parameter, for independent noise (black) and for correlated noise (blue), for the baseline hot Jupiter case.}
    \label{fig:error}
\end{figure*}

The distribution of retrieved errors is shown in Figure \ref{fig:error}. The effect of correlated noise is clearly visible for all parameters in that the retrieved error bars show a tendency to be underestimated. For instance, the retrieved error on log-metallicity is on average underestimated by $\sim 0.2$ dex. While this disparity is smaller still than typical constraints, it is worth bearing in mind as this is a statistical effect; the spread over the retrieved error is by itself broad enough that the actual effect of a given instance can be much larger than this value. Additionally, when \textit{JWST} allows for precise measurement of metallicity, this level of uncertainty may not be negligible when one considers analyzing archival data simultaneously. The same consideration applies to other parameters. As such, in this context we suggest that the retrieved constraints for parameters, in the face of the potential for correlated noise, are best understood as lower limits. 

\figsetstart
\figsetnum{1}
\figsettitle{Retrieved means}
\label{figset:means}

\figsetgrpstart
\figsetgrpnum{1.1}
\figsetgrptitle{Means Clear HJ}
\figsetplot{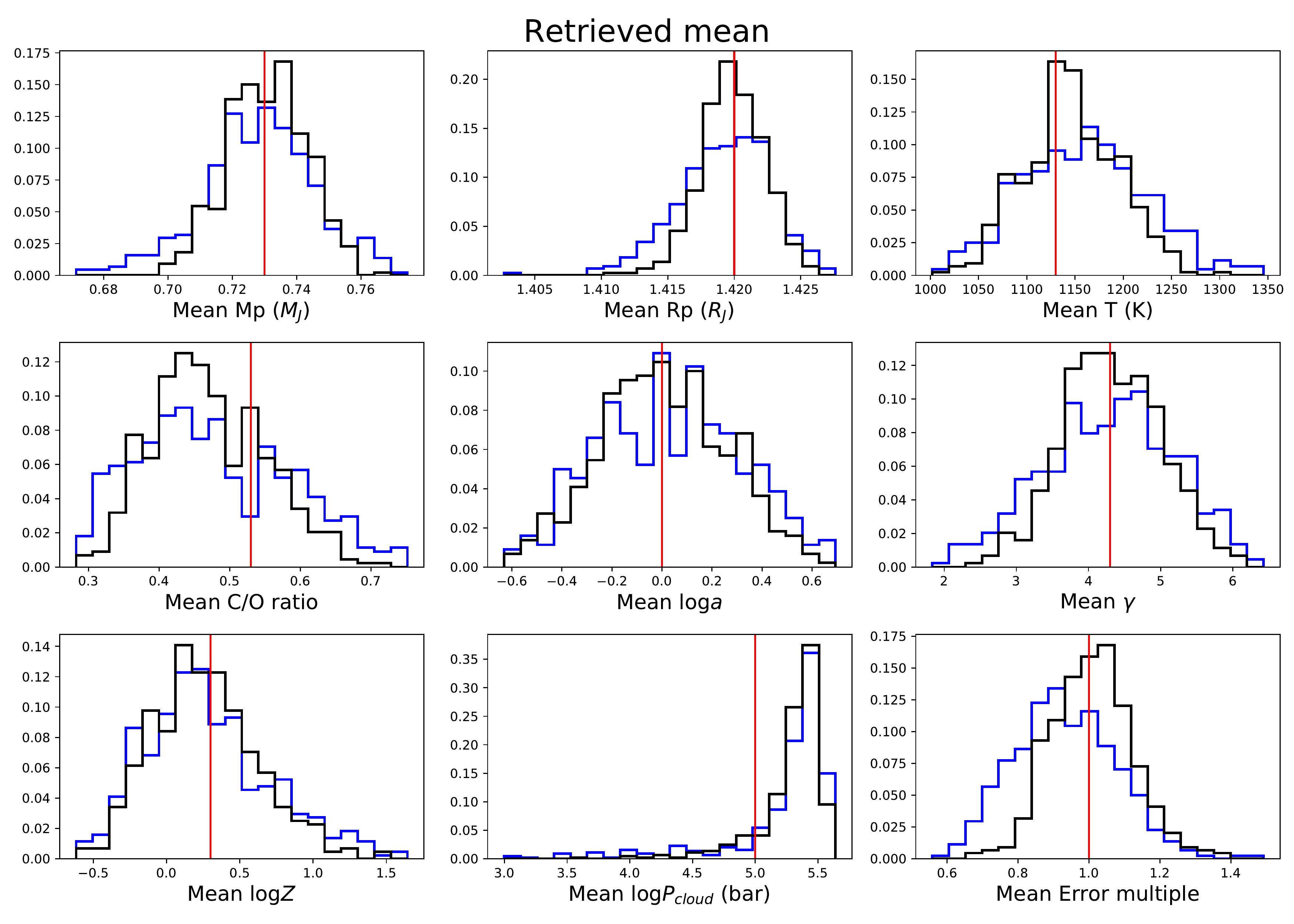}
\figsetgrpnote{The distribution of retrieved means per each parameter, for independent noise (black) and for correlated noise (blue), for the baseline hot Jupiter case. The input values used to generate the spectrum are shown in red.}
\figsetgrpend

\figsetgrpstart
\figsetgrpnum{1.2}
\figsetgrptitle{Means Clear HJ w. Offset}
\figsetplot{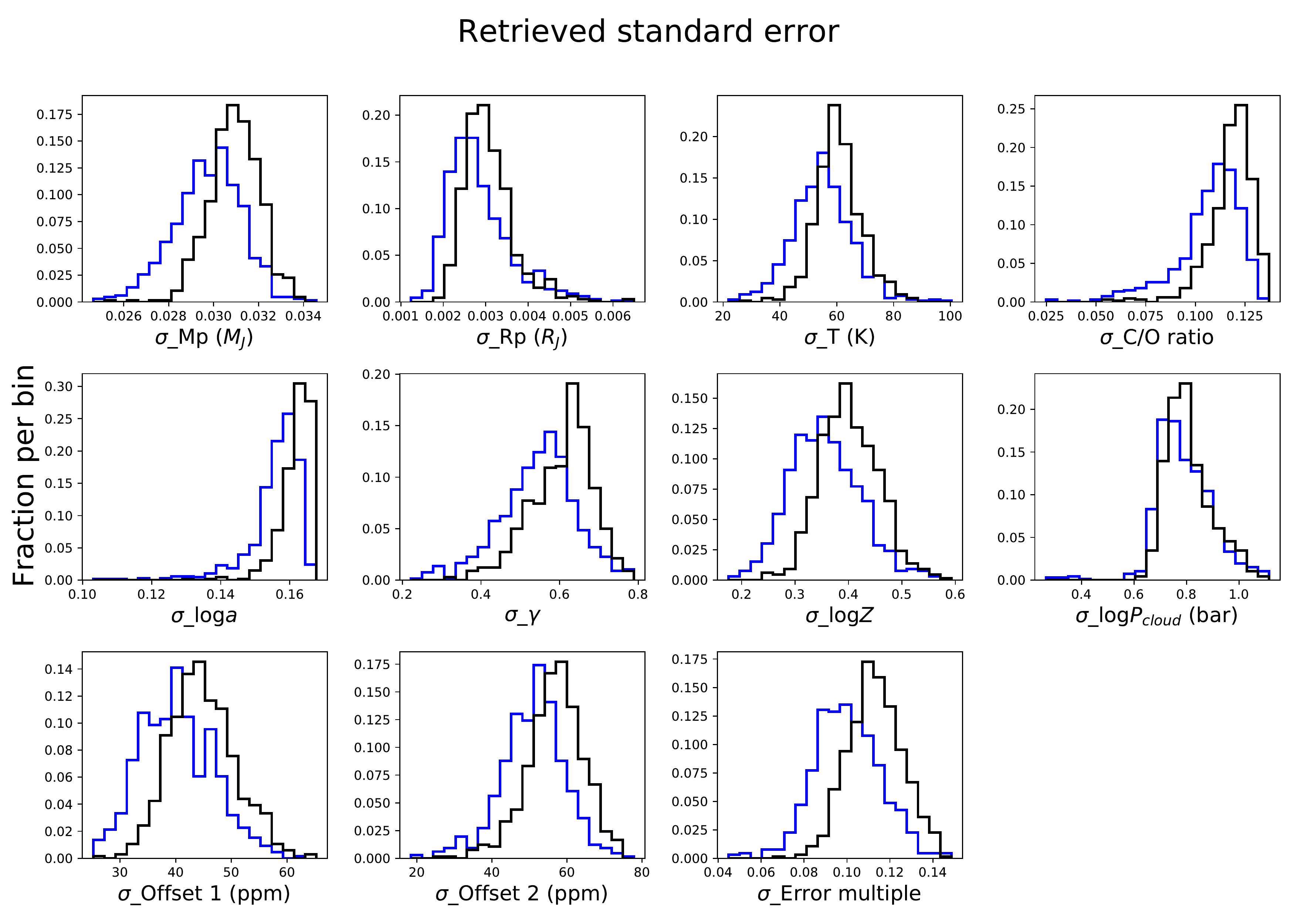}
\figsetgrpnote{Same as Figure \ref{fig:mean}, but for the hot Jupiter case including offsets This figure now includes panels for both instrumental offsets in addition to the original set of retrieved parameters.}
\figsetgrpend

\figsetgrpstart
\figsetgrpnum{1.3}
\figsetgrptitle{Means Cloudy HJ}
\figsetplot{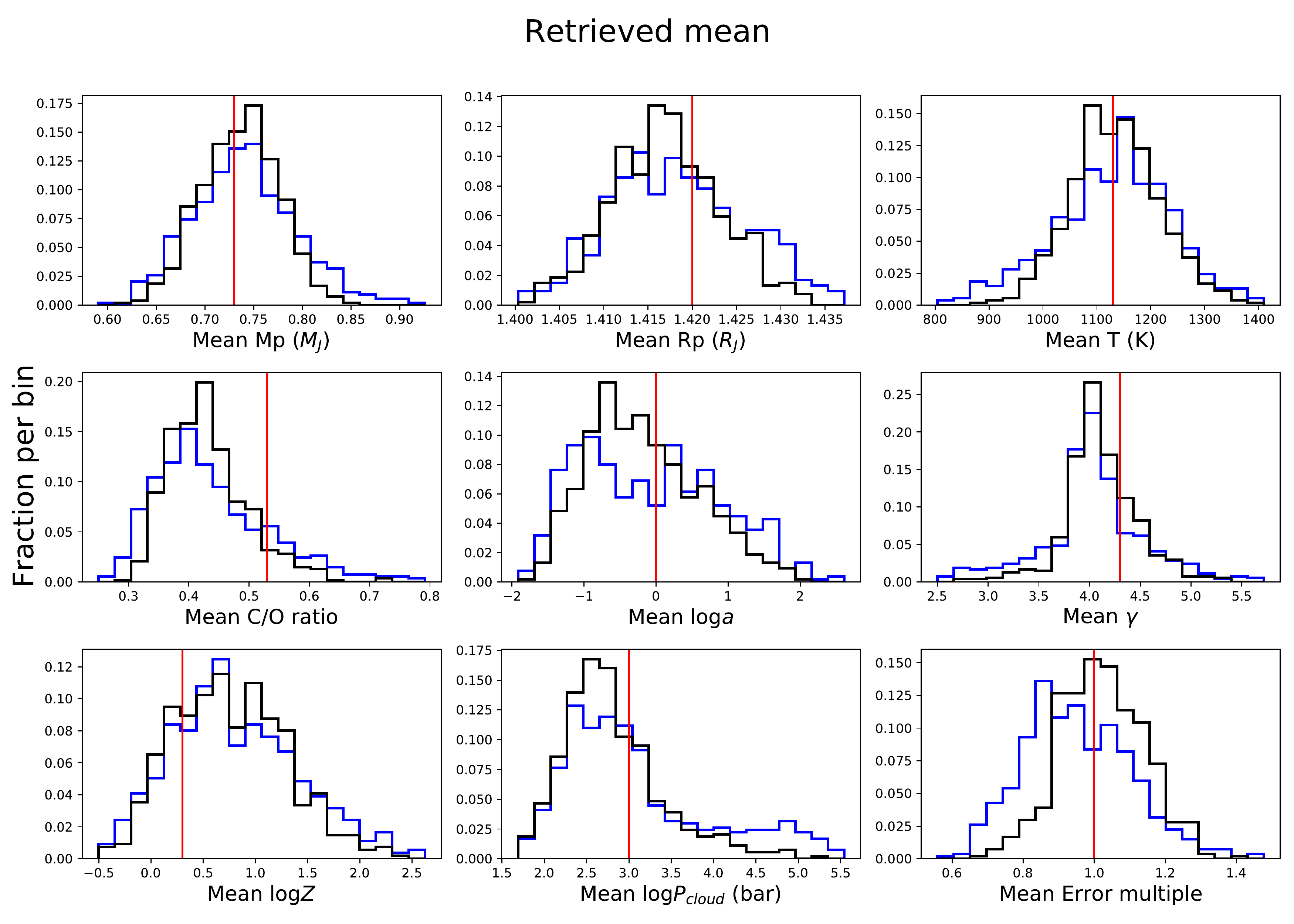}
\figsetgrpnote{Same as Figure \ref{fig:mean}, but for the cloudy hot Jupiter case.}
\figsetgrpend

\figsetgrpstart
\figsetgrpnum{1.4}
\figsetgrptitle{Means HJ High Precision}
\figsetplot{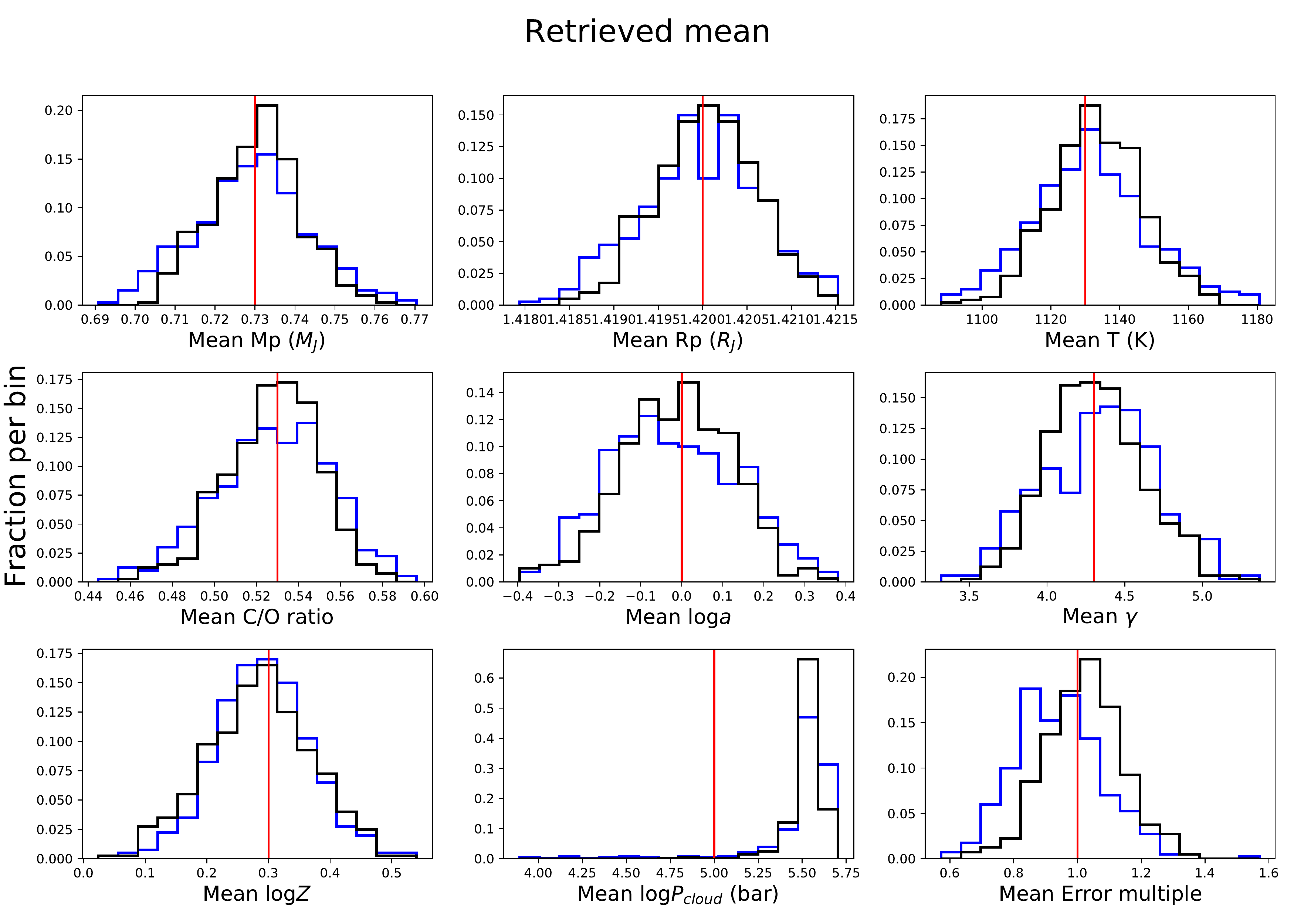}
\figsetgrpnote{Same as Figure \ref{fig:mean}, but for the high precision hot Jupiter case.}
\figsetgrpend

\figsetgrpstart
\figsetgrpnum{1.5}
\figsetgrptitle{Means Warm Neptune}
\figsetplot{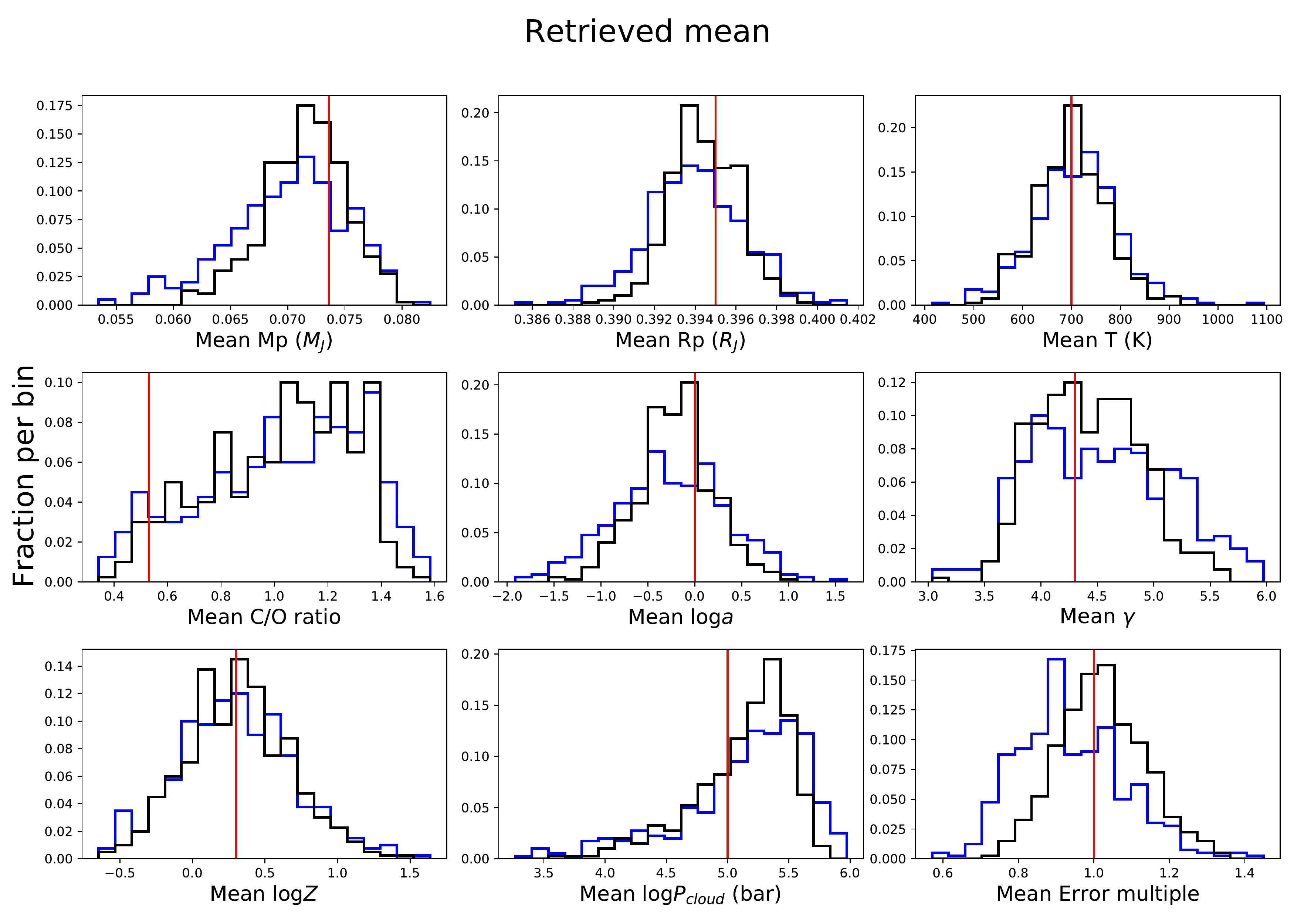}
\figsetgrpnote{Same as Figure \ref{fig:mean}, but for the warm Neptune case.}
\figsetgrpend

\figsetend

\figsetstart
\figsetnum{2}
\figsettitle{Retrieved errors}
\label{figset:errors}

\figsetgrpstart
\figsetgrpnum{2.1}
\figsetgrptitle{Errors Clear HJ}
\figsetplot{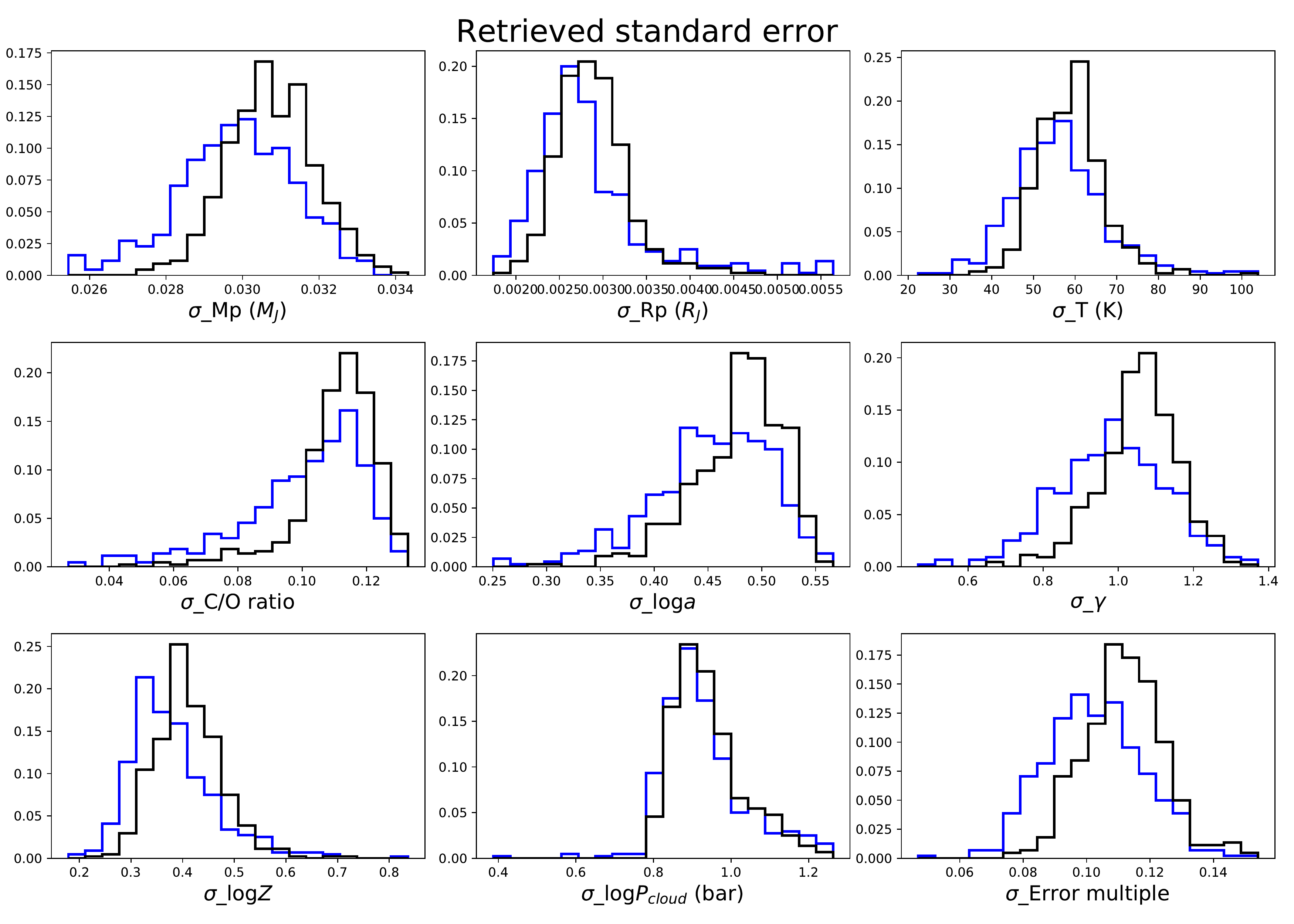}
\figsetgrpnote{The distribution of retrieved standard errors per each parameter, for independent noise (black) and for correlated noise (blue), for the baseline hot Jupiter case.}
\figsetgrpend

\figsetgrpstart
\figsetgrpnum{2.2}
\figsetgrptitle{Errors Clear HJ w. Offset}
\figsetplot{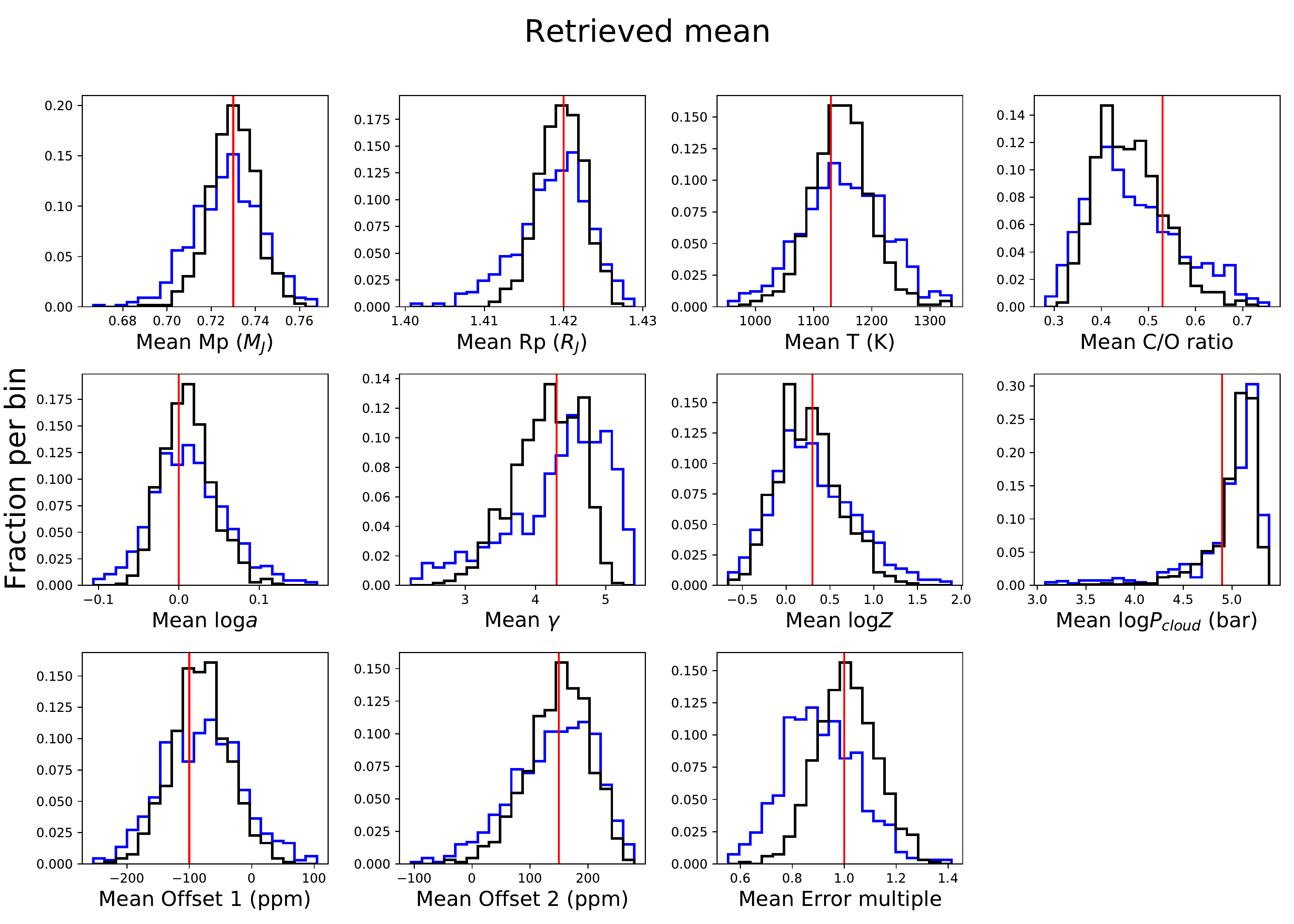}
\figsetgrpnote{Same as Figure \ref{fig:error}, but for the hot Jupiter case including offsets This figure now includes panels for both instrumental offsets in addition to the original set of retrieved parameters.}
\figsetgrpend

\figsetgrpstart
\figsetgrpnum{2.3}
\figsetgrptitle{Errors Cloudy HJ}
\figsetplot{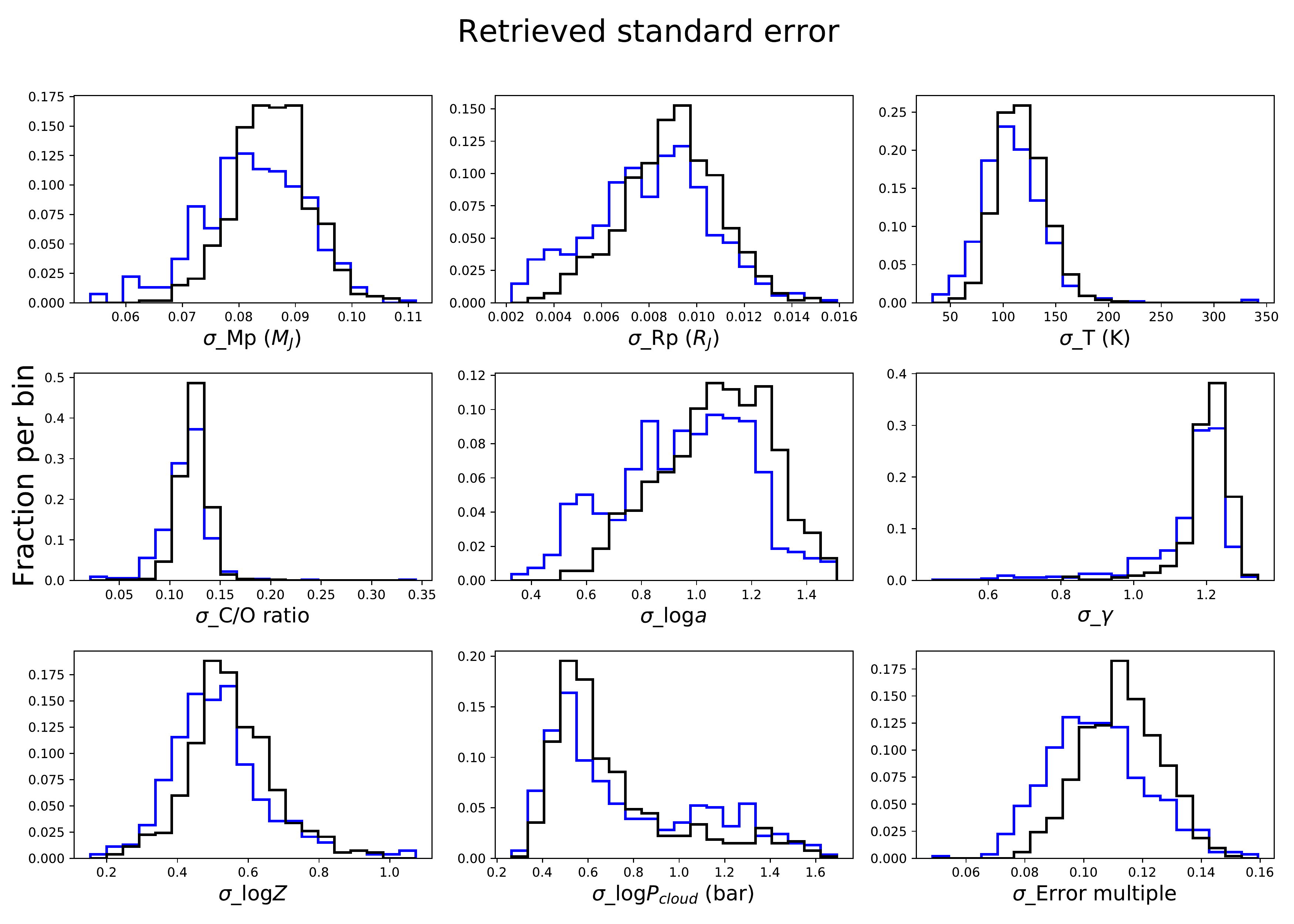}
\figsetgrpnote{Same as Figure \ref{fig:error}, but for the cloudy hot Jupiter case.}
\figsetgrpend

\figsetgrpstart
\figsetgrpnum{2.4}
\figsetgrptitle{Errors HJ High Precision}
\figsetplot{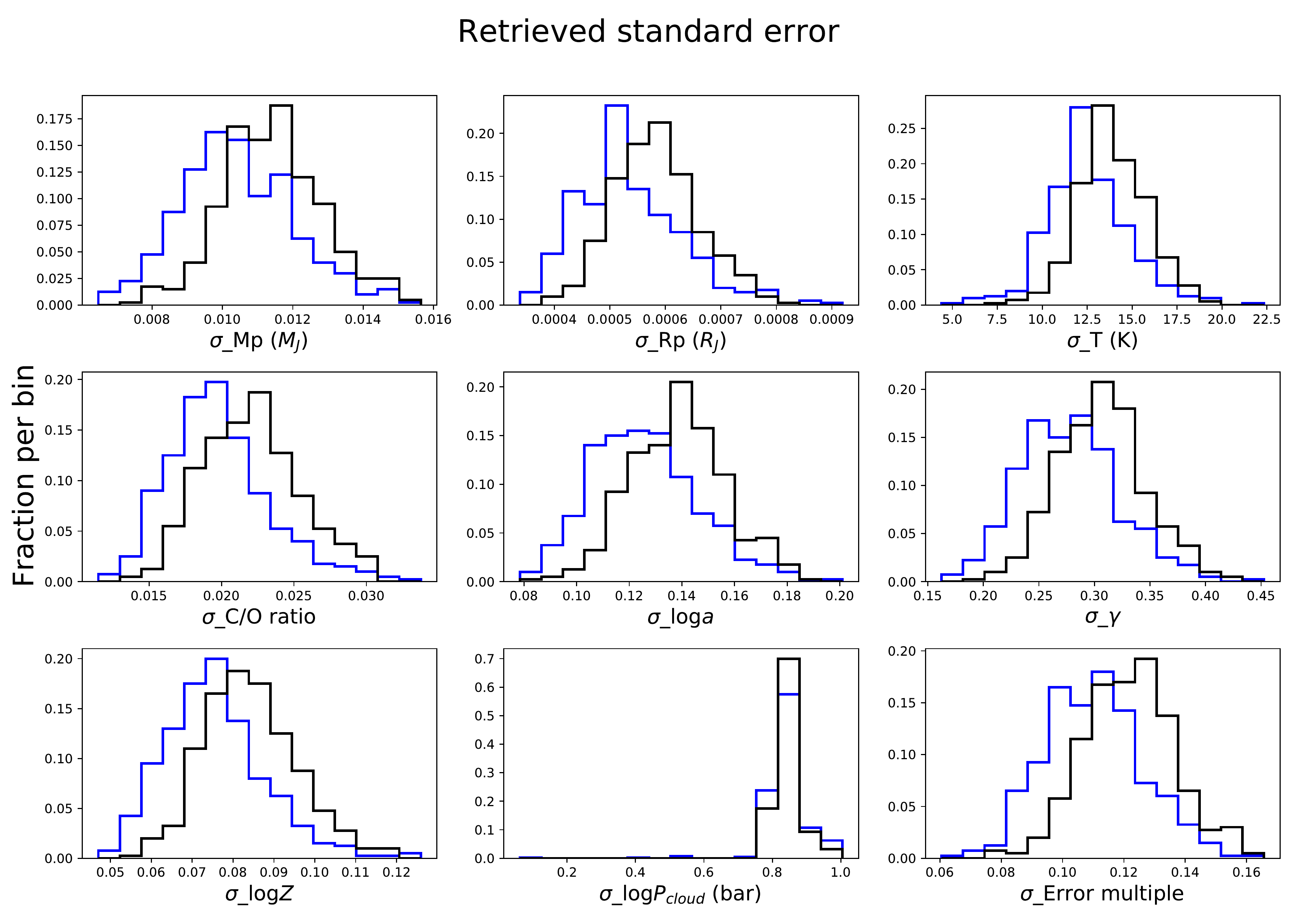}
\figsetgrpnote{Same as Figure \ref{fig:error}, but for the high precision hot Jupiter case.}
\figsetgrpend

\figsetgrpstart
\figsetgrpnum{2.5}
\figsetgrptitle{Errors Warm Neptune}
\figsetplot{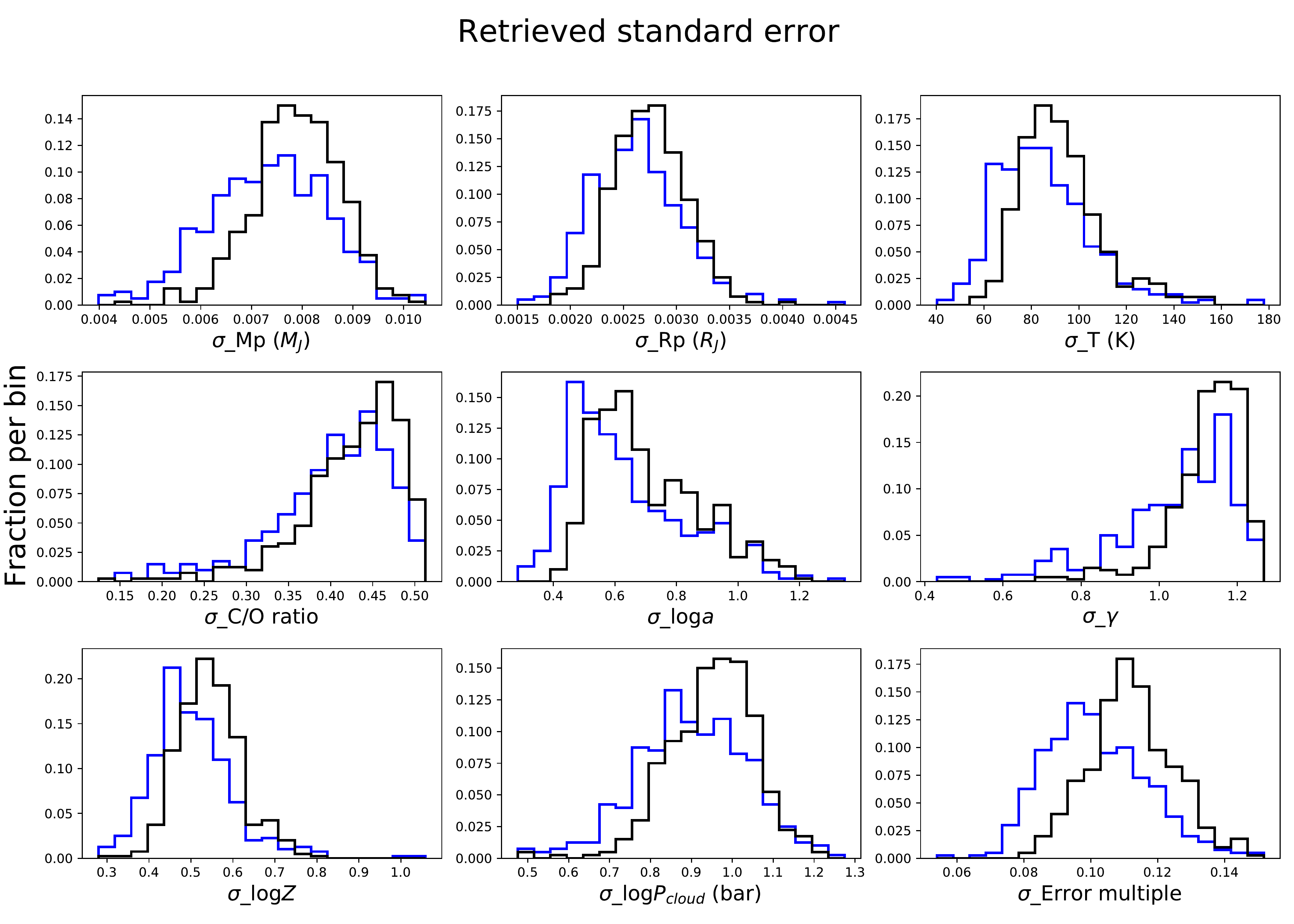}
\figsetgrpnote{Same as Figure \ref{fig:error}, but for the warm Neptune case.}
\figsetgrpend

\figsetend

\subsection{Extensions to Other Planet Parameters}

\begin{figure}[h]
    \centering
    \includegraphics[width=\columnwidth]{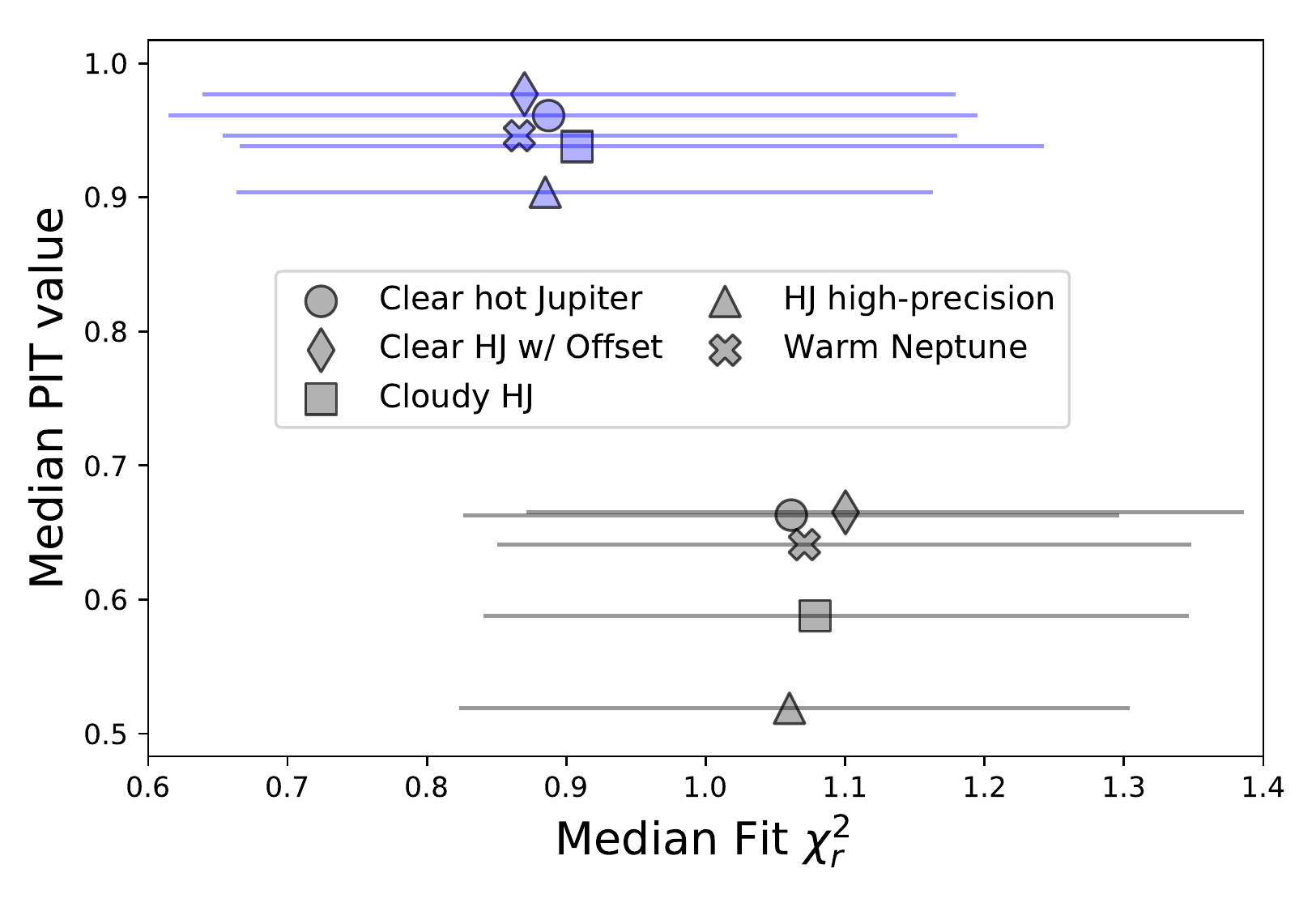}
    \caption{Summary of the goodness-of-fit and retrieval accuracy for the five planetary cases; black is for Gaussian noise, and blue is for correlated noise.  The filled symbols mark the peak in both distributions, and the error bars denote the 1-$\sigma$ spread. Values of $\chi^2_{\textrm{r}}$ closer to unity imply a better fit.  Similarly, lower  {PIT values} correspond to a more accurate retrieval result. This plot shows the information in bottom middle panel of Figure \ref{fig:compare_HJ} for all five cases. For the clear hot Jupiter with offset, the retrieval accuracy is measured as the error in an 11-parameter Gaussian distribution; the rest are quantified in 9 parameters. The correlated retrievals clearly have distinct distributions of goodness-of-fit and accuracy of retrieval, but have enough overlap such that one cannot discern whether a single retrieval instance has correlated noise.}
    \label{fig:summary_scatter}
\end{figure}

In this section, we present our results for planet scenarios other than the baseline clear hot Jupiter case to understand the sensitivities of our results to various system and dataset parameters.  We show histograms of the retrieved mean and retrieved standard error for each parameter in \sout{Figure Sets} Appendix for the remaining planetary scenarios (the hot Jupiter with offsets, cloudy hot Jupiter, high precision hot Jupiter, and warm Neptune).   We generally find that the main results stated so far  hold true for all cases: correlated noise causes both overfitting in $\chi^2_{\textrm{r}}$ and worsening of the accuracy-of-retrieval (i.e.\ larger  {PIT values}). This point is summarized in Figure~\ref{fig:summary_scatter}, in which we show the medians of the $\chi^2_{\textrm{r}}$ and  {PIT value} distributions for each planet realization, i.e.\ distilling down the results of Figure~\ref{fig:compare_HJ} and the like to values quantifying the peak and the spread.

To further quantify this point, we perform a Kol\-mo\-go\-rov-Smir\-nov (K-S) test for the goodness-of-fit and the accuracy-of-retrieval metrics to measure the discrepancy between the results of Gaussian and correlated noise. In Table \ref{tab:kstest}, we show the K-S statistics, $D$, for the fit $\chi^2_{\textrm{r}}$ and  {PIT value representing the }accuracy-of-retrieval. We find that the clear hot Jupiter happens to be the best-case for the smallest discrepancy of accuracy-of-retrieval between Gaussian and correlated noise, and that other cases generally result in further discrepancy between results with Gaussian and correlated noise. 

\begin{deluxetable}{cccccch}
\tablenum{2}
\label{tab:kstest}
\tablewidth{\columnwidth}
\tablecaption{K-S statistic of $\chi^2_r$ and PIT values.}
\tablehead{
\colhead{} & \colhead{$\chi^2_{\textrm{r}}$} && \multicolumn{4}{c}{ {PIT value}} \\
\cline{2-2}\cline{4-6}
\colhead{Case} & \colhead{G-C} && \colhead{$\mathcal{U}$-G} & \colhead{$\mathcal{U}$-C} & \colhead{{G-C}} & \nocolhead{$p$-value}
}
\startdata
Clear HJ & 0.28 && 0.19 & 0.60 & 0.47 & $2 \cdot 10^{-44}$\\
Clear HJ w/ offset & 0.36 && 0.19 & 0.67 & 0.58  & $9 \cdot 10^{-104}$ \\
Cloudy HJ & 0.27 &&0.12 & 0.54 & 0.48 &  $2 \cdot 10^{-60}$ \\
Clear HJ, high precision & 0.28 && 0.04 & 0.48 & 0.47 & $1 \cdot 10^{-15}$\\
Warm Neptune & 0.35 && 0.16 & 0.58 & 0.52 & $3 \cdot 10^{-49}$ \\
\enddata
\tablecomments{The K-S statistic, $D$, measures the maximum vertical discrepancy between cumulative distributions (see Figure \ref{fig:compare_HJ_cumul}), of goodness-of-fit and retrieval accuracy for each planet scenario. The first column shows the two-sample $D$ between the distributions of fit $\chi^2_{\textrm{r}}$ for the Gaussian and correlated noise. The next two columns contain the $D$ between the expected  {uniform} distribution and the PIT value distributions from retrievals with Gaussian and correlated noise, respectively. The final column shows the $D$ between the two distributions.
% , and the last column shows $D_{n,m}$ as a $p$-value.
In all cases, the discrepancy is due to overfitting and worsening of retrievals.}
\end{deluxetable}

\subsubsection{Bias in Non-Rayleigh Scattering Slope}

In the retrievals of the hot Jupiter with instrumental offsets and the warm Neptune, we find that the retrieved haze properties also show the potential to be biased. Correlated noise can bias the scattering slope, $\gamma$, away from the Rayleigh value of 4, misleading the retrieval to infer the presence of aerosols. This bias makes intuitive sense as, if a handful of points in the optical wavelengths align due to correlated noise, those points can mimic the behavior of non-Rayleigh slope \citep{may20}. As such, we caution that a spurious detection of haze can be possible in interpreting data in which the presence of correlated noise is either expected or suspected. 

We suspect that this bias happens more readily for the warm Neptune case compared to the hot Jupiter retrievals because the overall signal is smaller while the data error used to scramble the spectrum was held constant at 75 ppm, resulting in a larger relative error. The warm Neptune spectra consequently have greater potential for large (apparent) optical slopes to manifest.

\subsubsection{Retrieving Offsets}

We ran a set of retrievals that includes non-zero ``offset" parameters between datasets from different instruments. We find that, while the presence of correlated noise does cause underestimation of the uncertainty in the offset in an identical manner to other parameters, it does not worsen the retrieval of the means. The offsets are accurately retrieved in both Gaussian and correlated noise retrievals and do not pose any obvious degeneracies.

This is somewhat surprising as, in our formulation of correlated noise, offsets can be regarded as correlated noise with high correlation and long wavelength-order. For instance, in Figure \ref{fig:method_noise}, the data instances with correlated noise in the \textit{Spitzer} band mimics the presence of an offset. 

It should be obvious that the influence of offset data points will strongly depend on the specific wavelength those points occupy, as well as the offset magnitude and sign. As such, we present here only one possible manifestation of how real data could behave. For instance, we have only considered offsets between datasets disjointed in wavelength, but, say, merging data from ground-based and space-based observations can produce offset data with overlapping wavelength coverage. \citet{yip20} found that if there is overlapping data with non-zero offsets and if free retrievals are used, such offsets can be degenerate with the estimated abundances if equilibrium chemistry is not assumed.

\subsubsection{Effects of Clouds}

The main effect of adding gray clouds to the model, from the point of view of the retrieval, is washing out information contained in the spectrum that originates from the high-pressure portion of the atmosphere. In Figure \ref{fig:truth_HJ}, roughly half of the \textit{Hubble} points are covered by clouds, no longer constraining, say, a baseline radius or metallicity. We find that the broad effect of underestimating uncertainty and biasing means due to correlated noise still holds for cloudy hot Jupiter retrievals.

For the retrieved cloud-top pressure parameter, the main effect is a bias in the retrieved mean. Specifically, the presence of correlated noise disrupts the distribution of retrieved means of cloud-top pressure by extending the tail in the high-pressure direction. In other words, the spectrum is more likely to be understood as having a clear atmosphere. Upon examining the spectra for the retrievals that populate this tail, we find that the correlated noise happens to manifest as a number of data points dipping under the opaque cloud-top where the atmosphere is normally optically thick, thereby mimicking the behavior of a clear atmosphere. %(TODO: maybe add an example?)

\subsubsection{Effects of Higher Precision}

We find that in the hot Jupiter retrieval with high-precision (10 ppm) data, the broad conclusions again still hold. Correlated noise leads to an underestimation of retrieved uncertainty for all parameters. Compared to other cases however, correlated noise does not shift the retrieved means as much, which is to be expected since every noise instance only has a minor deviation from the unpolluted spectrum (specifically a factor of 7.5$\times$ smaller than in our baseline case), even with correlation.

\begin{figure*}[!t]
    \centering
    \includegraphics[width=0.8\textwidth]{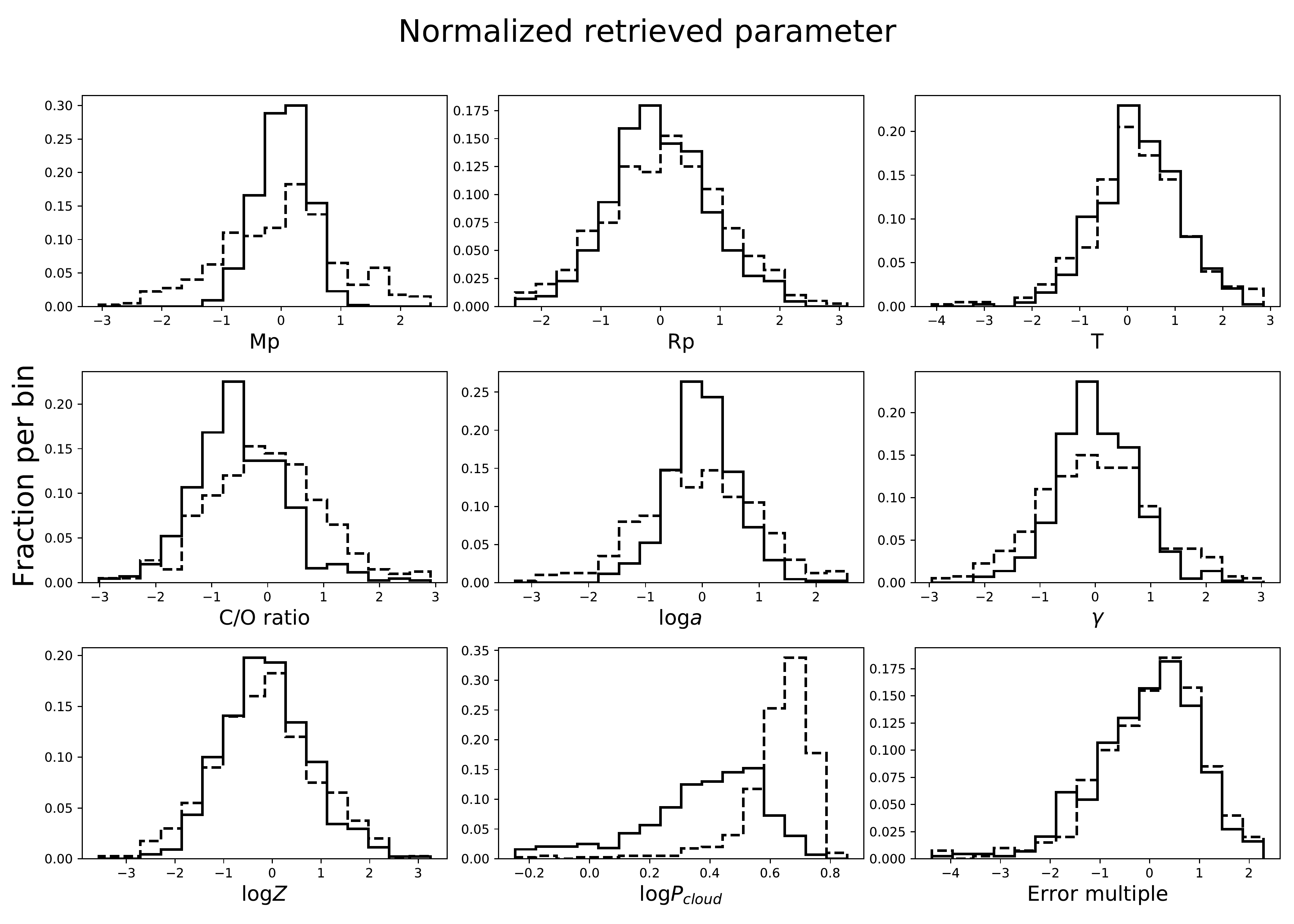}
    \caption{Distributions of each parameter, normalized by the retrieved uncertainty, for the hot Jupiter retrievals.  {A value of 0 indicates that the retrieved mean coincides with the input value.} The dashed lines show the high-precision (10 ppm) case, and the solid line shows the baseline (75 ppm) case. Only the Gaussian noise retrievals are shown here, but the correlated noise case displays similar behavior.}
    \label{fig:HJ_big_small_norm}
\end{figure*}

Comparing the high-precision case to the baseline case with 75 ppm errors, we find that, naturally, both the estimated means are retrieved closer to the input values and the retrieved parameter uncertainties are concurrently smaller. Interestingly, the uncertainties shrink more than the means approach the input values; consequently, in the high-precision case, the retrieval more readily rules out the input. This is shown in Figure \ref{fig:HJ_big_small_norm}, in which the retrieved means are normalized by their retrieved uncertainty to show the number of standard errors the input value is retrieved within for each parameter. The high-precision case (dashed line) actually has more retrievals \textit{further} from the input value when normalized. We find an identical trend for the retrievals with correlated noise. 

\begin{figure}
    \centering
    \includegraphics[width=\columnwidth]{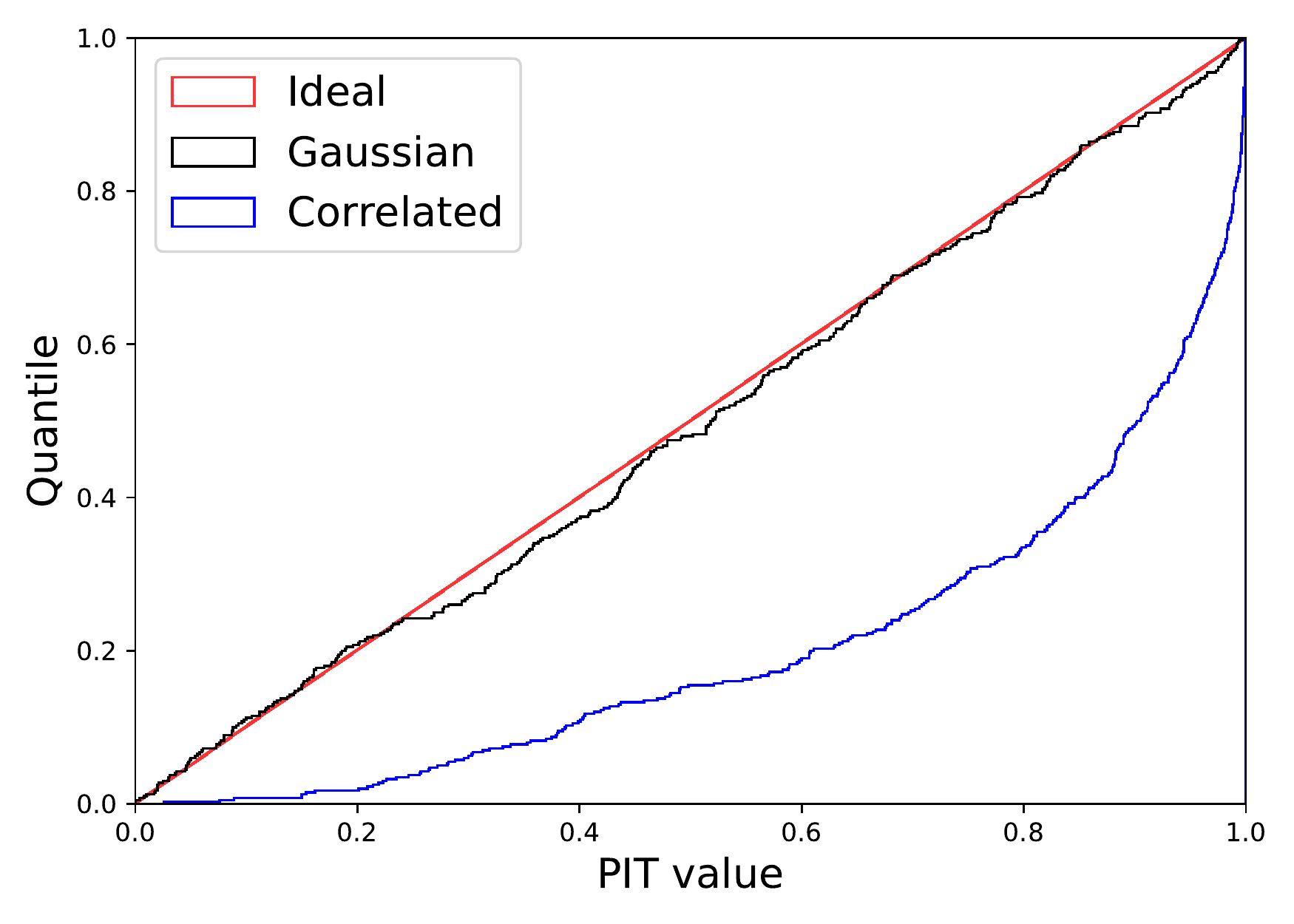}
    \caption{Same as Figure \ref{fig:compare_HJ_cumul}, but for the high-precision case. The smaller data error results in a better agreement between the ideal uniform distribution and the retrieval accuracy in the Gaussian noise retrievals.}
    \label{fig:compare_HJ_small_cumul}
\end{figure}

Additionally, we find the  {PIT value}  distribution for retrievals with Gaussian noise much more successfully follows the ideally expected  {uniform} distribution, with a \mbox{K-S} statistic of $D=0.04$ (see Table~\ref{tab:kstest} and Figure~\ref{fig:compare_HJ_small_cumul}). Given that the accuracy-of-retrieval for all other cases (with 75 ppm error bars instead) are at least somewhat discrepant from the expected distribution even for the Gaussian noise retrievals, this result suggests that, even for hot Jupiters, 75 ppm error bars are too large to assume \textit{a priori} that the retrieved posterior will follow a multivariate Gaussian. This has implications for parameter estimation methods that need this assumption of Gaussian posterior, such as optimal estimation or some of recent machine learning-based retrievals \citep{lin13, cob19}. These methods require that the data uncertainty is small enough such that the forward model behaves linearly over the parameter uncertainties. Retrievals using these methods must be trusted only when the data has exceptional SNR.

\section{Can We Tell if Systematics are Present?} \label{sec:if}
 {
In the motivating example spectrum of HD 97658b in Section \ref{sec:intro}, the presence of correlated noise was suspected based on the fact that no forward model can produce a satisfactory fit under the assumption of randomly scattered residuals. If we are to presume that the retrieval is indeed correct, and the residuals evidence correlated noise contaminating a genuine featureless spectrum, then we should also consider how prevalent unnoticed correlated noise can be in the observed data of other planets. The natural question then is to ask whether there is a more robust and comprehensive way of distinguishing correlated noise within the framework of a retrieval. Especially, given that correlated noise can give rise to overfitting, it is of special interest whether correlated noise can be distinguished from merely overestimated errorbars. A natural way to achieve this is to modify the likelihood function such that it can reward or penalize when residuals are correlated.
}

 {
To test this, we implement a parameterized covariance matrix and let the retrieval estimate the hyperparameter that measures the correlation strength. We use a nearest-neighbor correlated noise model as in \citet{siv96}, where the correlation strength is parameterized by $\epsilon$, such that the covariance matrix element between points at i and j is given by:
}
\begin{equation}
    K_{ij} = \sigma_{i} \sigma_{j} \epsilon^{|i - j|}
\end{equation}

 {
\noindent This differs slightly from the correlated noise model used in Section \ref{sec:methods} in that this model does not depend on the wavelength difference between two points but depends instead on the difference in indices. The two implementations would be identical if the wavelength grid was regularly spaced, with the exponential base giving correlation strength $\epsilon = \mathrm{e}^{-1} \sim 0.37$. While extension to accommodate wavelength-dependent correlation is certainly possible, as a first test this simplification provides a reasonable starting point for exploring whether correlated noise can be retrieved.
}

 {
This simplification allows for writing the likelihood function as:
}
\begin{equation}
    \ln{\mathcal{L}} = - \frac{1}{2} \left[ (N-1) \ln{(1-\epsilon^2)} + \sum_{i=1}^{N} \ln{2 \pi \sigma_i^2} + \frac{Q}{1-\epsilon^2} \right],
\end{equation}

 {
\noindent in which $Q$ is the modified chi-squared-like term related to the error-scaled residuals $R_i$ by:
}

\begin{equation}
    Q = (1+\epsilon^2) \sum_{i=1}^{N} R_i^2 - \epsilon^2 (R_1^2 + R_N^2) - 2 \epsilon \sum_{i=1}^{N-1} R_i R_{i+1}.
\end{equation}

 {
We perform a small grid of retrievals to study when the correlated noise can be distinguished from overestimated errorbars and correctly retrieved. The input parameters for the planet remain the same as Case 1 in Section \ref{sec:methods}, while we vary the noise properties. Our grid consists of three values of error multiple -- $\eta=$0.8, 1, 1.25, three values of correlation strength -- $\epsilon=$0, 0.25, 0.5, and two wavelength grids modeling HST WFC3 and JWST NIRSPEC prism grids. Our NIRSPEC grid consists of 133 linearly spaced points between 0.6 and 5 microns. From a few initial tests we find that the varying the absolute size of the error bar does not change the results, and hence we keep them fixed at 75 ppm. For each combination of noise parameters, we run three retrievals: one in which both error multiple and correlation strength were included as retrieved parameters, and two in which either was removed. We set a uniform prior between 0 and 1 for the correlation strength, as we do not expect an anti-correlation between neighboring points.
}
% \begin{figure*}[!t]
%     \centering
%     \includegraphics[width=0.8\textwidth]
\begin{figure*}
    \centering
    \includegraphics[width=\textwidth]{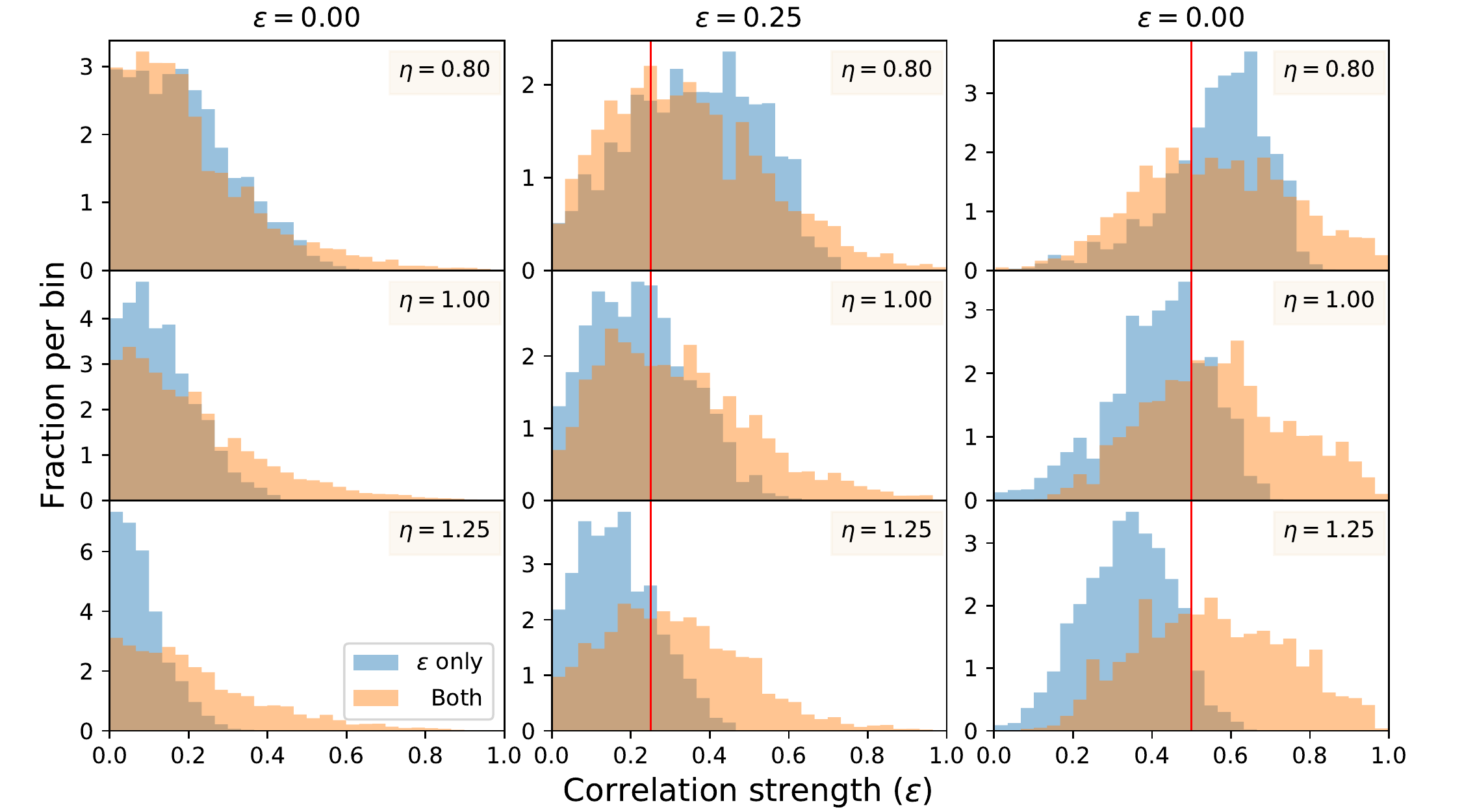}
    \caption{Histograms showing marginalized posterior distribution for the correlation strength parameter for a hot Jupiter on a WFC3-like wavelength grid. The relevant input values for error multiple and correlation strength for each run is shown in legend. Two posteriors are shown for each: one where only correlation strength was retrieved (blue) and one where both error multiple and correlation strength were retrieved (orange). The corresponding marginalized posteriors for the error multiple are not shown. We note that our original model in Section \ref{sec:methods} roughly corresponds to $\epsilon \sim$0.37 (see text).}
    \label{fig:corr_wfc3}
\end{figure*}

\begin{figure*}
    \centering
    \includegraphics[width=\textwidth]{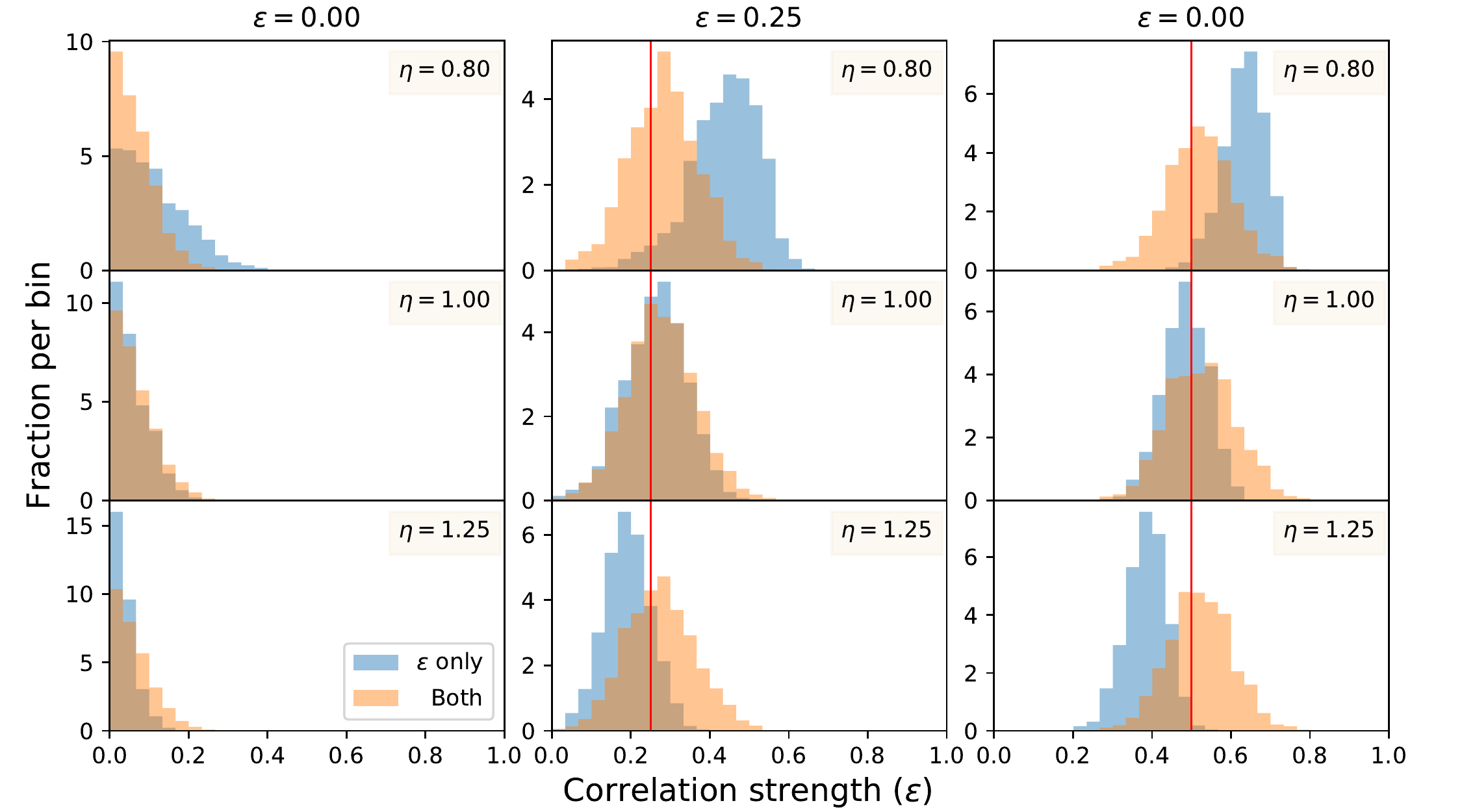}
    \caption{Same as Fig \ref{fig:corr_wfc3}, but for NIRSPEC-like wavelength grid. The horizontal scale has been kept the same as Fig \ref{fig:corr_wfc3} and reflects the full width of the uniform prior used.}
    \label{fig:corr_nirspec}
\end{figure*}

\begin{figure}
    \centering
    \includegraphics[width=\columnwidth]{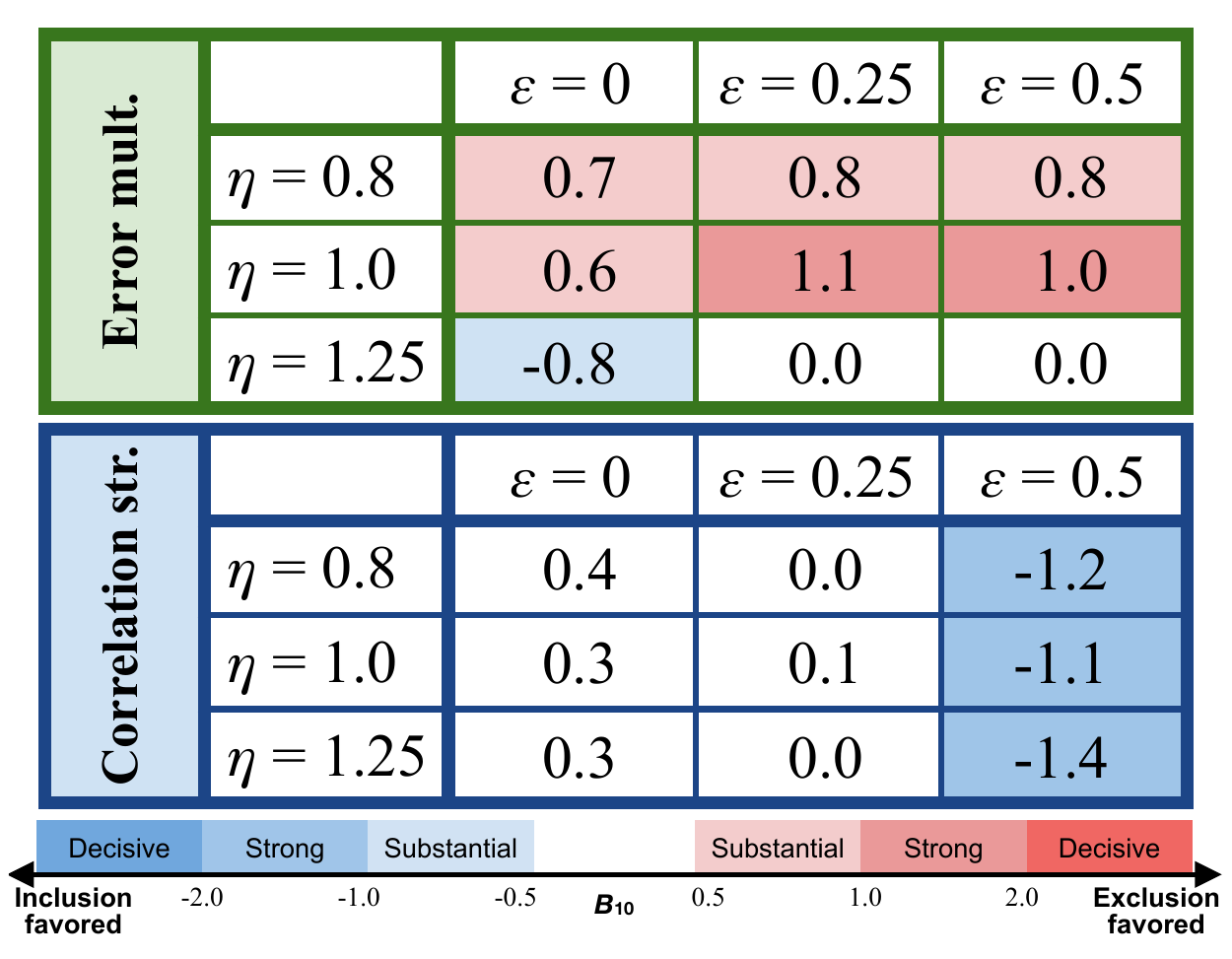}
    \caption{Bayes factors $B_{10}$ of retrieval models that exclude error multiple (top) or correlation strength (bottom) for WFC3-like data. The model that includes both hyperparameters is used as the baseline.  A positive value indicates that the model better fits the data while spanning a smaller prior volume and hence supports the removal of the parameter. The interpretation of the strength of evidence is shown in the colorbar. The Bayes factors are calculated as the difference in estimated Bayesian log-evidence; the values in the table have a resulting uncertainty of $\sim 0.3$.}
    \label{tab:bayes_wfc3}
\end{figure}

\begin{figure}
    \centering
    \includegraphics[width=\columnwidth]{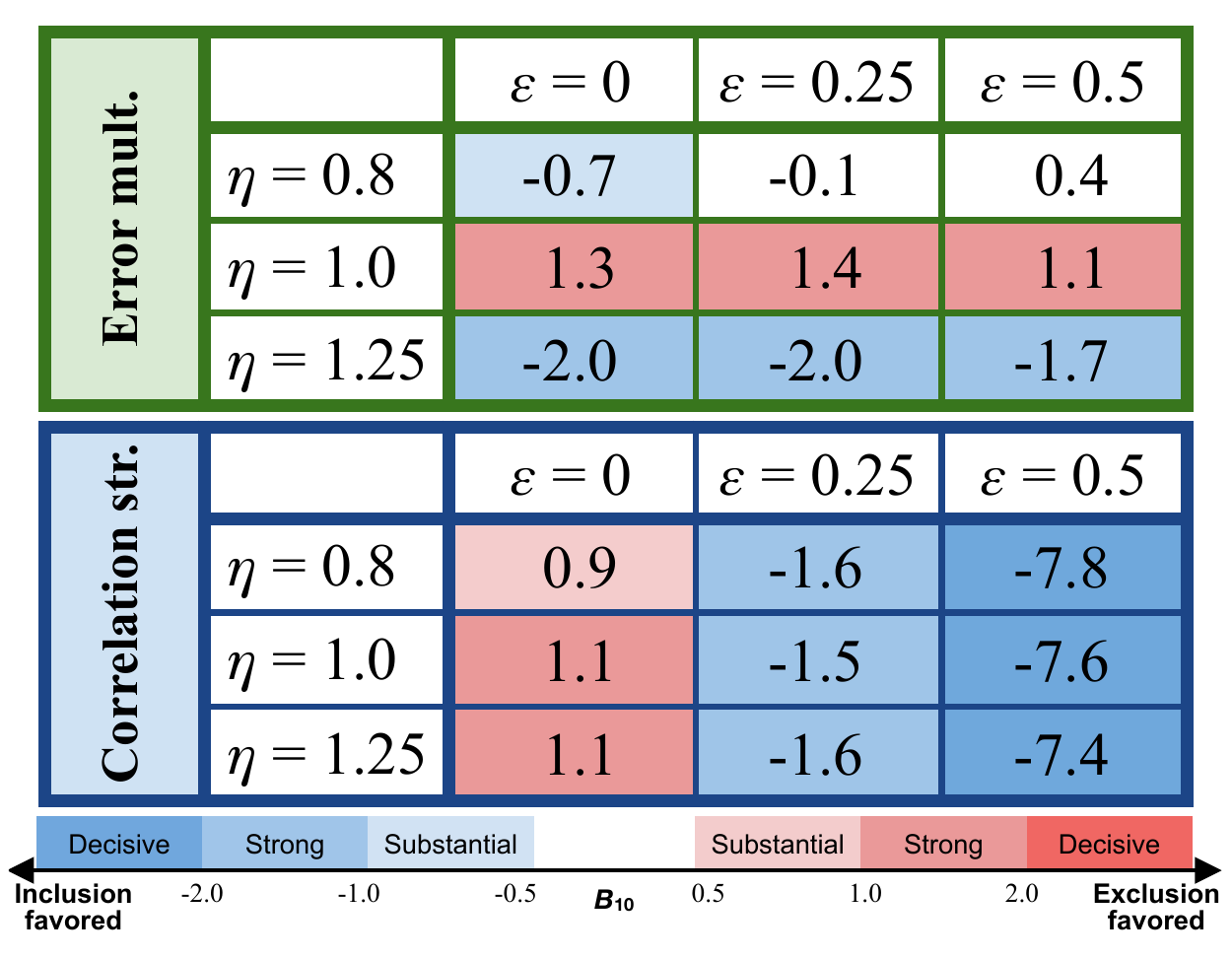}
    \caption{Same as Fig \ref{tab:bayes_wfc3}, but for NIRSPEC-like spectrum. }
    \label{tab:bayes_nirspec}
\end{figure}

 {
We show the marginalized posterior distribution for correlation strength in Figures \ref{fig:corr_wfc3} and \ref{fig:corr_nirspec} for the WFC3 and NIRSPEC grids, respectively. From the posterior distributions it is clear that the correlation strength can be adequately retrieved, and the posterior width is dependent on the number of points in the data, as expected. It can also be seen that underestimated error bars when unaccounted for can be mistaken for the presence of correlated noise, and vice versa, as demonstrated in the previous section. The main difference between the two instruments from the point of view of the retrieval is simply the number of points.
}

 {
Additionally, we calculate the Bayes factor among retrievals to determine whether the inclusion of each parameter is warranted. The results for WFC3- and JWST-like  data is shown in Figure \ref{tab:bayes_wfc3} and \ref{tab:bayes_nirspec}. The case in which both parameters were included is used as the baseline, and the log-ratio (in base 10) of Bayesian evidence is shown for each case in the grid. A positive value indicates that the model better fits the data while spanning a smaller prior volume, indicating that the parameter should be removed. 
}

 {
For WFC3-like data (Figure \ref{tab:bayes_wfc3}), the inclusion of correlation strength is supported with strong evidence only in the strong correlation ($\epsilon=0.5$) cases and is otherwise not easily ruled out either way. Interesetingly, the inclusion of error multiple is disfavored with substantial or stronger evidence not only when the errorbars are correct, but also even in the case of underestimated errorbars. This shows that, for WFC3-like data, the error multiple parameter is not warranted in general.
}

 {
For NIRSPEC-like data (Figure \ref{tab:bayes_nirspec}), the inlcusion of the correlation strength can be more robustly judged. Its inclusion is supported with strong to decisive evidence when correlation is present. Conversely, its removal is supported with substantial to strong evidence when it is not present. This indicates that, for JWST-like data, there is the possibility that we can characterize the correlation during the retrieval. The inclusion of the error multiple can be more robustly judged as well. Its removal is supported when the errorbars are correct, and its inclusion is favored when the errorbars are overestimated. However, when the errorbars are underestimated, its inclusion is supported with substantial evidence only when there is no correlation. This behavior matches with the results from the previous section that correlation can be mistaken for underestimated error bars. This shows that the error multiple is generally not effective at indicating accommodating underestimated errorbars. 
}

 {
The above results show that while it is difficult to conclusively infer the presence of correlated noise with \textit{HST} data, it is certainly within possibility that its presence and strength can be measured with \textit{JWST} data. A few caveats must be made. Our preliminary test presented here treats the data over the entire NIRSPEC wavelength range as sharing correlation, hence providing an abundant number of points for the correlation strength to be measured; in reality, if its discrete grisms are used, any correlation of instrumental origin will be per each wavelength range. Additionally, as we will discuss in Section \ref{sec:mod_lim}, missing physics in the model acts as a source of systematics, which we do not consider here. Furthermore, we used the same noise model to generate the observation instance as well as to retrieve its parameter. While doing so is obviously a gross simplification, especially considering that numerous sources of correlated noise can operate simultaneously, this provides a reasonable starting point towards using a more complex likelihood function to fit for correlated noise. Additionally, adding hyperparameters to a retrieval further dilutes the noise budget, broadening the retrieved uncertainties of other parameters. Ascertaining what degeneracy this incurs on the estimation of other parameters is left for future work. 
}

\section{Discussion} \label{sec:discussion}

\subsection{Model Limitations} \label{sec:mod_lim}
  
A major compounding issue is that, when retrieving on real data, model assumptions and unknowns contribute to and act as systematic errors in addition to the data systematics themselves. In short, bad data are degenerate with bad models. In our study we generated the synthetic observations using the identical forward model as that used in the retrieval in order to minimize any model-dependent effects and to isolate the effects of data systematics. In interpreting real data, the fact that our forward models are a simplified incomplete representation of complex atmospheric phenomena will act as a source of systematic error that will remain pervasive, even if the observed data were perfect and free from their own systematics. We therefore remain open to the possibility that the observed examples of potential systematic noise in the data are in fact due to unaccounted for obscure physics.

For the same reasons discussed above, this will adversely affect high SNR observations in particular, in which the fine (and the not-so-fine) details of the model become discernible. There has recently been a growing body of work that studies the biases incurred by model assumptions and parameterization. To list a few examples for demonstration, \citet{mac20} performed 1-D retrievals on 3-D synthetic spectra to show that the retrieval biases the limb temperature to few hundred Kelvins cooler than the actual day-night mean temperature. \citet{lac20} extended this study to cloudy atmospheres, finding that the presence of aerosols exacerbate the biases induced by 3-D effects when not accounted for. \citet{cha19b} found that using a vertically constant chemical abundance profiles may no longer be sufficient to fully capture signatures of disequilibrium processes in the spectrum. Perhaps most inconspicuously, \citet{bar20b} compared and performed cross-retrieval between retrieval codes developed by three groups and found that \textit{JWST}-quality data is now sensitive to rather rudimentary model unknowns such as the line lists used to generate the opacities and the precision of fundamental constants used.

In our framing of describing biases, these results generally effect shifts in the estimated means of retrieved parameters. Incorporating our conclusion that correlated noise generally leads to underestimated error of retrieved parameters means that biases due to model limitation now strike with a stronger statistical significance. Furthermore, the wavelength-dependent effect of both missing physics and systematics now leave possibility for degenerate interpretations.

\subsection{Instrumental Systematics}

To better understand the significance of our results, it would be useful to consider the different sources of how systematics can arise and evaluate their prevalence especially in the context of future space missions such as \textit{JWST}. While other sources of systematics are possible, such as starspots \citep{rac18, bru19}, inaccurate orbital parameters, or time-dependent telluric contamination in ground-based observations, the key source of systematics that we will discuss here is instrumental. However, we remind the reader that our formulation of correlated noise can be generalized to any effect that results in wavelength-dependent correlation.

While instrumental systematics are expected to be ubiquitous to some extent, the exact magnitude of their effect in generating wavelength-correlated noise has not been fully understood. Yet a handful of observations exist that hint at the existence of such systematics. \citet{col20} argued that, with current facilities, these systematics are visible at the highest level of precision ($\sim$15 ppm), inferred from an unusual behavior of residuals in the H$_2$O band. In the spectrum of HD~97658b in \citet{guo20}, our motivating example in \S \ref{sec:intro}, we inferred from the inability to fit the data as well as no obvious physics being missing that there must be some systematics present, even at a lower precision ($\sim$25 ppm). These examples indicate that some wavelength-correlated noise must be present, unless there is unaccounted for physics in the retrieval model. 

% Original text
% There is some reason to surmise that these systematics are more prevalent (or, at least, more noticeable) in the case of bright host stars, as alluded to in \S \ref{sec:intro}. While a brighter host star allows for a better SNR and higher precision and thereby naturally makes the presence of these systematics more conspicuous, given that the instrumental systematics can also behave differently with bright sources, it is not out of question that there is a separate effect at play here. Comparing the WFC3 spectra of GJ~1214b \citep{kre14} and of HD~97658b \citep{guo20}, for instance, while both planets are comparable sub-Neptunes with featureless spectra and have a similar level of precision, the spectrum of the latter displays an unusual upward trend in transit depth in the redder end. It may be that the relevant difference here is the host star brightness (9.8 versus 6.2 in J-band magnitude).

% One possible explanation for this discrepancy is that when the host star is brighter, fewer stacking of observations is required, thereby making the correlated noise instance more apparent. For a dimmer host star, by contrast, the stacking required to achieve the same SNR naturally averages out any non-repeatable correlated noise. The difference between the earlier and updated spectra of HD 97658b from \citet{knu14} and \citet{guo20} illustrates this point....

There is some reason to surmise that these systematics are more prevalent (or, at least, more noticeable) in the case of bright host stars. A brighter host star allows for a better SNR and higher precision and thereby naturally makes the presence of these systematics more conspicuous compared to a lower SNR data. Additionally, even at the same data quality, a brighter host star requires fewer stacking of observations; for a dimmer host star, by contrast, the number of stacking required to achieve the same SNR naturally averages out any \textit{non-repeatable} correlated noise. Finally, as alluded to in \S \ref{sec:intro}, given that instrumental systematics can also behave differently with bright sources, it is not out of question that there is a separate effect at play here beyond SNR which may persist through multiple observations in a repeatable fashion.

Comparing the WFC3 spectra of GJ~1214b \citep{kre14} and of HD~97658b \citep{guo20} illustrates this point. While both planets are comparable sub-Neptunes with featureless spectra and have a similar level of precision, the spectrum of the latter displays an unusual upward trend in transit depth in the redder end and other wavelength-correlated residuals throughout the WFC3 bandpass. The relevant difference here may be the host star brightness (9.8 versus 6.2 in J-band magnitude, respectively). The spectrum of GJ~1214b is the combination of stacking 15 transits, whereas that of HD~97658b has 4. Further, the spectrum of HD~97685b presented in \citet{knu14}, which only had the first 2 visits, shows the most obvious possible example of correlated noise due to systematics. % {add these spectra as figure?}

These types of systematics may be even more pernicious for future high SNR observations from \textit{JWST} for a few reasons. First, as the noise floor is lower,  {correlated} noise will be relatively more prominent even if it actually manifests at weaker levels. Then one can no longer reliably assume that the observed noise is strictly photon-dominated.  {This requires an additional step of modeling out now wavelength-dependent systematics during the data reduction}, which is necessarily (although perhaps not practically) incomplete. Secondly, the high SNR per transit means that stacking will be unnecessary for most targets. As such, non-repeating systematics do not get averaged out. Thirdly, we have demonstrated that higher precision leads to biases of stronger significance. This is true even in the absence of systematics in the sense that a retrieval will be more sensitive to the observational instance. Fourthly, we can predict that our understanding of the characteristics of \textit{JWST} instruments and their appropriate data reduction tools will be only partially correct, at least during the initial few cycles of \textit{JWST} before practical experience accumulates. Finally, as planets around bright host stars allow for achieving high SNR, they will make attractive targets for \textit{JWST} observation. However, if the above intuition that bright host stars can exacerbate instrumental systematics is true, it adds another dimension to consider when selecting targets for observation, in addition to the SNR. % {wording for last sentence.....}

%Then one can no longer reliably assume that the observed noise is photon-noise dominated for a broader range of observations, and must be modelled out during the data reduction, which is necessarily (although perhaps not practically) incomplete. Secondly, we have demonstrated that higher precision leads to biases of stronger significance. This is true even in the absence of systematics in the sense that a retrieval will be more sensitive to the observation instance. Thirdly, we can predict that our understanding of the instrumental characteristics of \textit{JWST} instruments and the appropriate data reduction tools will be only partially correct, at least during the initial few cycles of \textit{JWST} before practical experience accumulates. Finally, as planets around bright host stars allow for achieving good SNR, they will make attractive targets for \textit{JWST} observation. However, if the above intuition that bright host stars can exacerbate instrumental systematics is true, it adds another dimension to consider when selecting for observation targets in addition to the SNR.

 {Given that this is the case, it would be worthwhile to put the above heuristic that bright stars bring about correlated noise to a more formal test. This can be accomplished with \textit{JWST} if the correlation strength can actually be measured, and by marginalizing over the magnitude of the host star to obtain a trend. While this would not comprise a main scientific objective of any program, correlation strength is a parameter we can try to measure for all observations, so this is a test we can perform at no extra cost in observation time.}

\subsection{Data Outliers and Free Retrieval}

\textsc{platon}  {originally} supports equilibrium chemistry retrievals only. Using this method the molecular abundance of each species at a given temperature and pressure is set by the metallicity and carbon-to-oxygen ratio. A popular alternative method to constrain chemistry is to use ``free" retrievals, in which the abundances of each species are allowed to vary independently. This accounts for any non-equilibrium chemistry effects in the atmosphere, brought on by vertical mixing or photochemical interactions. To establish how wavelength-dependent systematics or outliers can bias certain species, we implement free retrievals in \textsc{platon} and perform some basic tests in addition to the suite of retrievals already presented that used equilibrium chemistry models.

 {\textsc{platon} already natively supports inputting custom chemical profiles to its forward models, but only accepts equilibrium chemistry parameters during retrievals. We extend its capability by allowing it to accept custom chemical abundances during retrievals as well. As such, all other details regarding how the spectrum is calculated during the retrieval remain exactly the same as the original implementation in \textsc{platon}.} We assume that each species has a vertically fixed mixing ratio. This is a reasonable approximation over the pressure range probed by transit spectroscopy at present data quality, and most current 1-D free retrieval codes parameterize the composition using this assumption \citep{cal19}. \textit{JWST}-quality data may merit a more complex prescription, such as a 2-part vertical abundance profile \citep{cha19b}, but for now we do not consider these possibilities.  %, since species with a single line correspond to only one photospheric pressure. (wording??) 
Additionally, for these tests we remove the simulated \textit{Spitzer} data points from our synthetic spectra to limit the wavelength range, thereby reducing the choice of necessary species to be included. We assume a H/He-dominated background atmosphere. We include H$_2$O, Na, K, TiO, and VO as they are the primary detectable species in the remaining wavelength and temperature range; we are mostly interested in how differently species with broad absorption features (e.g. H$_2$O) versus species with narrow ones (e.g. Na) can be biased.

\begin{figure}[h]
    \centering
    \includegraphics[width=\linewidth]{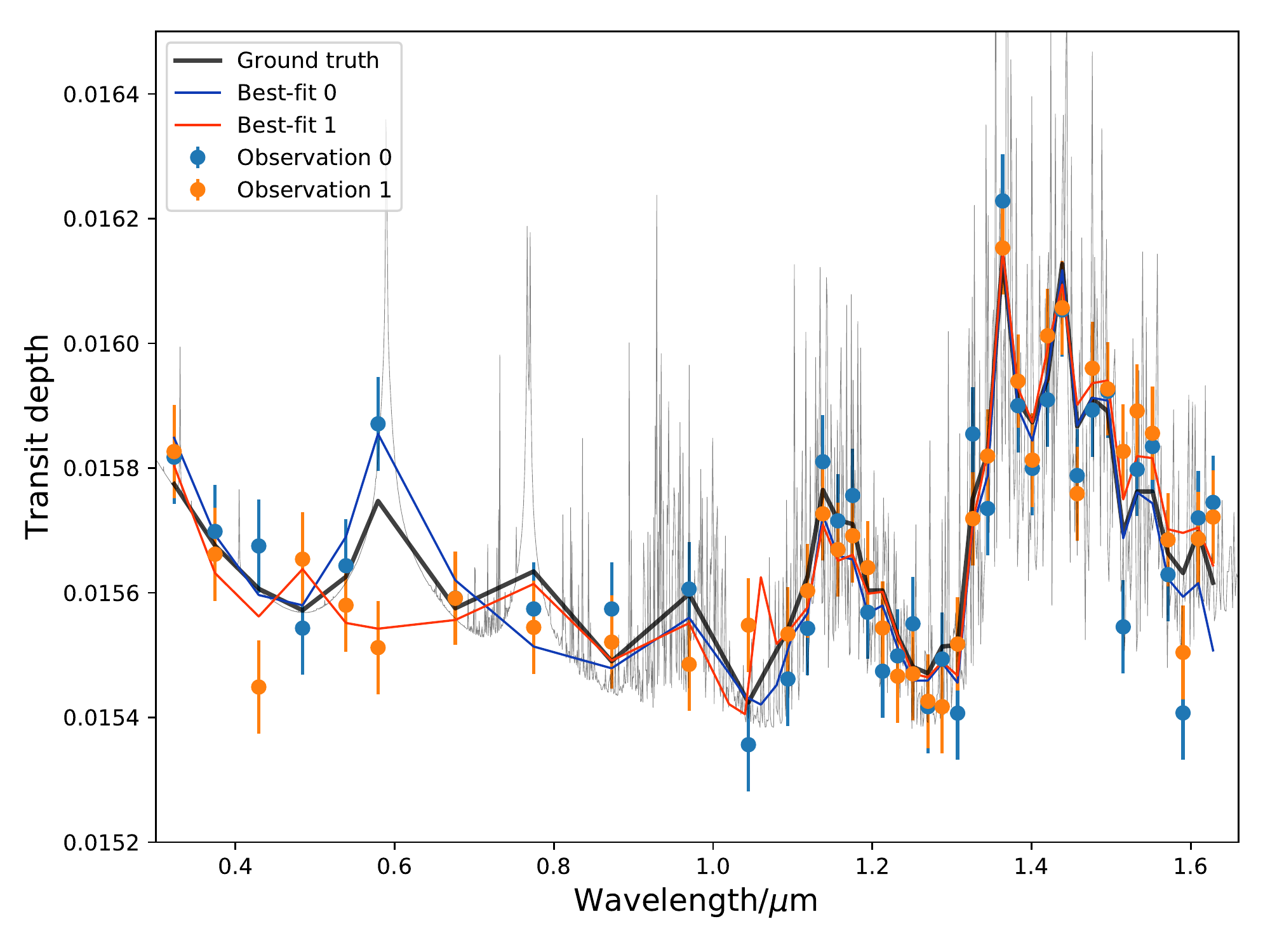}
    \caption{Two observation instances (blue and orange) of the ground-truth spectrum (black) for our baseline hot Jupiter case detailed in Table~\ref{tab:ground}. The best-fit spectrum for each observation are also shown.}
    \label{fig:free_input}
\end{figure}

\begin{figure*}[!t]
    \centering
    \includegraphics[width=6.5in]{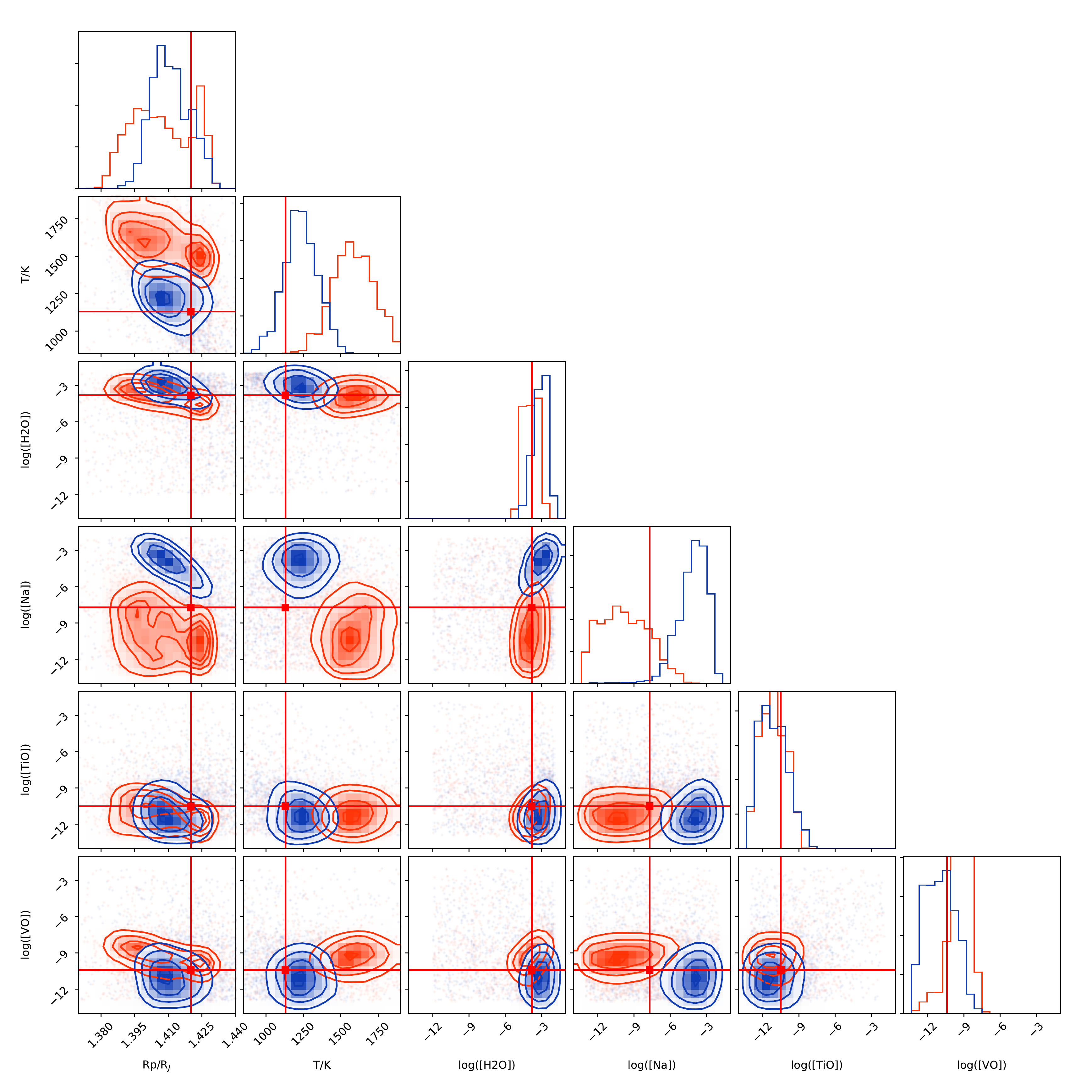}
    \caption{The retrieved posterior from the two input datasets from Figure \ref{fig:free_input}. We only show the relevant parameters. The colors corresponding to each dataset are the same as in Figure~\ref{fig:free_input}.  Black crosshairs indicate the ground-truth input values.
    }
    \label{fig:free_corner}
\end{figure*}

We show a free retrieval on two observational instances of the same baseline hot Jupiter as described in Section~\ref{sec:methods}, where we did not add correlated noise or any instrumental offsets. Figure \ref{fig:free_input} shows the two random input data realizations and the best fit spectra, and Figure \ref{fig:free_corner} shows the retrieved posteriors.

The point of interest here is how the retrieved abundances compare for species with broad spectral features (H$_2$O, $1-2$ $\mu$m) and with a narrow feature (Na, 0.58~$\mu$m). As one may expect, for both of the two random spectra, the retrieved posterior distribution for water shows a tighter constraint than for Na, which is dictated almost entirely by one data point. Consequently, between the two posteriors as well, the retrieved distributions for Na show little overlap, ruling out each other by $\geq$ 2$\sigma$. These are simply two randomly drawn samples, but it demonstrates the point that measurements of water abundance or metallicity are more robust compared to that of say Na or K abundances, because of the impact the former parameters have across a broader wavelength range. 

% \sout{ We are perhaps belaboring the obvious point that more data points provide a tighter constraint. The more subtle point we wish to make here is that any given posterior distribution is sensitive to the noise realization and is therefore itself a random instance drawn from an underlying distribution \citep{fen18}.  }    

\section{Summary and Future Work} \label{sec:summary}

Atmospheric retrieval provides a robust framework to interface theory and observations and is a key tool to furthering our understanding of exoplanets. One major outstanding issue is disentangling the effects of systematic biases that may be in operation, and in response there is a growing body of work in the literature that investigates the consequences of biases that arise from forward model assumptions.

This paper instead presents an assessment of biases that arise from systematic noise in data, while remaining agnostic as to the source of such systematics.  {We stress that, although our implementation of correlated noise (using Gaussian process) is just one mathematical option, the general results remain robust.} We find that the presence of correlated noise can mislead us in various ways. We are more likely on average to obtain better goodness-of-fit, but obtain worse retrieval accuracy overall. This is due to both the parameter mean being biased and the retrieved error being underestimated. Specifically, we observe that correlated noise can bias the retrieved aerosol properties, mimicking non-Rayleigh slopes or misrepresenting the location of a cloud deck. Additionally, we find that offsets between datasets can be correctly retrieved and are not degenerate with retrieved chemistry when equilibrium chemistry is assumed, so long as the forward model is an accurate depiction of the atmosphere.  {We also find that while correlated noise cannot be characterized during retrieval for \textit{HST} data, there is potential (and perhaps necessity) for \textit{JWST} data, even though our tests reflect optimistic conditions.} Additionally, we validate the intuition that retrievals are sensitive to individual noise instances, and, especially in the context of free retrievals, that statistical outliers can have significant effects on the retrieved chemistry.

\acknowledgements

We thank Jacob Bean, Drake Deming, and Kevin Stevenson for useful conversations about data systematics in exoplanet spectra and Cole Miller for discussions on the finer points of Bayesian statistics.  J.I.\ and E.M.-R.K.\ acknowledge support from the National Science Foundation under CAREER grant No.\ 1931736 and from the Research Corporation for Science Advancement through their Cottrell Scholar program.

%\begin{thebibliography}{}
\bibliography{ref}

\begin{thebibliography}{}
\expandafter\ifx\csname natexlab\endcsname\relax\def\natexlab#1{#1}\fi
\providecommand{\url}[1]{\href{#1}{#1}}

\bibitem[{{Andrae} {et~al.}(2010){Andrae}, {Schulze-Hartung}, \&
  {Melchior}}]{and10}
{Andrae}, R., {Schulze-Hartung}, T., \& {Melchior}, P. 2010, arXiv e-prints,
  arXiv:1012.3754

\bibitem[{{Asplund} {et~al.}(2009){Asplund}, {Grevesse}, {Sauval}, \&
  {Scott}}]{asp09}
{Asplund}, M., {Grevesse}, N., {Sauval}, A.~J., \& {Scott}, P. 2009, \araa, 47,
  481

\bibitem[{{Barstow} {et~al.}(2020){Barstow}, {Changeat}, {Garland}, {Line},
  {Rocchetto}, \& {Waldmann}}]{bar20b}
{Barstow}, J.~K., {Changeat}, Q., {Garland}, R., {et~al.} 2020, \mnras, 493,
  4884

\bibitem[{{Barstow} \& {Heng}(2020)}]{bar20a}
{Barstow}, J.~K., \& {Heng}, K. 2020, arXiv e-prints, arXiv:2003.14311

\bibitem[{{Benneke} {et~al.}(2019){Benneke}, {Knutson}, {Lothringer},
  {Crossfield}, {Moses}, {Morley}, {Kreidberg}, {Fulton}, {Dragomir}, {Howard},
  {Wong}, {D{\'e}sert}, {McCullough}, {Kempton}, {Fortney}, {Gilliland },
  {Deming}, \& {Kammer}}]{ben19}
{Benneke}, B., {Knutson}, H.~A., {Lothringer}, J., {et~al.} 2019, Nature
  Astronomy, 3, 813

\bibitem[{{Bruno} {et~al.}(2020){Bruno}, {Lewis}, {Alam}, {L{\'o}pez-Morales},
  {Barstow}, {Wakeford}, {Sing}, {Henry}, {Ballester}, {Bourrier}, {Buchhave},
  {Cohen}, {Mikal-Evans}, {Garc{\'\i}a Mu{\~n}oz}, {Lavvas}, \&
  {Sanz-Forcada}}]{bru19}
{Bruno}, G., {Lewis}, N.~K., {Alam}, M.~K., {et~al.} 2020, \mnras, 491, 5361

\bibitem[{{Caldas} {et~al.}(2019){Caldas}, {Leconte}, {Selsis}, {Waldmann},
  {Bord{\'e}}, {Rocchetto}, \& {Charnay}}]{cal19}
{Caldas}, A., {Leconte}, J., {Selsis}, F., {et~al.} 2019, \aap, 623, A161

\bibitem[{{Changeat} {et~al.}(2019){Changeat}, {Edwards}, {Waldmann}, \&
  {Tinetti}}]{cha19b}
{Changeat}, Q., {Edwards}, B., {Waldmann}, I.~P., \& {Tinetti}, G. 2019, \apj,
  886, 39

\bibitem[{{Cobb} {et~al.}(2019){Cobb}, {Himes}, {Soboczenski}, {Zorzan},
  {O'Beirne}, {G{\"u}ne{\textcommabelow s} Baydin}, {Gal}, {Domagal-Goldman},
  {Arney}, {Angerhausen}, \& {2018 NASA FDL Astrobiology Team}}]{cob19}
{Cobb}, A.~D., {Himes}, M.~D., {Soboczenski}, F., {et~al.} 2019, \aj, 158, 33

\bibitem[{{Col{\'o}n} {et~al.}(2020){Col{\'o}n}, {Kreidberg}, {Welbanks},
  {Line}, {Madhusudhan}, {Beatty}, {Tamburo}, {Stevenson}, {Mandell},
  {Rodriguez}, {Barclay}, {Lopez}, {Stassun}, {Angerhausen}, {Fortney},
  {James}, {Pepper}, {Ahlers}, {Plavchan}, {Awiphan}, {Kotnik}, {McLeod},
  {Murawski}, {Chotani}, {LeBrun}, {Matzko}, {Rea}, {Vidaurri}, {Webster},
  {Williams}, {Sheraden Cox}, {Tan}, \& {Gilbert}}]{col20}
{Col{\'o}n}, K.~D., {Kreidberg}, L., {Welbanks}, L., {et~al.} 2020, arXiv
  e-prints, arXiv:2005.05153

\bibitem[{{Deming} \& {Seager}(2017)}]{dem17}
{Deming}, D., \& {Seager}, S. 2017, arXiv e-prints, arXiv:1701.00493

\bibitem[{{Deming} {et~al.}(2011){Deming}, {Sada}, {Jackson}, {Peterson},
  {Agol}, {Knutson}, {Jennings}, {Haase}, \& {Bays}}]{dem11}
{Deming}, D., {Sada}, P.~V., {Jackson}, B., {et~al.} 2011, \apj, 740, 33

\bibitem[{{Deming} {et~al.}(2013){Deming}, {Wilkins}, {McCullough}, {Burrows},
  {Fortney}, {Agol}, {Dobbs-Dixon}, {Madhusudhan}, {Crouzet}, {Desert},
  {Gilliland}, {Haynes}, {Knutson}, {Line}, {Magic}, {Mandell}, {Ranjan},
  {Charbonneau}, {Clampin}, {Seager}, \& {Showman}}]{dem13}
{Deming}, D., {Wilkins}, A., {McCullough}, P., {et~al.} 2013, \apj, 774, 95

\bibitem[{{Dittmann} {et~al.}(2009){Dittmann}, {Close}, {Green}, {Scuderi}, \&
  {Males}}]{dit09}
{Dittmann}, J.~A., {Close}, L.~M., {Green}, E.~M., {Scuderi}, L.~J., \&
  {Males}, J.~R. 2009, \apj, 699, L48

\bibitem[{{Evans} {et~al.}(2015){Evans}, {Aigrain}, {Gibson}, {Barstow},
  {Amundsen}, {Tremblin}, \& {Mourier}}]{eva15}
{Evans}, T.~M., {Aigrain}, S., {Gibson}, N., {et~al.} 2015, \mnras, 451, 680

\bibitem[{Feroz \& Hobson(2008)}]{fer08}
Feroz, F., \& Hobson, M.~P. 2008, Monthly Notices of the Royal Astronomical
  Society, 384, 449–463.
\newblock \url{http://dx.doi.org/10.1111/j.1365-2966.2007.12353.x}

\bibitem[{{Foreman-Mackey} {et~al.}(2013){Foreman-Mackey}, {Hogg}, {Lang}, \&
  {Goodman}}]{for13}
{Foreman-Mackey}, D., {Hogg}, D.~W., {Lang}, D., \& {Goodman}, J. 2013, \pasp,
  125, 306

\bibitem[{{Fortney} {et~al.}(2020){Fortney}, {Visscher}, {Marley}, {Hood},
  {Line}, {Thorngren}, {Freedman}, \& {Lupu}}]{for20}
{Fortney}, J.~J., {Visscher}, C., {Marley}, M.~S., {et~al.} 2020, \aj, 160, 288

\bibitem[{{Garhart} {et~al.}(2020){Garhart}, {Deming}, {Mandell}, {Knutson},
  {Wallack}, {Burrows}, {Fortney}, {Hood}, {Seay}, {Sing}, {Benneke}, {Fraine},
  {Kataria}, {Lewis}, {Madhusudhan}, {McCullough}, {Stevenson}, \&
  {Wakeford}}]{gar20}
{Garhart}, E., {Deming}, D., {Mandell}, A., {et~al.} 2020, \aj, 159, 137

\bibitem[{{Gennaro}(2018)}]{gen18}
{Gennaro}, M. e.~a. 2018, {WFC3 Data Handbook, Version 4.0}, Baltimore: STScI

\bibitem[{{Guo} {et~al.}(2020){Guo}, {Crossfield}, {Dragomir}, {Kosiarek},
  {Lothringer}, {Mikal-Evans}, {Rosenthal}, {Benneke}, {Knutson}, {Dalba},
  {Kempton}, {Henry}, {McCullough}, {Barman}, {Blunt}, {Chontos}, {Fortney},
  {Fulton}, {Hirsch}, {Howard}, {Isaacson}, {Matthews}, {Mocnik}, {Morley},
  {Petigura}, \& {Weiss}}]{guo20}
{Guo}, X., {Crossfield}, I. J.~M., {Dragomir}, D., {et~al.} 2020, arXiv
  e-prints, arXiv:2004.03601

\bibitem[{{Harrington}(2016)}]{har16}
{Harrington}, J. 2016, {Atmospheric Retrievals from Exoplanet Observations and
  Simulations with BART}, NASA Proposal id.16-XPR16-106, ,

\bibitem[{{Hou Yip} {et~al.}(2020){Hou Yip}, {Changeat}, {Edwards}, {Morvan},
  {Chubb}, {Tsiaras}, {Waldmann}, \& {Tinetti}}]{yip20}
{Hou Yip}, K., {Changeat}, Q., {Edwards}, B., {et~al.} 2020, arXiv e-prints,
  arXiv:2009.10438

\bibitem[{{Knutson} {et~al.}(2007){Knutson}, {Charbonneau}, {Noyes}, {Brown},
  \& {Gilliland}}]{knu07}
{Knutson}, H.~A., {Charbonneau}, D., {Noyes}, R.~W., {Brown}, T.~M., \&
  {Gilliland}, R.~L. 2007, \apj, 655, 564

\bibitem[{{Knutson} {et~al.}(2014){Knutson}, {Dragomir}, {Kreidberg},
  {Kempton}, {McCullough}, {Fortney}, {Bean}, {Gillon}, {Homeier}, \&
  {Howard}}]{knu14}
{Knutson}, H.~A., {Dragomir}, D., {Kreidberg}, L., {et~al.} 2014, \apj, 794,
  155

\bibitem[{Kreidberg {et~al.}(2014)Kreidberg, Bean, Désert, Benneke, Deming,
  Stevenson, Seager, Berta-Thompson, Seifahrt, \& Homeier}]{kre14}
Kreidberg, L., Bean, J.~L., Désert, J.-M., {et~al.} 2014, Nature, 505,
  69–72.
\newblock \url{http://dx.doi.org/10.1038/nature12888}

\bibitem[{{Lacy} \& {Burrows}(2020)}]{lac20}
{Lacy}, B.~I., \& {Burrows}, A.~S. 2020, arXiv e-prints, arXiv:2006.06899

\bibitem[{{Line} {et~al.}(2011){Line}, {Vasisht}, {Chen}, {Angerhausen}, \&
  {Yung}}]{lin11}
{Line}, M.~R., {Vasisht}, G., {Chen}, P., {Angerhausen}, D., \& {Yung}, Y.~L.
  2011, \apj, 738, 32

\bibitem[{{Line} {et~al.}(2013){Line}, {Wolf}, {Zhang}, {Knutson}, {Kammer},
  {Ellison}, {Deroo}, {Crisp}, \& {Yung}}]{lin13}
{Line}, M.~R., {Wolf}, A.~S., {Zhang}, X., {et~al.} 2013, \apj, 775, 137

\bibitem[{{Lothringer} {et~al.}(2018){Lothringer}, {Benneke}, {Crossfield},
  {Henry}, {Morley}, {Dragomir}, {Barman}, {Knutson}, {Kempton}, {Fortney},
  {McCullough}, \& {Howard}}]{lot18}
{Lothringer}, J.~D., {Benneke}, B., {Crossfield}, I. J.~M., {et~al.} 2018, \aj,
  155, 66

\bibitem[{{MacDonald} {et~al.}(2020){MacDonald}, {Goyal}, \& {Lewis}}]{mac20}
{MacDonald}, R.~J., {Goyal}, J.~M., \& {Lewis}, N.~K. 2020, arXiv e-prints,
  arXiv:2003.11548

\bibitem[{{Madhusudhan}(2018)}]{mad18}
{Madhusudhan}, N. 2018, {Atmospheric Retrieval of Exoplanets}, 104

\bibitem[{{Marley} {et~al.}(2013){Marley}, {Ackerman}, {Cuzzi}, \&
  {Kitzmann}}]{mar13}
{Marley}, M.~S., {Ackerman}, A.~S., {Cuzzi}, J.~N., \& {Kitzmann}, D. 2013,
  {Clouds and Hazes in Exoplanet Atmospheres}, ed. S.~J. {Mackwell}, A.~A.
  {Simon-Miller}, J.~W. {Harder}, \& M.~A. {Bullock}, 367

\bibitem[{{May} {et~al.}(2020){May}, {Gardner}, {Rauscher}, \&
  {Monnier}}]{may20}
{May}, E.~M., {Gardner}, T., {Rauscher}, E., \& {Monnier}, J.~D. 2020, \aj,
  159, 7

\bibitem[{{Miller-Ricci Kempton} {et~al.}(2012){Miller-Ricci Kempton},
  {Zahnle}, \& {Fortney}}]{kem12}
{Miller-Ricci Kempton}, E., {Zahnle}, K., \& {Fortney}, J.~J. 2012, \apj, 745,
  3

\bibitem[{{Pinhas} {et~al.}(2018){Pinhas}, {Rackham}, {Madhusudhan}, \&
  {Apai}}]{pin18}
{Pinhas}, A., {Rackham}, B.~V., {Madhusudhan}, N., \& {Apai}, D. 2018, \mnras,
  480, 5314

\bibitem[{{Rackham} {et~al.}(2018){Rackham}, {Apai}, \& {Giampapa}}]{rac18}
{Rackham}, B.~V., {Apai}, D., \& {Giampapa}, M.~S. 2018, \apj, 853, 122

\bibitem[{Rodgers(2000)}]{rod00}
Rodgers, C.~D. 2000, Inverse Methods for Atmospheric Sounding (WORLD
  SCIENTIFIC), https://www.worldscientific.com/doi/pdf/10.1142/3171.
\newblock \url{https://www.worldscientific.com/doi/abs/10.1142/3171}

\bibitem[{{Sing} {et~al.}(2016){Sing}, {Fortney}, {Nikolov}, {Wakeford},
  {Kataria}, {Evans}, {Aigrain}, {Ballester}, {Burrows}, {Deming},
  {D{\'e}sert}, {Gibson}, {Henry}, {Huitson}, {Knutson}, {Lecavelier Des
  Etangs}, {Pont}, {Showman}, {Vidal-Madjar}, {Williamson}, \&
  {Wilson}}]{sin16}
{Sing}, D.~K., {Fortney}, J.~J., {Nikolov}, N., {et~al.} 2016, \nat, 529, 59

\bibitem[{{Sivia} \& {Skilling}(1996)}]{siv96}
{Sivia}, D., \& {Skilling}, J. 1996, {Data Analysis: A Bayesian Tutorial}
  ({Oxford University Press}).
\newblock
  \url{https://global.oup.com/academic/product/data-analysis-9780198568322?cc=us&lang=en&#}

\bibitem[{{Stevenson} \& {Fowler}(2019)}]{ste19}
{Stevenson}, K.~B., \& {Fowler}, J. 2019, {Analyzing Eight Years of Transiting
  Exoplanet Observations Using WFC3's Spatial Scan Monitor}, Space Telescope
  WFC Instrument Science Report, , , arXiv:1910.02073

\bibitem[{{Taylor} {et~al.}(2020){Taylor}, {Parmentier}, {Irwin}, {Aigrain},
  {Lee}, \& {Krissansen-Totton}}]{tay20}
{Taylor}, J., {Parmentier}, V., {Irwin}, P. G.~J., {et~al.} 2020, \mnras, 493,
  4342

\bibitem[{{Tsiaras} {et~al.}(2016){Tsiaras}, {Waldmann}, {Rocchetto}, {Varley},
  {Morello}, {Damiano}, \& {Tinetti}}]{tsi16}
{Tsiaras}, A., {Waldmann}, I.~P., {Rocchetto}, M., {et~al.} 2016, \apj, 832,
  202

\bibitem[{{Waldmann} {et~al.}(2015){Waldmann}, {Tinetti}, {Rocchetto},
  {Barton}, {Yurchenko}, \& {Tennyson}}]{wal15}
{Waldmann}, I.~P., {Tinetti}, G., {Rocchetto}, M., {et~al.} 2015, \apj, 802,
  107

\bibitem[{{Zhang} {et~al.}(2019){Zhang}, {Chachan}, {Kempton}, \&
  {Knutson}}]{zha19}
{Zhang}, M., {Chachan}, Y., {Kempton}, E.~M.-R., \& {Knutson}, H.~A. 2019,
  \pasp, 131, 034501

\end{thebibliography}

\appendix

\section{Retrieval Histograms for Various Planet Realizations}

\begin{figure*}[!h]
    \centering
    \includegraphics[width=0.7\textwidth]{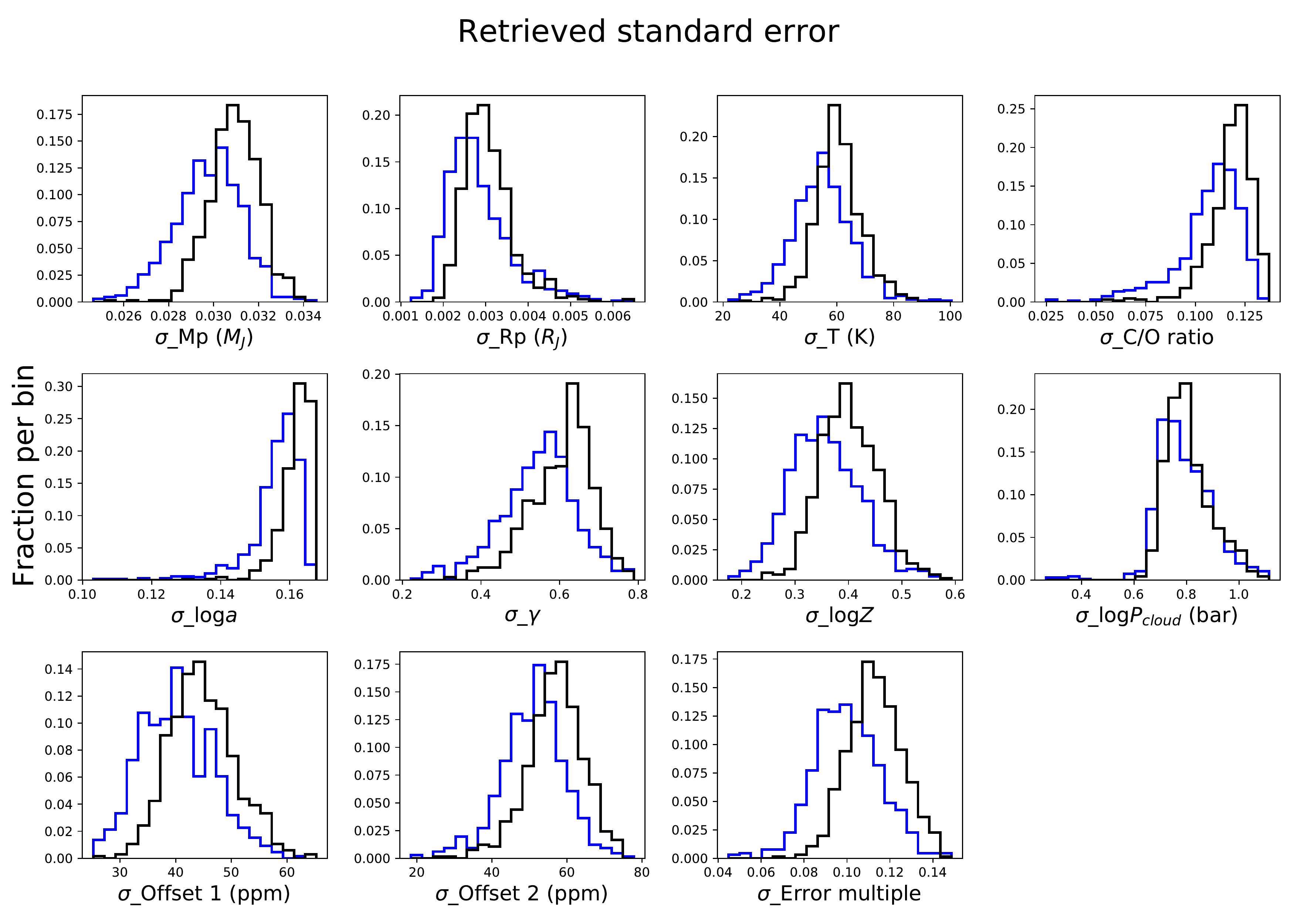}
    \caption{Same as Figure \ref{fig:mean}, but for the hot Jupiter case including offsets. This figure now includes panels for both instrumental offsets in addition to the original set of retrieved parameters.}
    \label{fig:offset_mean}
% \end{figure*}

% \begin{figure*}[!h]
    % \centering
    \includegraphics[width=0.75\textwidth]{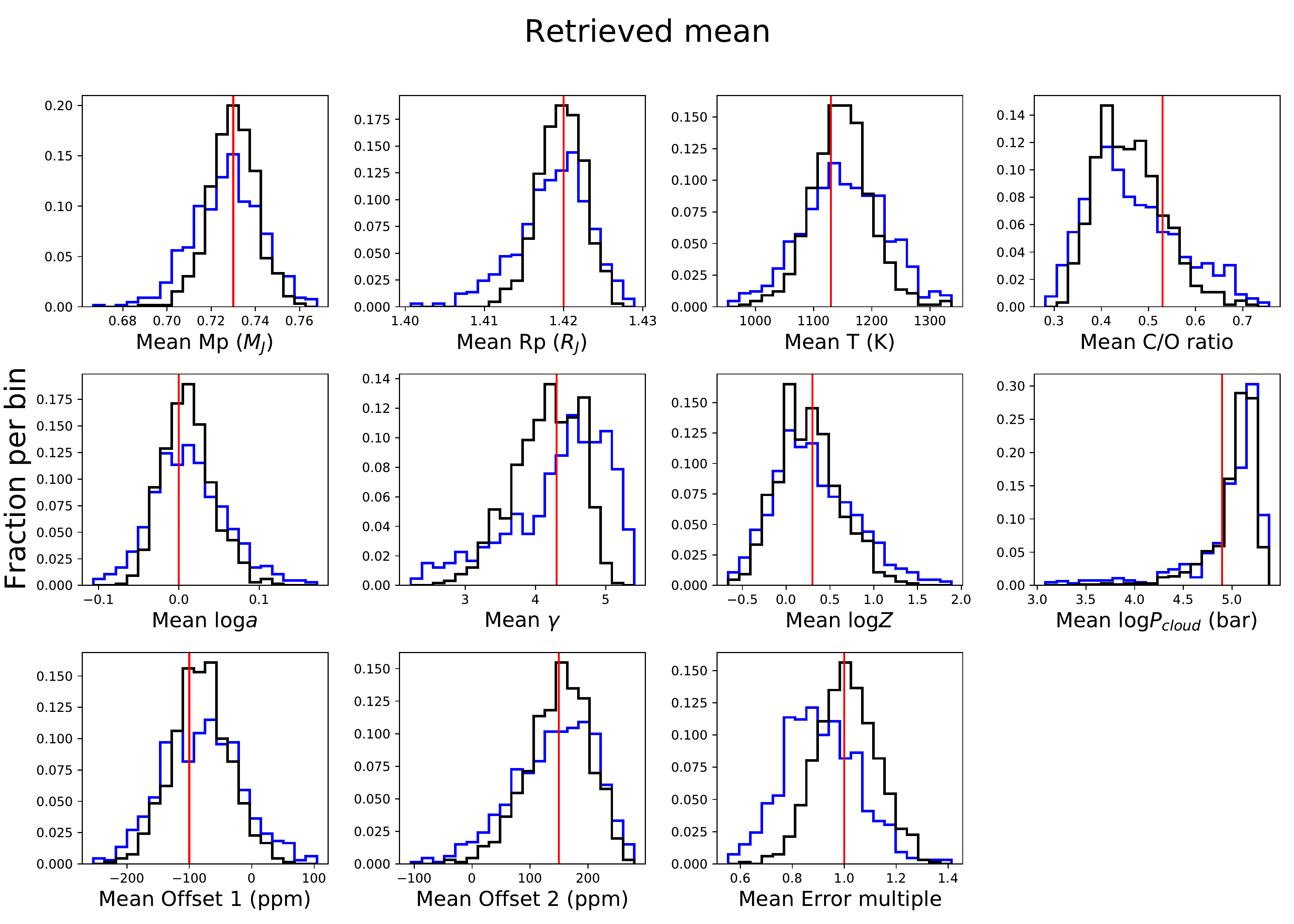}
    \caption{Same as Figure \ref{fig:error}, but for the hot Jupiter case including offsets. This figure now includes panels for both instrumental offsets in addition to the original set of retrieved parameters.}
    \label{fig:offset_error}
\end{figure*}

\begin{figure*}[!p]
    \centering
    \includegraphics[width=0.8\textwidth]{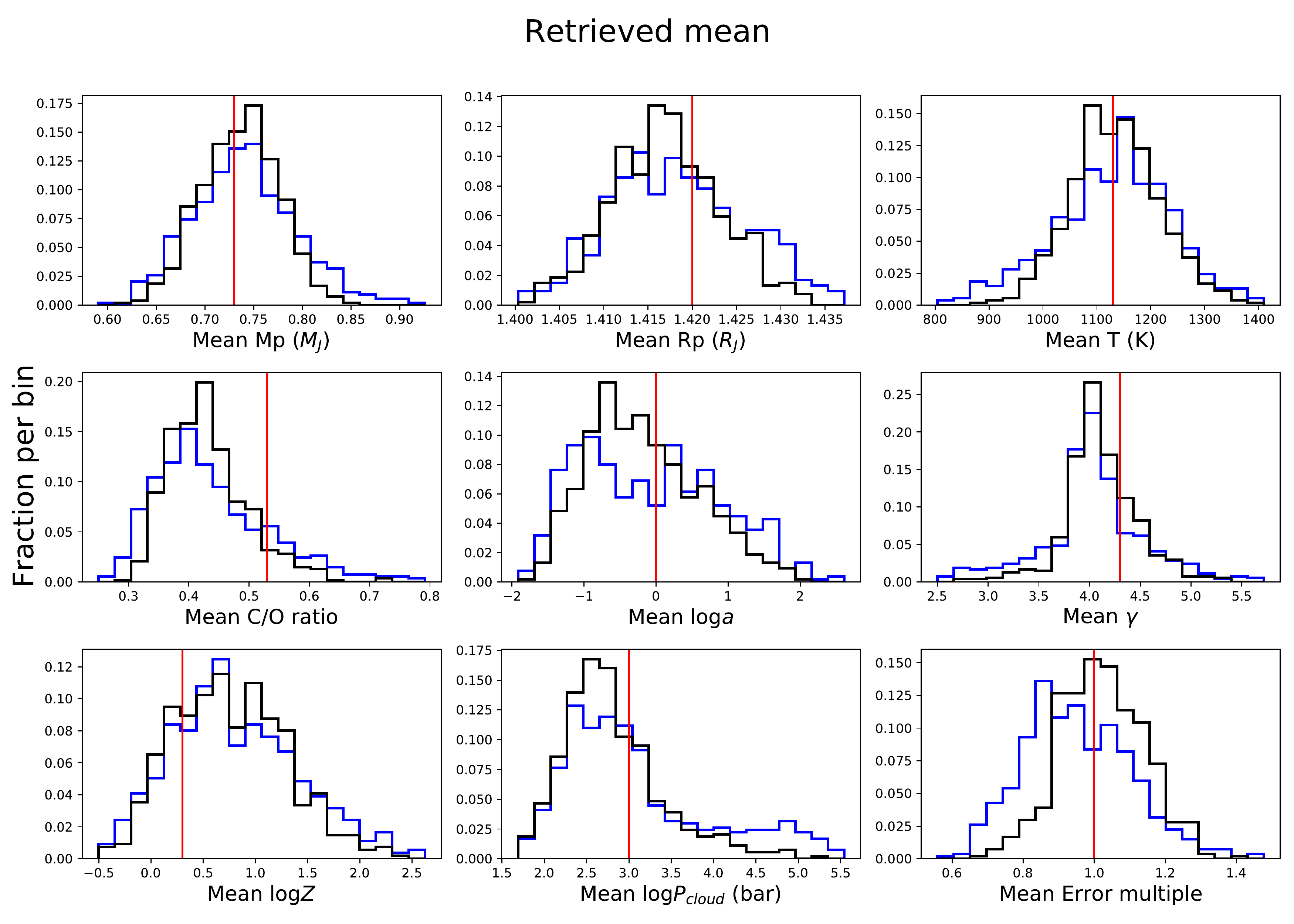}
    \caption{Same as Figure \ref{fig:mean}, but for the cloudy hot Jupiter case.}
    \label{fig:HJ_cloudy_mean}
% \end{figure*}

% \begin{figure*}[!h]
    % \centering
    \includegraphics[width=0.8\textwidth]{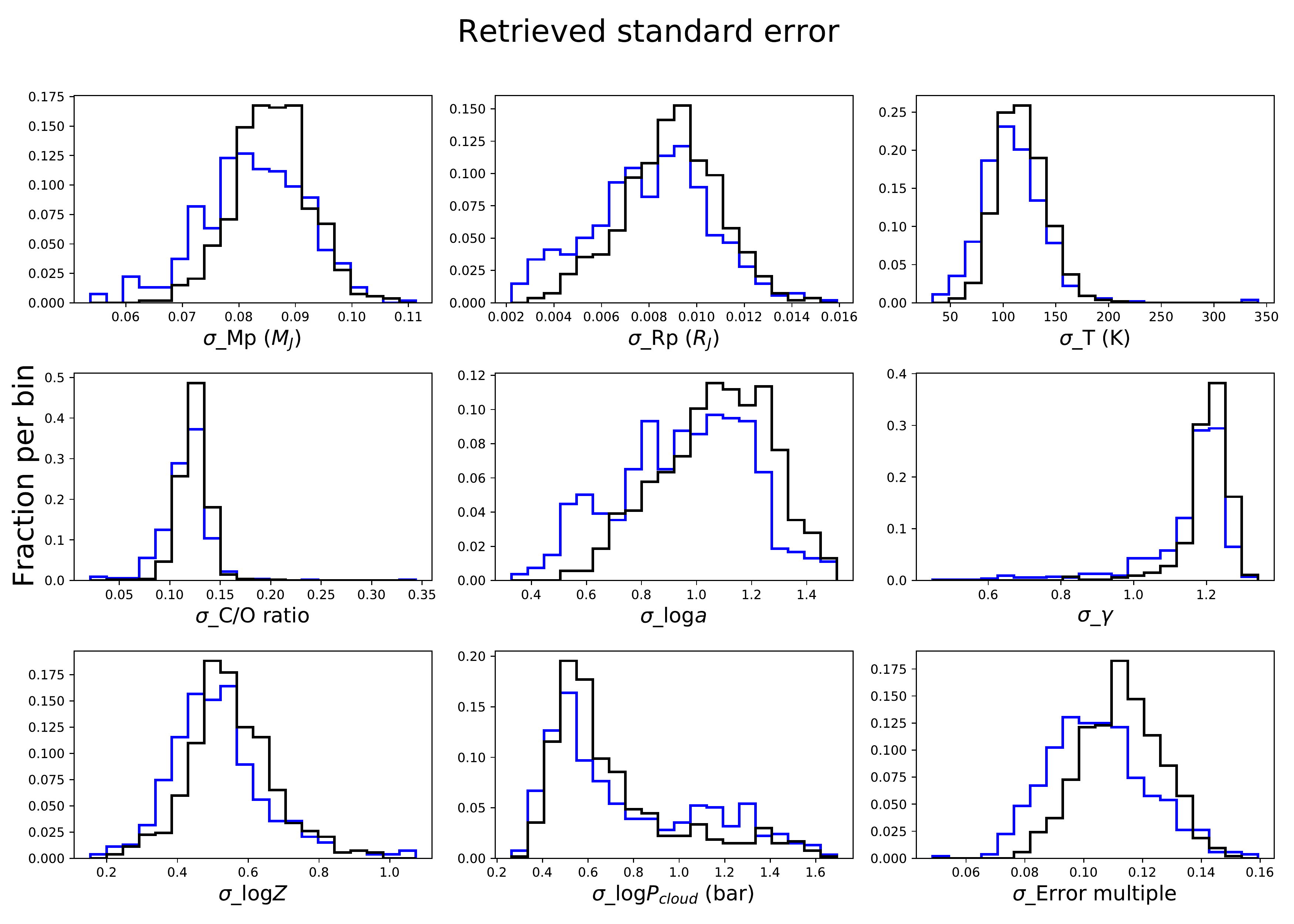}
    \caption{Same as Figure \ref{fig:error}, but for the cloudy hot Jupiter case.}
    \label{fig:HJ_cloudy_error}
\end{figure*}

\begin{figure*}[!p]
    \centering
    \includegraphics[width=0.8\textwidth]{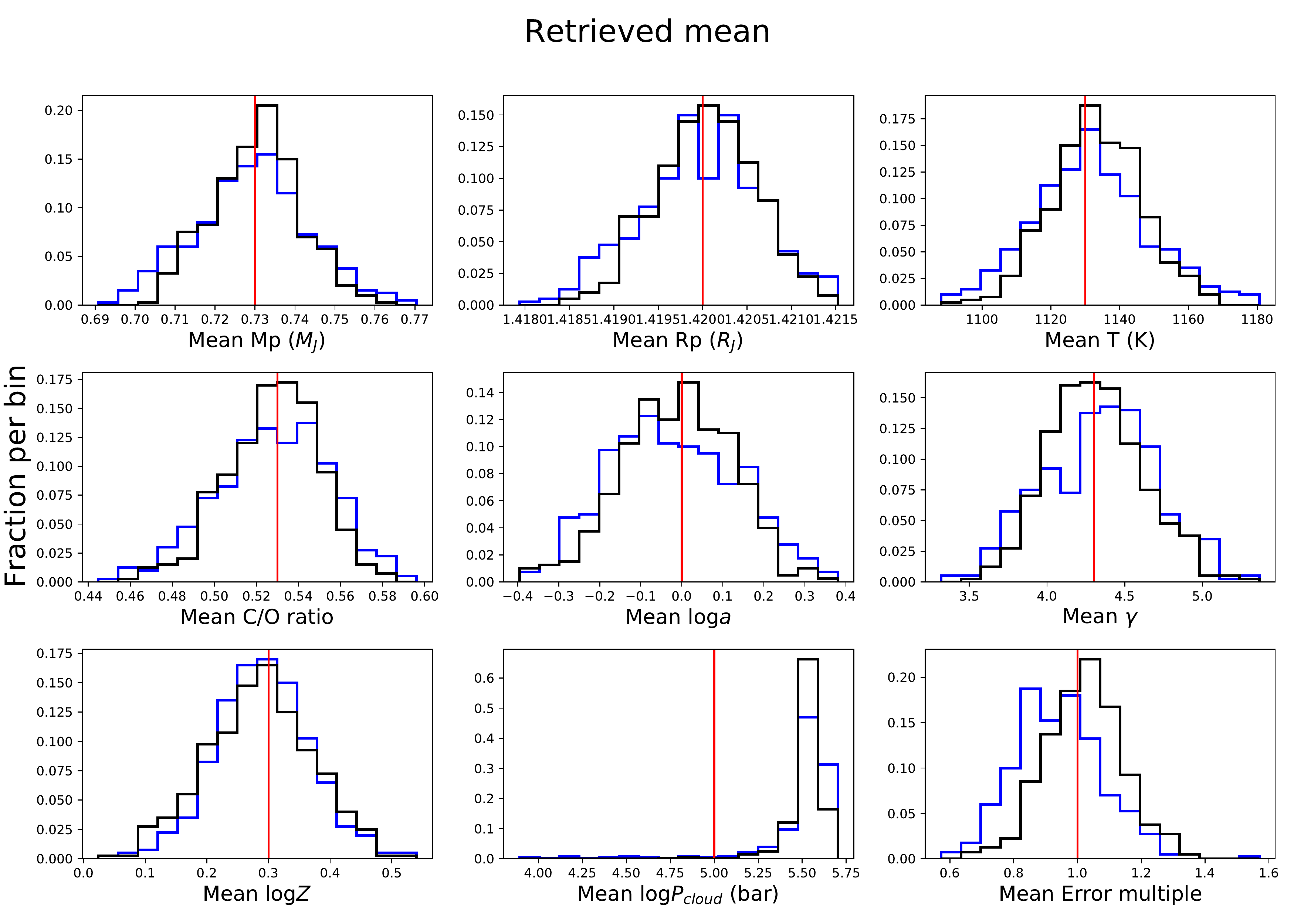}
    \caption{Same as Figure \ref{fig:mean}, but for the high-precision hot Jupiter case.}
    \label{fig:HJ_small_mean}
% \end{figure*}

% \begin{figure*}[!h]
    % \centering
    \includegraphics[width=0.8\textwidth]{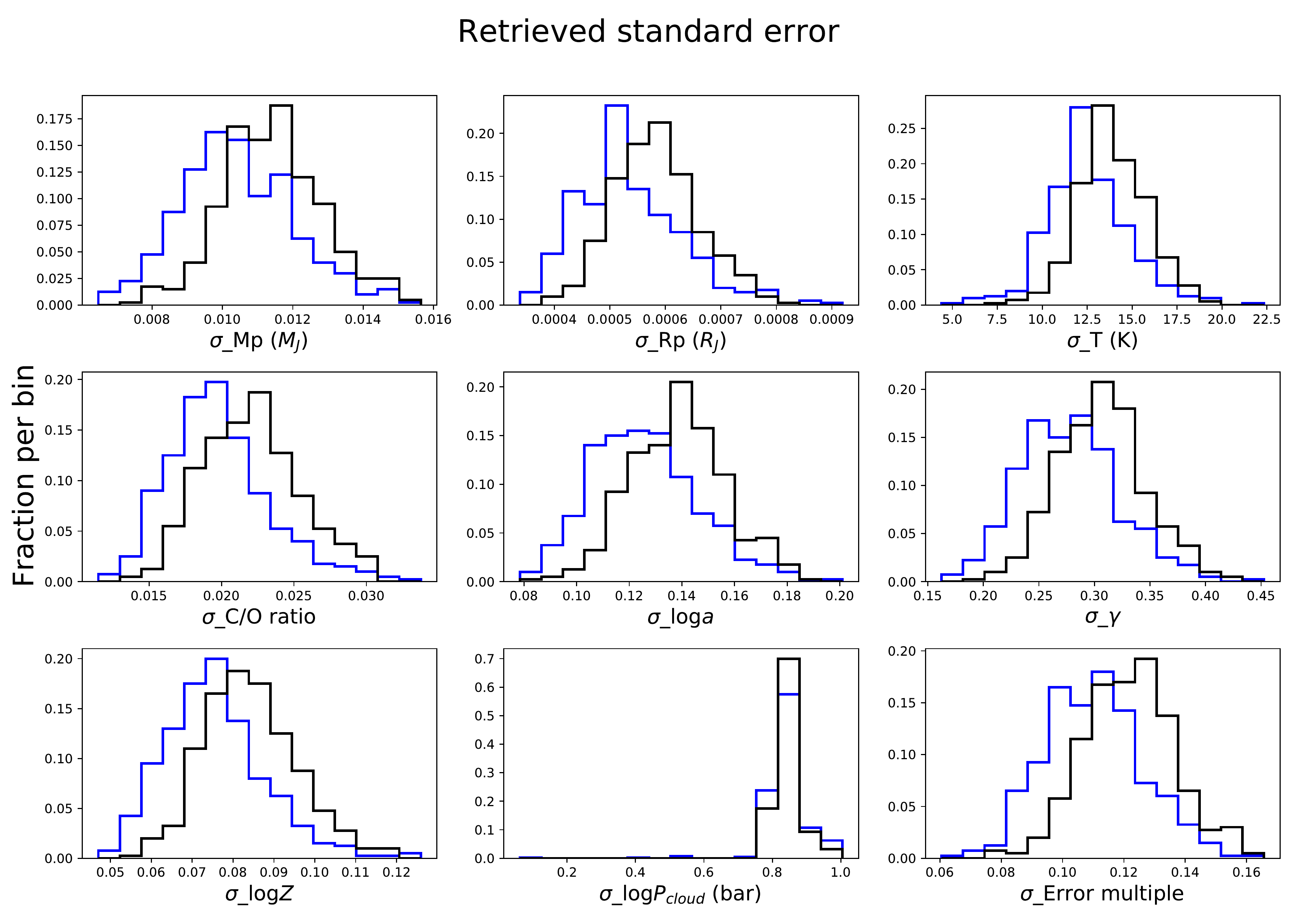}
    \caption{Same as Figure \ref{fig:error}, but for the high-precision hot Jupiter case.}
    \label{fig:HJ_small_error}
\end{figure*}

\begin{figure*}[!p]
    \centering
    \includegraphics[width=0.8\textwidth]{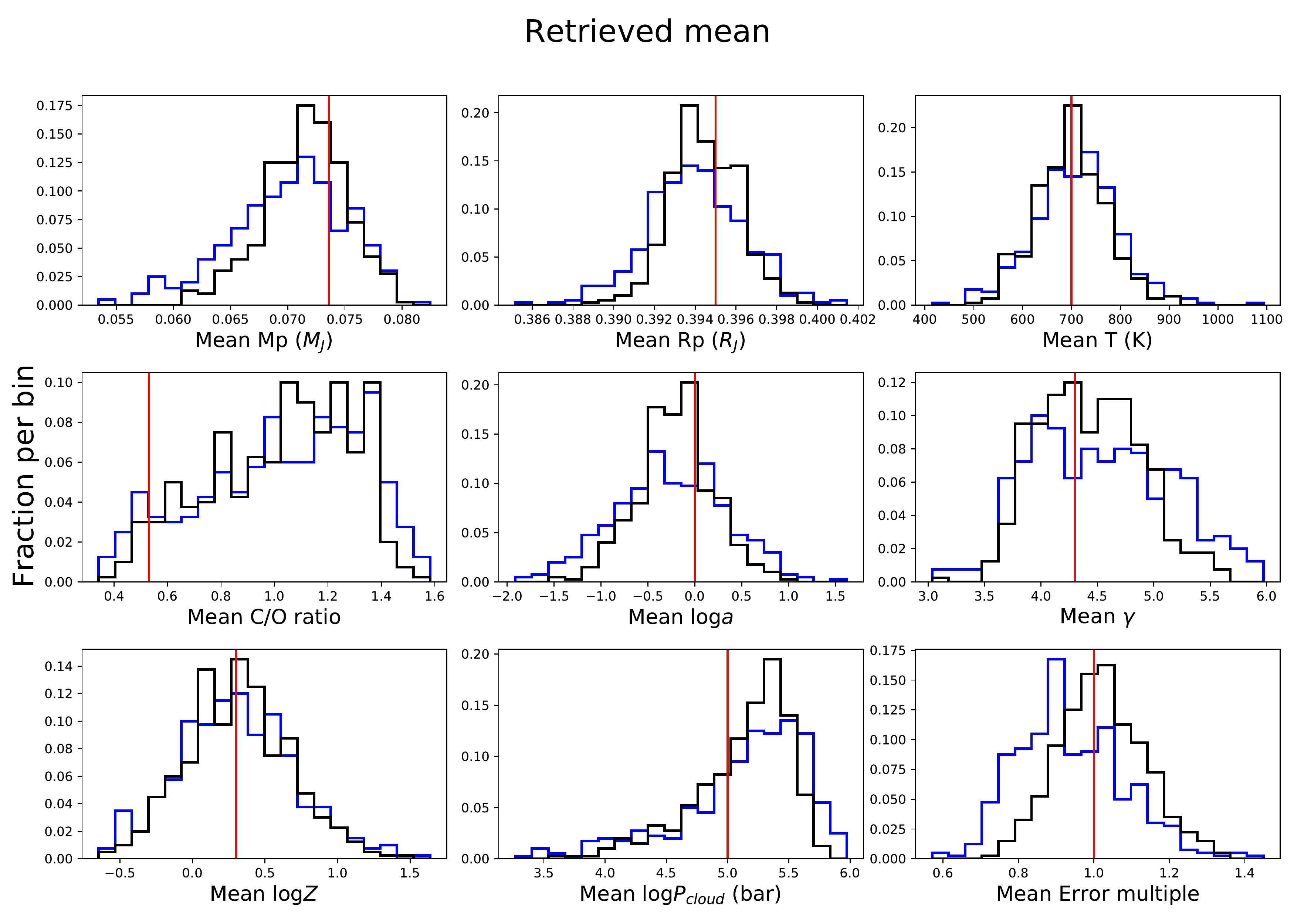}
    \caption{Same as Figure \ref{fig:mean}, but for the warm Neptune case.}
    \label{fig:SN_mean}
% \end{figure*}

% \begin{figure*}[!h]
    % \centering
    \includegraphics[width=0.8\textwidth]{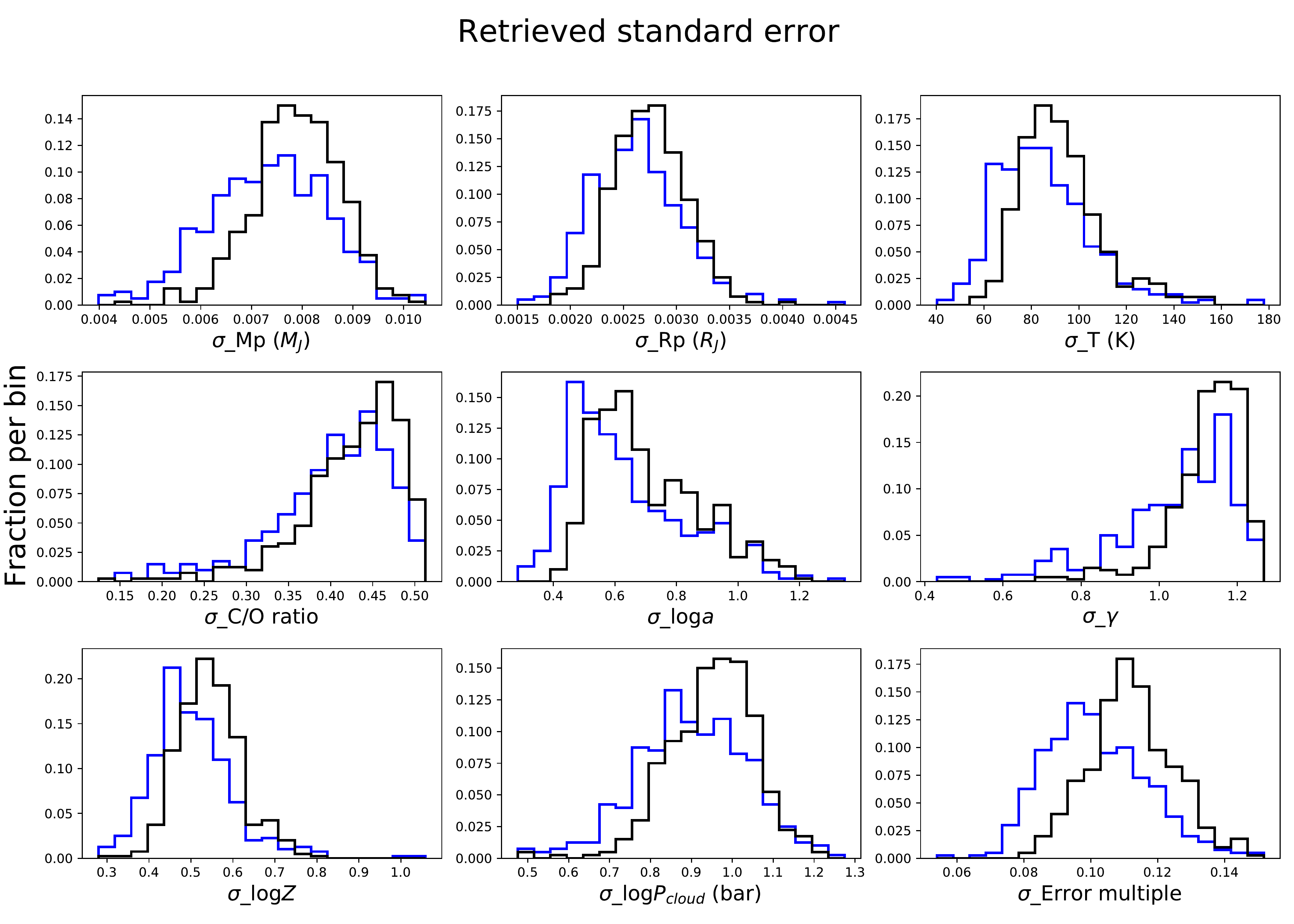}
    \caption{Same as Figure \ref{fig:error}, but for the warm Neptune case.}
    \label{fig:SN_error}
\end{figure*}

\end{document}